\documentclass[a4paper,11pt]{article}

\usepackage{authblk}
\usepackage{fancyhdr}
\usepackage{titlesec}

\usepackage{amsmath,amsfonts, amsthm}
\usepackage{bm}
\usepackage{txfonts}
\usepackage{eurosym}
\usepackage{float}
\usepackage{color}
\usepackage{typearea}
\typearea{12}
\usepackage[american]{babel}
\usepackage{csquotes}
\usepackage[style=apa]{biblatex}
\usepackage{geometry}
\usepackage{caption}
\usepackage[subrefformat=parens]{subcaption}
\usepackage{tabularx}
\usepackage{multirow}

\usepackage{booktabs}
\usepackage{here}
\usepackage[subrefformat=parens]{subcaption}
\usepackage{tabularx}
\usepackage{multirow}
\usepackage{setspace}
\usepackage{array}
\usepackage{makecell}
\usepackage{footmisc}
\usepackage{authblk}

\usepackage{graphicx}

\theoremstyle{definition}

\numberwithin{equation}{section}

\renewcommand{\arraystretch}{1}

\title{Estimating the Stochastic Discount Factor from Option Prices and Predicting the Equity Premium}

\author[1]{Kenichiro Shiraya\textsuperscript{\dag}}
\author[2]{Tomohisa Yamakami\textsuperscript{\dag}}
\author[3]{Akira Yamazaki\textsuperscript{\dag}}

\affil[1]{Graduate School of Economics, The University of Tokyo}
\affil[2]{Graduate School of Economics, The University of Tokyo \\ 
Mizuho-DL Financial Technology Co., Ltd.\footnote{The opinions expressed herein are only those of the authors and do not represent the oﬃcial views of the company.}}
\affil[3]{Graduate School of Business Administration, Hosei University}

\date{\today}

\begin{document}
\maketitle

\begingroup
\renewcommand\thefootnote{\dag}
\footnotetext[1]{Supported by Center for Advanced Research in Finance (CARF) at the University of Tokyo.}
\endgroup

\begin{abstract}
This paper proposes a stochastic discount factor (SDF) scaled by time-varying volatility.
By utilizing prices and market data implied solely from S\&P 500 options, the proposed framework recovers a stable, non-monotonic SDF that captures the pure forward-looking expectations of market participants while mitigating observation noise.
Our empirical analysis reveals that the SDF exhibits a distinctive hump on the shallow put side,
which transitions into a more clearly defined W-shape as the time to maturity increases, identifying maturity as a key factor influencing the intensity of the central hump.
We show that this structural feature can be theoretically rationalized by stochastic volatility dynamics under a constant market price of risk.
The equity premium derived from the time-varying volatility scaled SDF demonstrates superior out-of-sample predictive performance relative to existing benchmarks, such as the Martin bounds.
\end{abstract}


\section{Introduction}
The stochastic discount factor (SDF), also known as the pricing kernel, is a cornerstone of modern financial economics and provides a fundamental link between future uncertain payoffs and their current market values.
Mathematically, it represents the Radon-Nikodym derivative that transforms the physical probability measure into the risk-neutral measure.
From an equilibrium perspective, the SDF reflects the marginal rate of substitution of an intertemporal utility-maximizing investor,
thereby encoding essential information regarding market-wide risk aversion and economic expectations.

The SDF depends on the prices of all conceivable assets and is difficult to estimate.
Therefore, empirical work typically considers an SDF projected onto a specific asset.
In what follows, we also refer to this projected SDF simply as the SDF whenever the context makes the distinction clear.
The empirical literature has traditionally followed two distinct paths to recover the SDF.
The first approach treats the SDF as the ratio of the physical density to the risk-neutral density and estimates each distribution separately.
The risk-neutral density is available from option prices across multiple strikes following the method of \textcite{breeden1978prices}.
The physical distribution is estimated from the underlying asset price using techniques such as kernel density estimation or security price models like GARCH (e.g. \textcite{jackwerth2000recovering}, \textcite{rosenberg2002empirical}).
Since these methods estimating the physical distribution rely solely on historical asset prices, the resulting distribution has limited forward-looking applicability.
The second approach involves the direct estimation of the SDF through the relationship between payoffs and market values.
Parametric specifications of the SDF have been studied, for example, in \textcite{chabi2012pricing}, \textcite{song2016tale} and \textcite{bakshi2023recovery}.
These studies explored the factors contributing to the SDF by linking the parameters to risk premia.
In contrast, \textcite{bakshi2010returns} apply the mimicking portfolio approach representing the SDF as a linear combination of basis option payoffs, thereby estimating the SDF nonparametrically.
This method is particularly important because it allows the SDF to be constructed directly from market data without imposing a specific functional form.
These approaches allow to recover investors' subjective probability assessments
since a risk-neutral probability that reflect market expectations is transformed into a physical probability through the SDF.

Despite these extensive efforts, two questions remain at the heart of empirical asset pricing.
The first is {\it how to precisely estimate the functional shape of the SDF} from market data.
While the assumption of risk-averse investors implies that the SDF should be monotonically decreasing in asset returns,
empirical studies report a variety of shapes.
\textcite{cuesdeanu2016empirical} show that the SDF tends to be U-shaped during periods of high uncertainty with elevated variance risk premia,
and W-shaped during tranquil periods with low variance risk premia.
\textcite{bakshi2010returns} and \textcite{yamazaki2025subjective} theoretically demonstrate that the upward slope of the SDF in the region of deep-out-of-the-money call options is associated with the negative expected call returns observed in actual index option markets.
Furthermore, \textcite{bakshi2010returns} also argue that the presence of investors using calls as hedges can justify the U-shaped SDF.
\textcite{song2016tale} assume that the SDF depends not only on returns but also on future VIX,
while \textcite{chabi2012pricing} consider SDFs that depend on the volatility of market returns.
When the SDF depends on both returns and volatility, its projection onto returns alone can exhibit complex shapes.
\textcite{SCHREINDORFER2025104106} formulate the SDF as a polynomial of log returns scaled by volatility and demonstrate the validity of this scaling by comparing it to the unscaled case.
They also show that the level of the SDF in the negative return region differs significantly between periods of high and low volatility.
Furthermore, several studies (e.g., \textcite{bakshi2023recovery}, \textcite{christoffersen2013capturing}) assume an SDF whose shape depends on contemporaneously observed volatility,
highlighting the importance of understanding how fluctuations in volatility affect the form of the SDF.

The second open question is {\it whether the equity premium calculated using the estimated SDF possesses predictive power for future market returns}.
While the SDF theoretically determines the expected excess return, empirical evidence linking the information embedded in option-implied SDFs to the time-series of realized returns is still evolving.
The estimated SDF must be consistent with discount bond prices at each point in time.
However, few studies impose such conditions, which makes it difficult to compute the equity premium with precision.
As an alternative to using the SDF directly, \textcite{martin2017expected} and \textcite{chabi2020conditional} derive lower bounds on the equity premium from conditions and moment restrictions that the SDF must satisfy,
and argue that these bounds provide good approximations to the equity premium.

To deepen our understanding of the SDF, we aim to identify a specification that simultaneously addresses these two questions.
This paper uses such a specification by incorporating contemporaneously observed time-varying volatility as a scaling parameter.
Specifically, the SDF is expressed as a piecewise polynomial of the log return divided by volatility.
We argue that using volatility to scale log returns in this manner yields a more robust recovery of the SDF's functional form across different market regimes.
Furthermore, we construct, at each point in time, an SDF consistent with discount bond prices by applying the adjustment introduced in \textcite{yamazaki2022recovering} and \textcite{shiraya2025forecasting}, which allows for a more precise computation of the equity premium.
Our overall specification is similar to that of \textcite{SCHREINDORFER2025104106} and corroborates their findings in terms of its fit to market data.
As an extension to their work, we compare the shapes of the SDF across multiple maturities and attempt to explain the observed shapes by focusing on the SDF around the at-the-money (ATM) level.
Specifically, by considering a stochastic volatility model with a constant market price of risk, we demonstrate that a hump emerges on the shallow put side of the SDF.

Using a comprehensive dataset of S\&P 500 index options, we demonstrate that scaling the SDF by time-varying volatility yields a more precise and consistent estimate of the SDF than its non-time-varying counterpart.
Furthermore, we compute the equity premium implied by the estimated SDF and examine whether it has predictive power for realized returns.
For both the estimation and the forecasting exercises, we extract interest rates, dividend yields and underlying asset prices implied solely from S\&P 500 options for two reasons.
First, we aim to remove noise arising from discrepancies in observation timing across different data sources.
Second, relying exclusively on option prices allows us to capture the expectations of option market participants, thereby enhancing the accuracy of our empirical analysis.
Overall, this study contributes to the asset pricing literature by presenting a robust framework that reconstructs the SDF with precision from the cross-sectional information embedded in option prices, as well as a time-series forecasting approach under the physical measure that targets the risk premium.

The remainder of this paper is organized as follows.
Section 2 details our theoretical framework and estimation methodology, including the construction of the volatility-scaled SDF and the calculation of the equity premium.
Section 3 presents the empirical estimation results and out-of sample perfomance.
Section 4 discusses the resulting shape of the SDF and the evaluation of its consistency with the rational expectations hypothesis.
Section 5 concludes the paper.

\section{Methodology}

\subsection{Stochastic Discount Factor}
\label{sec:SDF}
We consider a filtered probability space 
$(\Omega,\mathcal{F},\mathbb{P},\{\mathcal{F}_{t}\}_{0\leq t\leq \tau})$.
For a contingent claim with payoff $X_{T}$ realized at time $T (<\tau)$,
we assume that the pricing operator assigning its value $v_{t}$ at time $t (<T)$ can be written as
\begin{equation}
v_{t}=\mathbb{E}_{t}\left[m_{t,T}X_{T}\right]
\label{formula:RelationBetweenSDFAndPrice}
\end{equation}
where $\mathbb{E}_{t}$ denotes conditional expectation under $\mathbb{P}$ given $\mathcal{F}_{t}$,
and $m_{t,T}$ is the random variable known as the SDF.
Let $r_{s}$ denote the risk-free rate and define the money market account as
$B_{t,T}=\exp\left\{\int_{t}^{T}r_{s}\mathrm{d}s\right\}$.
In financial economics with complete markets, the SDF can be expressed using the Radon--Nikodym derivative
with respect to the risk-neutral measure $\mathbb{Q}$:
\begin{equation}
m_{t,T}=B_{t,T}^{-1}\mathbb{E}_{T}\left[\frac{\mathrm{d}\mathbb{Q}}{\mathrm{d}\mathbb{P}}\right].
\end{equation}
Standard asset pricing texts also derive the SDF from expected utility maximization
and the optimal consumption problem:
\begin{equation}
m_{t,T}=\check{\beta}_{t,T}\frac{U^{(1)}(C_{T})}{U^{(1)}(C_{t})}
\label{formula:UtilitySDFBase}
\end{equation}
where $U(c)$ is the utility function, $C_{t}$ denotes optimal consumption at time $t$,
$\check{\beta}_{t,T}$ is a discount factor in this context, and the superscript $\cdot^{(i)}$ denotes the $i$-th derivative.

In practice, it is difficult to estimate $m_{t,T}$ since it incorporates information about all assets up to time $T$.
Empirical studies therefore often treat the SDF as a projected
random variable with respect to the asset under consideration.
If the payoff can be written as $X_{T}=f(S_{T})$ using the asset price $S_{T}$,
then applying conditional expectation with respect to $S_{T}$ and the law of iterated expectations yields
\begin{equation}
v_{t}=\mathbb{E}_{t}\left[\mathbb{E}_{t}\left[m_{t,T}\mid S_{T}\right]f(S_{T})\right].
\end{equation}
Here we define the projected SDF as
\begin{equation}
m_{t,T}(S_{T})=\mathbb{E}_{t}\left[m_{t,T}\,\middle|\,S_{T}\right],
\end{equation}
which can be treated as a function of $S_{T}$.
Let $\mathrm{d}\mathbb{P}_{t}(S_{T})$ denote the physical probability density of $S_{T}$ at time $t$,
and $\mathrm{d}\mathbb{Q}_{t}^{T}(S_{T})$ the $T$-forward risk-neutral density.
Then the projected SDF can be expressed as
\begin{equation}
m_{t,T}(S_{T})=D_{t,T}\frac{\mathrm{d}\mathbb{Q}_{t}^{T}(S_{T})}{\mathrm{d}\mathbb{P}_{t}(S_{T})},
\end{equation}
where $D_{t,T}$ is the zero-coupon bond price at time $t$ and maturing at $T$.
From the utility-based perspective, under the assumption of a power utility function with risk aversion parameter $\gamma$, and substituting asset prices for optimal consumption,
we obtains
\begin{equation}
m_{t,T}(S_{T})=\tilde{\beta}_{t,T}\exp\left\{-\gamma \log\left(\frac{S_{T}}{F_{t,T}}\right)\right\}
\label{formula:UtilitySDF}
\end{equation}
where $F_{t,T}$ is the forward price at time $t$ for maturity $T$,
and $\tilde{\beta}_{t,T}$ is some appropriate discount factor in this context.
Since this paper assumes assets pay continuous dividends, we extend existing studies to the dividend-paying case.
In general, under the assumption of risk-averse investors,
the SDF as a function of $S_{T}$ should be monotonically decreasing.
However, numerous empirical studies report non-monotonic shapes such as U-shaped,
tilde-shaped, or W-shaped forms (see, e.g., \textcite{cuesdeanu2018pricing}), indicating that simple models based on power utility
fail to match observed data.

Next, to consider the SDF in a stochastic processes,
we model the asset price $S_{t}$ under the physical measure $\mathbb{P}$ using a Brownian motion $W_{t}^{\mathbb{P}}$:
\begin{equation}
\frac{\mathrm{d}S_{t}}{S_{t}}=\mu_{t}\,dt+\sigma_{t}\,\mathrm{d}W_{t}^{\mathbb{P}}.\label{eq:StochasticProcess}
\end{equation}
Here, $\mu_{t}$ and $\sigma_{t}$ denotes the instantaneous return and volatility on the asset price,
and we allow each variable to be time-varying random variable rather than constant.
The Brownian motion under the risk-neutral measure $\mathbb{Q}$, denoted $W_{t}^{\mathbb{Q}}$,
is related to $W_{t}^{\mathbb{P}}$ by
\begin{equation}
\mathrm{d}W_{t}^{\mathbb{Q}}=\lambda_{t}\,\mathrm{d}t+\mathrm{d}W_{t}^{\mathbb{P}},
\end{equation}
where $\lambda_{t}=\frac{\mu_{t}-r_{t}+q_{t}}{\sigma_{t}}$ is referred to as the market price of risk and $q_{t}$ denote the dividend yield.
Here, we assume that the risk-free rate and the dividend yield are non-stochastic,
and that the forward spot relationship $d\log F_{t,T}=d\log S_{t} + (r_{t}-q_{t})dt$ $(t < T)$ holds.
From the relationship between the SDF and the Radon--Nikodym derivative,
together with Girsanov's theorem, we obtain
\begin{equation}
m_{t,T}=B_{t,T}^{-1}\exp\left\{-\int_{t}^{T}\lambda_{s}\,\mathrm{d}W_{s}^{\mathbb{P}}
-\frac{1}{2}\int_{t}^{T}\lambda_{s}^{2}\,\mathrm{d}s\right\},
\end{equation}
which can be rewritten as
\begin{equation}
m_{t,T}=B_{t,T}^{-1}\exp\left\{\int_{t}^{T}\frac{1}{2}\lambda_{s}\left(\lambda_{s}-\sigma_{s}\right)-\int_{t}^{T}\left(\frac{\lambda_{s}}{\sigma_{s}}-\frac{\lambda_{t}}{\sqrt{V_{t,T}}}\right)\,\mathrm{d} \log F_{s,T}-\frac{\lambda_{t}}{\sqrt{V_{t,T}}}\log\frac{S_{T}}{F_{t,T}}\right\},
\label{formula:SPSDF}
\end{equation}
where
\begin{eqnarray}
V_{t,T}=\frac{D_{t,T}^{-1}}{T-t}\mathbb{E}_{t}^{\mathbb{Q}}\left[B_{t,T}^{-1}\int_{t}^{T}\sigma_{s}^{2}ds\right]=\frac{2D_{t,T}^{-1}}{(T-t)}\left(\int_{0}^{F_{t,T}}\frac{1}{K^2}P_{t,T}(K)dK+\int_{F_{t,T}}^{\infty}\frac{1}{K^2}C_{t,T}(K)dK\right)
\end{eqnarray}
is the variance swap rate. $P_{t,T}(K)$ and $C_{t,T}(K)$ denote, respectively, the prices at time \(t\) of put and call options with maturity \(T\) and strike \(K\).

Furthermore, under the assumption of constant volatility and constant market price of risk
$(\lambda_{t}=\lambda, \sigma_{t}=\sigma, V_{t,T}=\sigma^{2})$, the second term vanishes, yielding
\begin{equation}
m_{t,T}=B_{t,T}^{-1}\exp\left\{\frac{1}{2}\lambda(\lambda-\sigma)(T-t)\right\}
\exp\left\{-\frac{\lambda}{\sigma}\log\frac{S_{T}}{F_{t,T}}\right\}.
\label{formula:BSSDF}
\end{equation}
This shows that the constant-coefficient Black--Scholes model implicitly embeds a power utility function.
Comparing \eqref{formula:UtilitySDF} with \eqref{formula:BSSDF}, we obtain the relationship
$\gamma=\tfrac{\lambda}{\sigma}$.

In actual markets, volatility is time-varying, and therefore the SDF is expected to be adjusted according to the observed volatility at each point in time.
\textcite{christoffersen2013capturing} assumes an SDF of the form $\left(\frac{S_{T}}{F_{t,T}}\right)^{-\gamma} e^{g(\sigma_{t})}$,
in which the asset price \(S_{T}\) and volatility \(\sigma_{t}\) are separable.
This model can be viewed as one in which the level of the SDF fluctuates with volatility.
However, in the stochastic process representation \eqref{formula:SPSDF} with the assumptions that the second term is negligible, the SDF can be approximated as a function with $z_{t,T}=\frac{1}{\sqrt{V_{t,T}(T-t)}}\log\left(\frac{S_{T}}{F_{t,T}}\right)$ in the exponent.
In other words, the horizontal scaling of the SDF changes with the level of volatility.
From this perspective, Bakshi (2023) considers an SDF that includes the factor of variance premium
$\exp\left\{\frac{1}{\mathbb{E}^{\mathbb{Q}}\left[B_{t,T}^{-1}\left\{\log\left(\frac{S_{T}}{S_{t}}\right)\right\}^{2}\right]}\left\{\log\left(\frac{S_{T}}{S_{t}}\right)\right\}^{2}-D_{t,T}^{-1}\right\}$,
and examines the significance of this term.
Since $\mathbb{E}^{\mathbb{Q}}\left[B_{t,T}^{-1}\left\{\log\left(\frac{S_{T}}{S_{t}}\right)\right\}^{2}\right]\simeq D_{t,T}V_{t,T}(T-t)$,
this corresponds to approximating the squared term $z_{t,T}^{2}$ that appears in the exponent of the SDF.
\textcite{SCHREINDORFER2025104106} also formulate the SDF as a polynomial of log returns divided by the volatility estimated using a heterogeneous autoregressive (HAR) model, empirically demonstrating the validity of scaling by volatility.
We apply the idea that the square root of the synthetic variance swap rate horizontally scales the SDF to a more flexible functional form and evaluate its effectiveness.
Econometrically, this scaling acts as a normalization for heteroskedasticity.
By conditioning on the level of volatility, this formulation is expected to extract option-market information contained in higher-order moments
without it being obscured by the time-varying magnitude of market variance.

Viewed from another angle, the SDF must satisfy the restriction that $\mathbb{E}_{t}[m_{t,T}]$ equals the price at time $t$ of a zero-coupon bond maturing at $T$.
Suppose that $m_{t}(S_{T})$ can be decomposed into a component independent of the state at $T$ and a component dependent on it, i.e.,
\begin{eqnarray}
m_{t}(S_{T})=\beta_{t,T}n_{t}(S_{T}).\label{formula:AdjustedSDF}
\end{eqnarray}
By considering the payoff \(\frac{1}{m_{t,T}(S_{T})}\) under this assumption, we obtain
\begin{equation}
\mathbb{E}\left[m_{t,T}\frac{1}{m_{t,T}(S_{T})}\right]
=\mathbb{E}\left[\mathbb{E}_{t}\left[m_{t,T}\mid S_{T}\right]\frac{1}{m_{t,T}(S_{T})}\right]=1.
\end{equation}
Under the risk-neutral measure, the pricing formula also requires that the value of this claim equals one:
\begin{equation}
\mathbb{E}^{\mathbb{Q}}\left[B_{t,T}^{-1}\frac{1}{m_{t,T}(S_{T})}\right]=1.
\end{equation}
For a contingent claim with twice-differentiable payoff $g(S_{T})$, \textcite{Carr01012001} give
\begin{equation}
\mathbb{E}^{\mathbb{Q}}\left[B_{t,T}^{-1}g(S_{T})\right]
=D_{t,T}g(F_{t,T})
+\int_{0}^{F_{t,T}}g^{(2)}(K)P_{t,T}(K)\,\mathrm{d}K
+\int_{F_{t,T}}^{\infty}g^{(2)}(K)C_{t,T}(K)\,\mathrm{d}K.
\label{formula:Carr}
\end{equation}
Combining these results yields
\begin{equation}
\beta_{t,T}=\frac{D_{t,T}}{n_{t}(F_{t,T})}
+\int_{0}^{F_{t,T}}\left(\frac{1}{n_{t}(K)}\right)^{(2)}P_{t,T}(K)\,\mathrm{d}K
+\int_{F_{t,T}}^{\infty}\left(\frac{1}{n_{t}(K)}\right)^{(2)}C_{t,T}(K)\,\mathrm{d}K.
\label{formula:SDFBeta}
\end{equation}
\textcite{yamazaki2022recovering} and \textcite{shiraya2025forecasting} construct consistent SDF using this formula.
\textcite{SCHREINDORFER2025104106} also applies similar adjustment to the SDF.
When computing expectations under the empirical measure using the SDF, it is crucial to ensure consistency with the interest rate.  
Accordingly, we impose this alignment to guarantee that the valuation remains coherent with the interest rates. See also Section \ref{subsec:EP}

\subsection{Estimation Methods of Stochastic Discount Factor}

There are two main strands of research on estimating the SDF from market data.  
The first approach estimates the physical probability distribution solely from the underlying asset price and then takes the ratio with the risk-neutral distribution.  
The physical distribution is typically estimated using kernel density methods or GARCH-type models, while the risk-neutral distribution is obtained from option prices across multiple strikes following \textcite{breeden1978prices}.
However, since the physical distribution in this approach is inferred only from asset prices, the resulting statistics are limited to indicators based solely on asset price information.

The second approach directly estimates the SDF by exploiting the pricing relation \eqref{formula:RelationBetweenSDFAndPrice} between payoffs and security prices, using option prices and their realized payoffs.
Let $\mathbf{f}_{t}=(f_{1,t},\cdots,f_{M,t})'$ denote the vector of payoffs written in terms of $S_{T}$, and let $\mathbf{v}_{t}=(v_{1,t},\cdots,v_{M,t})'$ denote the corresponding vector of prices.  
Then the condition
\begin{eqnarray}
\mathbb{E}\left[m_{t,T}\mathbf{f}_{t}-\mathbf{v}_{t}\right]=0
\end{eqnarray}
must hold.  
Defining the SDF with parameter $\theta\in\Theta$ as $m_{t,T,\theta}$, and setting $\mathbf{h}_{t}(\theta)=m_{t,T,\theta}\mathbf{f}_{t}-\mathbf{v}_{t}$,
we obtain the moment condition
\begin{equation}
\mathbb{E}\left[\mathbf{h}_{t}(\theta)\right]=0.
\label{formula:GMMMomentCondition}
\end{equation}
Estimating the parameter $\theta$ that satisfies this condition yields the SDF.  
This method does not require explicit estimation of either the physical or risk-neutral distributions, and thus remains applicable even when these distributions are time-dependent.

A specific estimation technique is to apply the generalized method of moments (GMM) using \eqref{formula:GMMMomentCondition} as the moment condition.  
This involves minimizing the squared norm with respect to a positive definite weighting matrix $W$:
\begin{equation}
\left\|\mathbb{E}\left[\mathbf{h}_{t}(\theta)\right]\right\|_{W}^{2}
=\mathbb{E}\left[\mathbf{h}_{t}(\theta)\right]'W\mathbb{E}\left[\mathbf{h}_{t}(\theta)\right].
\end{equation}
Replacing expectations with time averages over sample points $t=t_{1},\cdots,t_{N}$, the estimator is given by
\begin{equation}
\hat{\theta}=\arg\min_{\theta\in\Theta}
\left(\frac{1}{N}\sum_{n=1}^{N}\mathbf{h}_{t_{n}}(\theta)\right)'\hat{W}
\left(\frac{1}{N}\sum_{n=1}^{N}\mathbf{h}_{t_{n}}(\theta)\right),
\end{equation}
where $\hat{W}$ is an estimate of the weighting matrix $W$ computed from available information.
GMM estimators are consistent and asymptotically normal.  
That is, under suitable regularity conditions, for the true parameter $\theta_{0}$,
the estimator $\hat{\theta}$ satisfies
\begin{align}
G&=\mathbb{E}\left[\nabla_{\theta}\mathbf{h}_{t}(\theta)\right],\\
\Omega&=\mathbb{E}\left[\mathbf{h}_{t}(\theta)\mathbf{h}_{t}'(\theta)\right],\\
C&=G'WG,\\
D&=G'W\Omega W'G,\\
\Sigma&=C^{-1}DC^{-1},\\
\sqrt{N}\left(\hat{\theta}-\theta_{0}\right)&\xrightarrow{\mathrm{d}}N\left(0,\Sigma\right).
\label{formula:GMMAsymptoticNormal}
\end{align}
Here $N(\mu,\Sigma)$ denotes the normal distribution with mean $\mu$ and covariance matrix $\Sigma$.
Asymptotic normality holds for any positive definite weighting matrix $W$,
but efficiency is achieved when $W\propto\Omega^{-1}$.
However, when using a weighting matrix $W$ that yields efficient estimation, difficulties arise due to noise in the SDF.
To address this, an alternative weighting scheme is used based on the second moments of payoffs \parencite{hansen2009pricing, chabi2012pricing}.
This approach is closely related to considering the minimization of the Hansen-Jagannathan (HJ) distance for moment condition of the GMM framework.

The HJ distance, introduced by \textcite{hansen1997assessing}, measures the distance between the true SDF $m$ and a proxy SDF $m(\theta)$.
We define $\mathbf{f}_{t}$ and $\mathbf{v}_{t}$ be a set of basis payoffs and their prices.
Therefore, admissible SDFs $m\in\mathcal{M}$ satisfy $\mathbb{E}\left[m\mathbf{f}_{t}\right]=\mathbb{E}\left[\mathbf{v}_{t}\right]$.
The HJ distance is then defined as
\begin{eqnarray}
\delta=\min_{m\in\mathcal{M}}\left|m(\theta)-m\right|,
\end{eqnarray}
and can be expressed as
\begin{equation}
\delta^{2}=\mathbb{E}\left[m(\theta)\mathbf{f}_{t}-\mathbf{v}_{t}\right]'
\mathbb{E}\left[\mathbf{f}_{t}\mathbf{f}_{t}'\right]^{-1}
\mathbb{E}\left[m(\theta)\mathbf{f}_{t}-\mathbf{v}_{t}\right].
\label{formula:HJDistance}
\end{equation}
Thus, minimizing the HJ distance is equivalent to GMM estimation with the weighting matrix $W=\left(\mathbb{E}\left[\mathbf{f}_{t}\mathbf{f}_{t}'\right]\right)^{-1}$.

Next, we discuss the construction of confidence intervals for the estimated function.
Let the function $f_{\theta}(x)=\theta'\mathbf{\psi}(x)$ be represented as a linear combination of the parameter vector $\theta$ and the basis function vector $\mathbf{\psi}(x)$.  
Since the GMM estimator $\hat{\theta}$ is asymptotically normally distributed by \eqref{formula:GMMAsymptoticNormal}, the function $f_{\hat{\theta}}(x)$ is also normally distributed at each point.
Its variance is given by
\begin{equation}
\begin{aligned}
\mathbb{V}\left[f_{\hat{\theta}}(x)\right]
&=\mathbf{\psi}'(x)\,\mathbb{E}\left[(\hat{\theta}-\theta_{0})(\hat{\theta}-\theta_{0})'\right]\mathbf{\psi}(x) \\
&=\frac{1}{N}\mathbf{\psi}'(x)\Sigma\mathbf{\psi}(x).\label{formula:FuncVariance}
\end{aligned}
\end{equation}

\subsection{Equity Premium}
\label{subsec:EP}
The equity premium (EP) is defined as the expected excess return relative to the risk-free rate.  
There are several variations of its definition, such as those based on log returns or gross returns:
\begin{equation}
\mathrm{EPL}_{t,T}=\mathbb{E}_{t}\left[\log\left(\frac{S_{T}}{S_{t}}\right)+(q-r)(T-t)\right]
=\mathbb{E}_{t}\left[\log\left(\frac{S_{T}}{F_{t,T}}\right)\right],
\end{equation}
\begin{equation}
\mathrm{EP}_{t,T}=\mathbb{E}_{t}\left[\frac{S_{T}}{S_{t}e^{-q(T-t)}}\right]-D_{t,T}^{-1}
=D_{t,T}^{-1}\mathbb{E}_{t}\left[\frac{S_{T}}{F_{t,T}}-1\right].
\end{equation}

In general, the expectation of a function $g(S_{T})$, i.e. $\mathbb{E}_{t}[g(S_{T})]$, can be computed using the SDF and Carr--Madan's spanning formula \eqref{formula:Carr}:
\begin{equation}
\begin{aligned}
\mathbb{E}_{t}[g(S_{T})]
&=\mathbb{E}_{t}\left[m_{t,T}\frac{g(S_{T})}{m_{t,T}(S_{T})}\right]
=\mathbb{E}_{t}^{\mathbb{Q}}\left[B_{t,T}^{-1}\frac{g(S_{T})}{m_{t,T}(S_{T})}\right] \\
&=D_{t,T}\frac{g(F_{t,T})}{m_{t,T}(F_{t,T})}
+\int_{0}^{F_{t,T}}\left(\frac{g(K)}{m_{t,T}(K)}\right)^{(2)}P_{t,T}(K)\,dK
+\int_{F_{t,T}}^{\infty}\left(\frac{g(K)}{m_{t,T}(K)}\right)^{(2)}C_{t,T}(K)\,dK.
\end{aligned}
\label{formula:PhysicalExpectation}
\end{equation}
By setting $g(S_{T})=1$, this simplifies to
\begin{equation}
D_{t,T}=m_{t,T}(F_{t,T})\left\{1
-\int_{0}^{F_{t,T}}\left(\frac{1}{m_{t,T}(K)}\right)^{(2)}P_{t,T}(K)\,dK
-\int_{F_{t,T}}^{\infty}\left(\frac{1}{m_{t,T}(K)}\right)^{(2)}C_{t,T}(K)\,dK\right\},
\label{formula:DFSDFOptionRelation}
\end{equation}
which implies that unless the SDF satisfies this condition,
the expectation computed via \eqref{formula:PhysicalExpectation} will contain inconsistencies.
In contrast, the SDF expressed as \eqref{formula:AdjustedSDF} and \eqref{formula:SDFBeta} satisfies condition \eqref{formula:DFSDFOptionRelation} for any discount bond price.
Therefore, when an estimated SDF $n_{t}(S_{T})$ which is not consistent with \eqref{formula:DFSDFOptionRelation} is obtained, we use the adjusted SDF
\begin{eqnarray}
m_{t,T}(S_{T})=\beta_{t,T}n_{t}(S_{T}),
\end{eqnarray}
with $\beta_{t,T}$ determined by \eqref{formula:SDFBeta}, in the computation of expectations via \eqref{formula:PhysicalExpectation}.

Numerous studies have investigated the estimation and bounds of the EP.
\textcite{hansen1991implications} derive an upper bound for the EP from the mathematical properties of the SDF:
\begin{equation}
\mathrm{EP}_{t,T}\leq D_{t,T}^{-1}\sqrt{\mathbb{V}_{t}\left[m_{t,T}(S_{T})\right]}
\sqrt{\mathbb{V}_{t}\left[\frac{S_{T}D_{t,T}}{F_{t,T}}\right]}.
\end{equation}
In the original formulation, projected SDF $m_{t,T}(S_{T})$ is replaced by SDF $m_{t,T}$, but considering conditional expectations with respect to $S_{T}$ yields this tighter bound.
The computation of this bound requires expectations under the physical probability measure.

\textcite{martin2017expected} derive a lower bound for the EP, known as the M-Bound, under assumption of the negative correlation condition (NCC).
He also claimed that this bound can be regarded as a precise estimate of the EP.
Here, we extend the derivation of the M-Bound to incorporate dividend yields.
Under the $T$-forward risk-neutral measure $\mathbb{Q}^{T}$, we have
\begin{equation}
\begin{aligned}
\mathbb{V}_{t}^{\mathbb{Q}^{T}}\left[\frac{S_{T}}{F_{t,T}}\right]
&=\mathbb{E}_{t}^{\mathbb{Q}^{T}}\left[\left(\frac{S_{T}}{F_{t,T}}\right)^{2}\right]
-\left(\mathbb{E}_{t}^{\mathbb{Q}^{T}}\left[\frac{S_{T}}{F_{t,T}}\right]\right)^{2} \\
&=D_{t,T}^{-1}\mathbb{E}_{t}\left[m_{t,T}\left(\frac{S_{T}}{F_{t,T}}\right)^{2}\right]-1.
\end{aligned}
\end{equation}
Using this relation, the EP can be expressed as
\begin{equation}
\begin{aligned}
\mathrm{EP}_{t,T}
&=D_{t,T}^{-1}\left(D_{t,T}^{-1}\mathbb{E}_{t}\left[m_{t,T}\left(\frac{S_{T}}{F_{t,T}}\right)^{2}\right]-1\right) \\
&\quad -D_{t,T}^{-1}\left(D_{t,T}^{-1}\mathbb{E}_{t}\left[m_{t,T}\left(\frac{S_{T}}{F_{t,T}}\right)^{2}\right]
-\mathbb{E}_{t}\left[\frac{S_{T}}{F_{t,T}}\right]\right) \\
&=D_{t,T}^{-1}\mathbb{V}_{t}^{\mathbb{Q}^{T}}\left[\frac{S_{T}}{F_{t,T}}\right]
-D_{t,T}^{-2}\mathbb{C}_{t}\left[m_{t,T}\frac{S_{T}}{F_{t,T}},\frac{S_{T}}{F_{t,T}}\right] \\
&\geq D_{t,T}^{-1}\mathbb{V}_{t}^{\mathbb{Q}^{T}}\left[\frac{S_{T}}{F_{t,T}}\right],
\end{aligned}
\end{equation}
where $\mathbb{C}_{t}$ is the operator to calculate covariance, and the final inequality follows from the NCC assumption,
$\mathbb{C}_{t}\left[m_{t,T}\frac{S_{T}}{F_{t,T}},\frac{S_{T}}{F_{t,T}}\right]\leq 0$.
Expectations of $f(S_{T})$ under the $T$-forward risk-neutral measure can be computed using the measure transformation
\begin{eqnarray}
\mathbb{E}^{\mathbb{Q}}\left[B_{t,T}^{-1}f(S_{T})\right]
=D_{t,T}\mathbb{E}^{\mathbb{Q}^{T}}\left[f(S_{T})\right],
\end{eqnarray}
and thus can be evaluated via Carr--Madan's spanning formula \eqref{formula:Carr}.  
Therefore, M-Bound can be obtained without explicit modeling of the SDF.

\textcite{chabi2020conditional} extend the idea of \textcite{martin2017expected} by deriving bounds and estimators for the EP using moments up to order four under the risk-neutral measure.
In this paper, we employ the lower bound formula (CYL-Bound) that incorporates continuous dividends as the estimator of the EP:
\begin{equation}
\mathbb{E}_{t}\left[\frac{S_{T}}{S_{t}e^{-\int_{t}^{T}q_{s}ds}}-D_{t,T}^{-1}\right]
\geq \frac{D_{t,T}M_{t,T}^{(2)}-D_{t,T}^{2}M_{t,T}^{(3)}-D_{t,T}^{3}M_{t,T}^{(4)}}
{1-D_{t,T}^{2}M_{t,T}^{(2)}+D_{t,T}^{3}M_{t,T}^{(3)}}.
\end{equation}
Here, the moments are defined as
\begin{equation}
M_{t,T}^{(k)}=\mathbb{E}_{t}^{\mathbb{Q}}\left[\left(\frac{S_{T}}{S_{t}e^{-\int_{t}^{T}q_{s}ds}}-D_{t,T}^{-1}\right)^{k}\right]=D_{t,T}^{-k}\,\mathbb{E}^{\mathbb{Q}}\left[\left(\frac{S_{T}}{F_{t,T}}-1\right)^{k}\right],
\end{equation}
and further expressed in terms of option prices:
\begin{equation}
M_{t,T}^{(k)}=\frac{k(k-1)D_{t,T}^{-(k+1)}}{F_{t,T}^{2}}
\left\{\int_{0}^{F_{t,T}}\left(\frac{K}{F_{t,T}}-1\right)^{k-2}P_{t,T}(K)\,dK
+\int_{F_{t,T}}^{\infty}\left(\frac{K}{F_{t,T}}-1\right)^{k-2}C_{t,T}(K)\,dK\right\}.
\end{equation}
This formula is straightforwardly derived from the expressions in \textcite{chabi2020conditional} work by replacing \(\tfrac{S_{T}}{S_{t}}\) with \(\tfrac{S_{T}}{S_{t}e^{-\int_{t}^{T}q_{s}ds}}\).
This bound is derived by imposing the condition that the odd moments of returns under the risk-neutral measure are negative,
and that the parameter restrictions $\frac{1}{\tau}\geq 1, \frac{1-\rho}{\tau^{2}}\leq -1, \frac{1-2\rho+\kappa}{\tau^{3}}\geq 1$.
Here, $\tau=\tfrac{1}{\mathcal{A}}, \rho=\tfrac{\mathcal{P}}{\mathcal{A}}, \kappa=\tfrac{\mathcal{P}\mathcal{T}}{6\mathcal{A}^{2}}$ denote risk-tolerance, skewness-tolerance, kurtosis-tolerance
and $\mathcal{A}, \mathcal{P}, \mathcal{T}$ are relative risk aversion, relative prudence, relative temperance.

While both the M-bound and the CYL-bound provide EP estimates without assuming a specific functional form for the SDF,
they do not simultaneously achieve the two primary objectives of this study:
the precise estimation of the SDF's shape and the predictability of the EP.

\section{Estimation of Stochastic Discount Factor and Equity Premium}
\label{sec:EstimaionOfSDF}
\subsection{Data}

For the numerical analysis, we focus on options on the S\&P 500 index.  
We employ a dataset obtained from Option Metrics
\footnote{Supported by Center for Advanced Research in Finance (CARF) at the University of Tokyo.}
covering the period from January 1996 to February 2023.
The dataset contains daily prices of the S\&P 500 index as well as bid and ask quotes for options.

Estimating the SDF necessitates the underlying asset price, the risk-free rate, and the dividend yield in addition to option prices.
In this study, these parameters are extracted directly from option prices themselves without relying on external data sources.
There are two primary motivations for this approach.
The first motivation is to maintain internal consistency within the model and to eliminate issues arising from lack of synchronization.
For example, the underlying asset market does not always move in perfect lockstep with the options market; temporal lags and pricing discrepancies frequently occur.
Moreover, the effective interest rates and dividend yields perceived by option market participants can differ substantially from those observed in the government bond market or in dividend derivative markets.
By relying on implied parameters, the estimated SDF remains internally consistent with the specific market environment under analysis, preventing distortions that may arise from inconsistencies across data sources.
The second motivation is that using information extracted solely from option prices enables the most direct and unadulterated measurement of expectations held by option market participants.
By empirically focusing on this particular investor group, the estimated SDF is expected to be largely free from the heterogeneous and extraneous noise embedded in the broader market.
This approach offers the possibility of uncovering distinctive features of the SDF that reflect the unique risk preferences, funding constraints, and leverage constraints faced by these sophisticated market participants.
Ultimately, this purified perspective is expected to yield a more refined and granular understanding of how risk is priced in derivative markets.
Details of the estimation procedure are provided in the Appendix \ref{sec:RateEstimation}.

To reduce data errors and enhance the reliability of the numerical analysis of estimating SDF and calculationg EP, we apply several screening criteria to the option data.  
First, observations where the ask price is less than twice the bid price are excluded.  
We also remove contracts with zero open interest or trading volume, as well as those with fewer than ten days to maturity.  
To satisfy no-arbitrage conditions, strikes that violate monotonicity of option prices across strikes are excluded.
For the monotonicity verification procedure, we follow Step 2-1 of \textcite{fukasawa2011model}.

When applying Carr--Madan's spanning formula \eqref{formula:Carr}, the numerical integration is carried out as follows.  
The integration intervals are set to $[F_{t,T}e^{-4\sigma_{p,t,T}\sqrt{T-t}},F_{t,T}]$ for puts and $[F_{t,T},F_{t,T}e^{4\sigma_{c,t,T}\sqrt{T-t}}]$ for calls.
The values $\sigma_{p,t,T}$ and $\sigma_{c,t,T}$ are defined as the averages of the implied volatility at $F_{t,T}$ and the available deepest OTM volatility on the put and call sides, respectively.
Each interval is divided into 1000, and the trapezoidal rule is applied.
For the second derivative $g^{(2)}(S_{T})$, we use a finite-difference scheme based on the grid points:
\begin{eqnarray}
g^{(2)}(S_{T}) \approx \frac{g(S_{T}+\Delta)-2g(S_{T})+g(S_{T}-\Delta)}{\Delta^{2}},
\end{eqnarray}
where $\Delta$ is the grid interval.
Available option prices $P_{t,T}(K)$ and $C_{t,T}(K)$ are first converted into implied volatilities, which are then linearly interpolated in the volatility domain, and subsequently reconverted into option prices.
For intervals without data, extrapolation is performed by holding the endpoint implied volatility constant.  
This numerical integration scheme provides a reasonable approximation as long as $f$, $C_{t,T}$, and $P_{t,T}$ are once differentiable except finite points, even if $g$ is not twice differentiable.  
Further details are given in the Appendix \ref{sec:NumericalIntegrationScheme}.

Here, $B_{t,t+\tau}$ is approximated by the risk-free rate observed at time $t$ for maturity $t+\tau$.
This calculation is performed for each reference date and maturity.
Furthermore, using data up to December 18, 2009 (option maturity date) as the in-sample period, we define $\bar{\sigma}_{\tau}^{\ast}$
as the average of squared root of variance swap rates across this period for each maturity $\tau$.

We then extract the dataset used for SDF estimation and examine its statistical properties.  
Reference dates are defined as the days on which monthly option maturity date.
We consider maturities of 30, 60, 90, 120, 180, 270, and 360 days.  
For each maturity, strikes are determined based on deviations measured in units of the standard deviation of log prices.  
Specifically, let $z_{1},\cdots,z_{6}=-1.0,-0.6,-0.2,0.2,0.6,1.0$ and define the strikes as
\begin{equation}
K_{i,t,\tau}=\left\lceil F_{t,\tau}e^{z_{i}\sigma_{t,t+\tau}\sqrt{\tau}}\right\rceil,\quad i=1,\cdots,6,
\end{equation}
where $\lceil\cdot\rceil$ denotes the function that selects the nearest available strike in the dataset.

For $\sigma_{t,t+\tau}$, we consider two specifications:
(i) using the same value $\bar{\sigma}_{\tau}^{\ast}$ across the entire period (constant volatility scaled (CVS) strike set), and  
(ii) using time-varying values $\sqrt{V_{t,T}}$ at each date (time-varying volatility scaled (TVS) strike set).
When considering an SDF that is scaled by time-varying volatility, it is natural to estimate it using option contracts whose strikes are themselves scaled by the same time-varying volatility.
Accordingly, these strike sets are specified and used for estimating each SDF model.

\begin{table}[htbp]
\centering
\scriptsize
\caption{Option Return Statistics (CVS strike set, full period)}\label{table:ActReturnCVS}
\begin{tabular}{cccccccc}
\toprule
Days & Type & Put -1.0 & Put -0.6 & Put -0.2 & Call 0.2 & Call 0.6 & Call 1.0 \\
\midrule
30&Avg&-48.7\%&-39.7\%&-30.0\%&11.9\%&-1.9\%&-37.1\%\\
&StdV&17.7\%&13.7\%&10.7\%&9.1\%&16.9\%&19.7\%\\
&CI Upper&-16.9\%&-16.3\%&-11.9\%&27.6\%&28.1\%&-2.0\%\\
&CI Lower&-73.8\%&-60.5\%&-46.8\%&-2.6\%&-27.9\%&-66.6\%\\
\midrule
60&Avg&-49.0\%&-46.9\%&-37.9\%&12.1\%&3.0\%&-47.1\%\\
&StdV&18.5\%&13.7\%&10.2\%&8.6\%&26.3\%&15.9\%\\
&CI Upper&-16.1\%&-23.2\%&-20.3\%&26.8\%&52.2\%&-18.4\%\\
&CI Lower&-76.1\%&-67.4\%&-53.9\%&-1.7\%&-33.3\%&-71.0\%\\
\midrule
90&Avg&-47.1\%&-46.4\%&-40.7\%&14.9\%&-1.4\%&-47.5\%\\
&StdV&14.6\%&11.9\%&9.5\%&8.5\%&16.6\%&15.6\%\\
&CI Upper&-21.7\%&-25.8\%&-24.6\%&29.0\%&27.6\%&-19.5\%\\
&CI Lower&-69.8\%&-65.1\%&-55.8\%&1.2\%&-26.4\%&-70.0\%\\
\midrule
120&Avg&-61.9\%&-53.5\%&-46.3\%&16.5\%&-1.6\%&-57.8\%\\
&StdV&10.0\%&9.1\%&7.8\%&8.2\%&13.2\%&10.2\%\\
&CI Upper&-44.7\%&-38.0\%&-33.1\%&30.5\%&21.3\%&-40.2\%\\
&CI Lower&-77.5\%&-67.9\%&-59.0\%&3.2\%&-22.2\%&-73.6\%\\
\midrule
180&Avg&-62.2\%&-54.7\%&-46.9\%&32.8\%&5.7\%&-50.2\%\\
&StdV&9.6\%&8.6\%&7.5\%&8.2\%&12.4\%&10.9\%\\
&CI Upper&-45.6\%&-39.6\%&-34.0\%&46.5\%&26.7\%&-31.2\%\\
&CI Lower&-77.2\%&-68.4\%&-59.0\%&19.4\%&-14.5\%&-67.2\%\\
\midrule
270&Avg&-64.5\%&-51.9\%&-45.7\%&50.9\%&40.0\%&-23.0\%\\
&StdV&8.3\%&8.1\%&7.4\%&9.4\%&15.3\%&15.1\%\\
&CI Upper&-50.3\%&-38.1\%&-33.2\%&66.5\%&65.9\%&2.7\%\\
&CI Lower&-77.2\%&-64.6\%&-57.5\%&36.0\%&15.8\%&-46.4\%\\
\midrule
360&Avg&-71.8\%&-58.3\%&-48.2\%&68.3\%&71.9\%&16.2\%\\
&StdV&6.7\%&6.9\%&6.8\%&10.3\%&18.6\%&22.8\%\\
&CI Lower&-82.3\%&-69.2\%&-59.0\%&51.7\%&42.2\%&-18.9\%\\
&CI Upper&-60.6\%&-46.6\%&-36.7\%&85.5\%&103.8\%&55.5\%\\
\bottomrule
\end{tabular}
\begin{quote}
\small
\textit{Notes}: This table reports the historical return statistics for the CVS strike set over the full sample period.
Strikes are determined as $K_{i,t,\tau}=\left\lceil F_{t,\tau}e^{z_{i}\bar{\sigma}_{\tau}^{\ast}\sqrt{\tau}}\right\rceil$, where $\bar{\sigma}_{\tau}^{\ast}$ is the average of square root of variance swap for maturity $\tau$ across the in-sample period.
``Avg'' and ``StdV'' denote the sample average and standard deviation of option returns, respectively.
``CI Lower'' and ``CI Upper'' represent the lower and upper bounds of the 95\% confidence interval computed by 1,000 bootstrap replications.
The columns ``Put/Call $z_i$'' indicate the option type and the strike index $z_{i}$, which determines the corresponding strike level, with $z_i \in \{-1.0, -0.6, -0.2, 0.2, 0.6, 1.0\}$.
\end{quote}
\end{table}

\begin{table}[htbp]
\centering
\scriptsize
\caption{Option Return Statistics (TVS strike set, full period)}\label{table:ActReturnTVS}
\begin{tabular}{cccccccc}
\toprule
Days & Type & Put -1.0 & Put -0.6 & Put -0.2 & Call 0.2 & Call 0.6 & Call 1.0 \\
\midrule
30&Avg&-40.8\%&-35.8\%&-29.2\%&10.2\%&12.5\%&-13.8\%\\
&StdV&15.7\%&12.8\%&10.4\%&8.2\%&13.5\%&22.7\%\\
&CI Upper&-13.6\%&-13.9\%&-11.7\%&24.3\%&35.8\%&25.9\%\\
&CI Lower&-64.4\%&-55.5\%&-45.4\%&-3.1\%&-8.9\%&-48.6\%\\
\midrule
60&Avg&-47.5\%&-45.3\%&-37.8\%&11.2\%&4.2\%&-29.4\%\\
&StdV&15.7\%&12.3\%&9.8\%&7.9\%&13.2\%&24.5\%\\
&CI Upper&-20.2\%&-24.0\%&-21.2\%&24.7\%&27.3\%&16.0\%\\
&CI Lower&-71.4\%&-64.4\%&-53.3\%&-1.4\%&-16.5\%&-63.9\%\\
\midrule
90&Avg&-47.7\%&-46.5\%&-40.6\%&14.4\%&7.3\%&-45.9\%\\
&StdV&14.3\%&11.2\%&9.2\%&7.8\%&13.4\%&16.7\%\\
&CI Upper&-22.5\%&-27.4\%&-25.1\%&27.4\%&29.6\%&-16.0\%\\
&CI Lower&-69.7\%&-64.3\%&-55.2\%&1.7\%&-13.6\%&-70.9\%\\
\midrule
120&Avg&-60.2\%&-52.9\%&-45.4\%&17.4\%&12.1\%&-59.9\%\\
&StdV&10.0\%&9.0\%&7.9\%&7.8\%&12.5\%&11.4\%\\
&CI Upper&-42.9\%&-37.4\%&-32.0\%&30.7\%&33.4\%&-39.5\%\\
&CI Lower&-75.8\%&-67.1\%&-58.0\%&4.8\%&-7.7\%&-77.6\%\\
\midrule
180&Avg&-61.3\%&-55.1\%&-46.3\%&32.3\%&20.6\%&-60.3\%\\
&StdV&9.9\%&8.6\%&7.6\%&7.9\%&12.4\%&9.9\%\\
&CI Upper&-44.2\%&-40.0\%&-33.3\%&45.5\%&41.5\%&-43.1\%\\
&CI Lower&-76.7\%&-68.8\%&-58.4\%&19.4\%&0.5\%&-75.7\%\\
\midrule
270&Avg&-59.6\%&-51.1\%&-45.7\%&49.9\%&55.2\%&-19.0\%\\
&StdV&8.7\%&8.3\%&7.4\%&8.9\%&15.7\%&23.5\%\\
&CI Upper&-44.6\%&-37.0\%&-33.2\%&64.6\%&81.5\%&23.5\%\\
&CI Lower&-72.9\%&-64.2\%&-57.5\%&35.6\%&30.4\%&-52.9\%\\
\midrule
360&Avg&-71.4\%&-56.8\%&-48.2\%&66.6\%&83.1\%&24.4\%\\
&StdV&6.9\%&7.1\%&6.8\%&9.8\%&18.0\%&31.1\%\\
&CI Lower&-59.7\%&-45.0\%&-36.8\%&83.0\%&113.7\%&79.1\%\\
&CI Upper&-82.2\%&-68.0\%&-59.0\%&50.7\%&54.2\%&-22.3\%\\
\bottomrule
\end{tabular}
\begin{quote}
\small
\textit{Notes}: This table reports the historical return statistics for the TVS strike set,
where strikes are determined using time-varying volatility $\sigma_{t,t+\tau}^{\ast}$ at each date $t$ as $K_{i,t,\tau}=\left\lceil F_{t,\tau}e^{z_{i}\sigma_{t,t+\tau}^{\ast}\sqrt{\tau}}\right\rceil$.
All other specifications, including the calculation of confidence intervals via 1,000 bootstrap replications and the definitions of row labels, are identical to those described in Table~\ref{table:ActReturnCVS}.
\end{quote}
\end{table}

For each dataset used in the estimation, the historical option returns by strike are reported in Tables \ref{table:ActReturnCVS} and \ref{table:ActReturnTVS}.
Table~\ref{table:ActReturnCVS} presents the results for the CVS strike set,
while Table~\ref{table:ActReturnTVS} shows the results for the TVS strike set.
Each column corresponds to an option type and strike.
For example, ``Put -1.0'' refers to the aggregated results for put options at the strike associated with $z_{1}=-1.0$.
The row labeled ``Avg'' reports the average return, ``StdV'' the standard deviation of returns, and ``CI Lower'' and ``CI Upper'' the lower and upper bounds of the 95\% confidence interval obtained via 1,000 bootstrap replications.
For both strike sets, the returns peak for calls in the range ``Call 0.2''--``Call 0.6''.
Under a monotonically decreasing SDF, option returns on the call side should increase monotonically with strike.
However, the aggregated returns decline in the deep OTM region of call side, implying the presence of a non-monotonic SDF.
\textcite{bakshi2010returns} reports a non-monotonic SDF under this consideration for the CVS strike set, and our results provide confirming evidence for their analysis.
Moreover, we newly report that the same pattern holds under the TVS strike set specification.

\subsection{Estimation of Stochastic Discount Factor}
\label{sec:EstimationOfPSDF}

In this section, we estimate the SDF with parametric formulation.
We take the expressions \eqref{formula:BSSDF} and \eqref{formula:SPSDF} derived from the stochastic process as our conceptual starting point
and specify the SDF by extending them to a polynomial in returns scaled by time-varying volatility.
We define the piecewise polynomial $f_{p,n}(y)$, which is a polynomial of degree $n$ on the interval $[y_{\min},y_{\max}]$ and outside this interval is extended linearly so that both the value and the first derivative are continuous at the endpoints:
\begin{equation}
f_{p,n}(y)=
\begin{cases}
\sum_{i=0}^{n}a_{i}y^{i} & (y_{\min} \leq y \leq y_{\max}), \\[6pt]
f^{(1)}_{p,n}(y_{\min})(y-y_{\min})+f_{p,n}(y_{\min}) & (y<y_{\min}), \\[6pt]
f^{(1)}_{p,n}(y_{\max})(y-y_{\max})+f_{p,n}(y_{\max}) & (y>y_{\max}).
\end{cases}
\end{equation}
Note that the function is expressed as a linear combination with coefficients $a_{i}$.
We then define the constant volatility scaled (CVS) SDF, $n_{t,t+\tau,n,CVS}(S_{t+\tau})$, and time-varying volatility scaled (TVS) SDF, $n_{t,t+\tau,n,TVS}(S_{t+\tau})$, as
\begin{equation}
n_{t,t+\tau,n,CVS}(S_{T})=\exp\left\{f_{p,n}\left(\frac{1}{\bar{\sigma}_{\tau}^{\ast}\sqrt{\tau}}
\log\left(\frac{S_{t+\tau}}{F_{t,t+\tau}}\right)\right)\right\},
\label{eq:CVSSDF}
\end{equation}
\begin{equation}
n_{t,t+\tau,n,TVS}(S_{t+\tau})=\exp\left\{f_{p,n}\left(\frac{1}{\sigma_{t,t+\tau}^{\ast}\sqrt{\tau}}
\log\left(\frac{S_{t+\tau}}{F_{t,t+\tau}}\right)\right)\right\}.
\label{eq:TVSSDF}
\end{equation}

After here, we refer to the arguments
$\tfrac{1}{\bar{\sigma}_{\tau}^{\ast}\sqrt{\tau}}\log\left(\tfrac{S_{t+\tau}}{F_{t,t+\tau}}\right)$
and $\tfrac{1}{\sigma_{t,t+\tau}^{\ast}\sqrt{\tau}}\log\left(\tfrac{S_{t+\tau}}{F_{t,t+\tau}}\right)$
as normalized log returns.
The scaling applied in the CVS is introduced solely to facilitate comparison of its parameters and the resulting SDF with those of the TVS.
In substantive terms, it is equivalent to performing no scaling at all.

The parameters of the SDF are estimated by minimizing the HJ distance given in \eqref{formula:HJDistance}.  
The basis assets used to compute the HJ distance include the implied discount bond price and forward price estimated from option data.
The strike prices of options in the basis assets are chosen as follows: when estimating the CVS SDF, we use the CVS strike set, and when estimating the TVS SDF, we use the TVS strike set.
These basis assets are scaled by adjusting their notionals so that their prices become $\mathbf{v}_{t}=1$ at each date.
For polynomial degrees $n=2,\ldots,6$, we estimate the parameters by minimizing the HJ distance.
Since Hansen's $J$-test in the GMM framework introduced by \textcite{hansen1982large} requires the model to be overidentified, the maximum polynomial order is set to six.
Also, we exclude the case $n=1$, which yields a monotonic SDF and is therefore inconsistent with the excess-return patterns observed in the option return statistics.

\begin{table}[htbp]
\centering
\scriptsize
\caption{Estimated SDF by CVS Using Full-Sample Data }\label{table:SDFEstimationFullCVS} 
\begin{tabular}{llllllllllllll}
\toprule
Days & Order & $a_{0}$ & $a_{1}$ & $a_{2}$ & $a_{3}$ & $a_{4}$ & $a_{5}$ & $a_{6}$ & $f_{p,n}^{(1)}(-1)$ & $f_{p,n}^{(1)}(1)$ & $J$-test & HC test & Data num \\
\midrule
30 & 2 & -0.037 & -0.104 & 0.065 &  &  &  &  & -0.234 & 0.025 & 0.134 & 0.555 & 324 \\
& 3 & -0.049 & -0.238 & 0.111 & 0.087 &  &  &  & -0.200 & 0.245 & 0.161 & 0.464 & 324 \\
& 4 & 0.041 & -0.312 & -0.367 & 0.142 & 0.273 &  &  & -0.242 & 0.473 & 0.131 & 0.505 & 324 \\
& 5 & 0.041 & -0.245 & -0.380 & -0.046 & 0.281 & 0.102 &  & -0.237 & 0.495 & 0.053 & 0.907 & 324 \\
& 6 & 0.164 & -0.461 & -2.802 & 0.652 & 5.454 & -0.290 & -2.662 & -0.194 & 0.283 & 0.083 & 0.154 & 324 \\
\midrule
60 & 2 & -0.055 & -0.110 & 0.100 &  &  &  &  & -0.309 & 0.089 & 0.000 & 0.345 & 323 \\
& 3 & -0.106 & -0.525 & 0.270 & 0.289 &  &  &  & -0.200 & 0.882 & 0.009 & 0.059 & 323 \\
& 4 & 0.044 & -0.579 & -0.590 & 0.366 & 0.505 &  &  & -0.321 & 1.362 & 0.028 & 0.173 & 323 \\
& 5 & 0.042 & 0.443 & -0.712 & -2.579 & 0.591 & 1.608 &  & -0.192 & 1.689 & 0.048 & 0.120 & 323 \\
& 6 & 0.192 & -0.273 & -3.448 & -0.469 & 6.597 & 0.432 & -3.141 & -0.167 & 1.127 & 0.030 & 0.203 & 323 \\
\midrule
90 & 2 & 0.033 & -0.287 & -0.032 &  &  &  &  & -0.224 & -0.351 & 0.000 & 0.761 & 322 \\
& 3 & -0.011 & -0.595 & 0.121 & 0.235 &  &  &  & -0.131 & 0.352 & 0.002 & 0.164 & 322 \\
& 4 & -0.036 & -0.575 & 0.249 & 0.218 & -0.075 &  &  & -0.117 & 0.278 & 0.001 & 0.809 & 322 \\
& 5 & -0.032 & -0.081 & 0.111 & -1.159 & 0.016 & 0.754 &  & -0.073 & 0.495 & 0.000 & 0.516 & 322 \\
& 6 & 0.157 & -0.690 & -3.193 & 0.647 & 7.020 & -0.255 & -3.615 & -0.028 & -0.020 & 0.001 & 0.077 & 322 \\
\midrule
120 & 2 & -0.146 & -0.154 & 0.249 &  &  &  &  & -0.651 & 0.343 & 0.000 & 0.045 & 321 \\
& 3 & -0.151 & -0.533 & 0.331 & 0.247 &  &  &  & -0.456 & 0.870 & 0.000 & 0.234 & 321 \\
& 4 & 0.098 & -0.726 & -0.876 & 0.364 & 0.687 &  &  & -0.629 & 1.361 & 0.000 & 0.038 & 321 \\
& 5 & 0.103 & -2.049 & -0.784 & 3.985 & 0.627 & -1.963 &  & -0.852 & 1.033 & 0.000 & 0.077 & 321 \\
& 6 & -5.142 & -62.490 & -110.567 & 125.772 & 238.616 & -63.008 & -122.182 & -0.454 & 0.025 & 0.000 & 0.754 & 321 \\
\midrule
180 & 2 & -0.080 & -0.260 & 0.135 &  &  &  &  & -0.531 & 0.010 & 0.000 & 0.384 & 318 \\
& 3 & -0.124 & -0.566 & 0.275 & 0.240 &  &  &  & -0.397 & 0.703 & 0.000 & 0.284 & 318 \\
& 4 & 0.331 & -0.936 & -1.895 & 0.489 & 1.243 &  &  & -0.652 & 1.716 & 0.000 & 0.006 & 318 \\
& 5 & 0.327 & -1.228 & -1.805 & 1.363 & 1.188 & -0.483 &  & -0.693 & 1.588 & 0.000 & 0.722 & 318 \\
& 6 & 0.559 & -2.811 & -7.910 & 5.403 & 14.538 & -2.660 & -6.950 & -0.534 & 0.730 & 0.000 & 0.027 & 318 \\
\midrule
270 & 2 & -0.084 & -0.344 & 0.120 &  &  &  &  & -0.583 & -0.105 & 0.000 & 0.571 & 315 \\
& 3 & 0.288 & 0.045 & -1.368 & -1.207 &  &  &  & -0.839 & -6.311 & 0.000 & 0.040 & 315 \\
& 4 & 0.543 & -1.031 & -2.901 & 0.511 & 1.807 &  &  & -0.922 & 1.926 & 0.097 & 0.000 & 315 \\
& 5 & 0.575 & -2.179 & -2.779 & 3.751 & 1.721 & -1.769 &  & -1.099 & 1.556 & 0.028 & 0.207 & 315 \\
& 6 & 0.525 & -1.978 & -1.825 & 3.222 & -0.340 & -1.483 & 1.069 & -1.130 & 1.678 & 0.022 & 0.646 & 315 \\
\midrule
360 & 2 & -0.219 & -0.358 & 0.260 &  &  &  &  & -0.877 & 0.161 & 0.000 & 0.298 & 312 \\
& 3 & 0.416 & 0.652 & -2.158 & -2.140 &  &  &  & -1.451 & -10.084 & 0.000 & 0.001 & 312 \\
& 4 & 0.683 & -0.612 & -3.679 & 0.239 & 2.251 &  &  & -1.541 & 1.752 & 0.820 & 0.000 & 312 \\
& 5 & 0.730 & -1.441 & -3.640 & 2.584 & 2.211 & -1.285 &  & -1.677 & 1.449 & 0.395 & 0.406 & 312 \\
& 6 & 0.683 & -1.281 & -2.736 & 2.111 & 0.177 & -1.020 & 1.072 & -1.719 & 1.618 & 0.467 & 0.637 & 312 \\
\bottomrule
\end{tabular}
\begin{quote}
\small
\textit{Notes}: This table presents the estimation results of CVS SDF using the full sample period from January 1996 to February 2023.
The parameters are estimated by minimizing the HJ distance.
``Days'' indicates the option maturity for which the SDF is estimated.
``Order'' denotes the polynomial degree $n$, and ``$a_{i}$'' ($i=0,\dots,6$) are the estimated coefficients $a_i$ of the polynomial SDF.
``$f_{p,n}^{(1)}(-1)$'' and ``$f_{p,n}^{(1)}(1)$'' represent the derivative values at the lower and upper endpoints of the interpolation interval, respectively, and correspond to the slopes of the function in the extrapolation region.
``$J$-test'' reports the $p$-value for Hansen's $J$-test of overidentifying restrictions.
``HC test'' provides the $p$-value for the null hypothesis that the highest-order coefficient is zero ($a_n = 0$).
``Data num'' indicates the number of monthly observations.
\end{quote}
\end{table}

\begin{table}[htbp]
\centering
\scriptsize
\caption{Estimated SDF by TVS Using Full-Sample Data }\label{table:SDFEstimationFullTVS} 
\begin{tabular}{llllllllllllll}
\toprule
Days & Order & $a_{0}$ & $a_{1}$ & $a_{2}$ & $a_{3}$ & $a_{4}$ & $a_{5}$ & $a_{6}$ & $f_{p,n}^{(1)}(-1)$ & $f_{p,n}^{(1)}(1)$ & $J$-test & HC test & Data num \\
\midrule
30 & 2 & -0.013 & -0.117 & 0.025 &  &  &  &  & -0.166 & -0.067 & 0.024 & 0.743 & 324 \\
& 3 & -0.028 & -0.184 & 0.068 & 0.057 &  &  &  & -0.151 & 0.123 & 0.190 & 0.647 & 324 \\
& 4 & -0.014 & -0.201 & -0.005 & 0.077 & 0.049 &  &  & -0.156 & 0.215 & 0.120 & 0.845 & 324 \\
& 5 & -0.032 & 0.622 & -0.175 & -2.084 & 0.210 & 1.200 &  & -0.117 & 0.863 & 0.349 & 0.128 & 324 \\
& 6 & -0.129 & 0.711 & 1.389 & -2.532 & -3.081 & 1.495 & 1.712 & -0.135 & 1.315 & 0.391 & 0.279 & 324 \\
\midrule
60 & 2 & 0.003 & -0.182 & 0.005 &  &  &  &  & -0.192 & -0.172 & 0.001 & 0.956 & 323 \\
& 3 & -0.056 & -0.417 & 0.177 & 0.212 &  &  &  & -0.134 & 0.573 & 0.076 & 0.126 & 323 \\
& 4 & -0.072 & -0.396 & 0.263 & 0.187 & -0.059 &  &  & -0.127 & 0.455 & 0.036 & 0.838 & 323 \\
& 5 & -0.076 & 0.750 & -0.056 & -2.970 & 0.221 & 1.776 &  & -0.050 & 1.495 & 0.517 & 0.031 & 323 \\
& 6 & -0.030 & 0.743 & -0.833 & -2.841 & 1.855 & 1.683 & -0.847 & -0.038 & 1.312 & 0.334 & 0.526 & 323 \\
\midrule
90 & 2 & 0.022 & -0.245 & -0.018 &  &  &  &  & -0.208 & -0.281 & 0.000 & 0.861 & 322 \\
& 3 & -0.094 & -0.635 & 0.306 & 0.386 &  &  &  & -0.091 & 1.134 & 0.006 & 0.008 & 322 \\
& 4 & 0.029 & -0.731 & -0.270 & 0.483 & 0.350 &  &  & -0.142 & 1.576 & 0.005 & 0.227 & 322 \\
& 5 & -0.064 & 0.957 & -0.332 & -3.974 & 0.452 & 2.422 &  & 0.000 & 2.288 & 0.338 & 0.019 & 322 \\
& 6 & -0.052 & 0.953 & -0.515 & -3.942 & 0.838 & 2.401 & -0.199 & 0.004 & 2.256 & 0.133 & 0.914 & 322 \\
\midrule
120 & 2 & -0.124 & -0.155 & 0.207 &  &  &  &  & -0.568 & 0.259 & 0.000 & 0.124 & 321 \\
& 3 & -0.282 & -0.622 & 0.571 & 0.464 &  &  &  & -0.373 & 1.913 & 0.000 & 0.026 & 321 \\
& 4 & 0.172 & -0.952 & -1.623 & 0.804 & 1.352 &  &  & -0.702 & 3.622 & 0.005 & 0.000 & 321 \\
& 5 & 0.138 & 0.075 & -1.617 & -1.837 & 1.380 & 1.431 &  & -0.566 & 4.006 & 0.043 & 0.210 & 321 \\
& 6 & -0.026 & 0.175 & 1.528 & -2.332 & -5.414 & 1.757 & 3.546 & -0.714 & 4.639 & 0.749 & 0.048 & 321 \\
\midrule
180 & 2 & -0.056 & -0.287 & 0.104 &  &  &  &  & -0.494 & -0.079 & 0.000 & 0.569 & 318 \\
& 3 & -0.347 & -0.821 & 0.679 & 0.622 &  &  &  & -0.313 & 2.404 & 0.000 & 0.024 & 318 \\
& 4 & 0.355 & -1.019 & -2.267 & 0.873 & 1.723 &  &  & -0.760 & 3.958 & 0.065 & 0.000 & 318 \\
& 5 & 0.312 & 0.826 & -2.433 & -4.285 & 1.890 & 2.852 &  & -0.464 & 4.925 & 0.716 & 0.028 & 318 \\
& 6 & 0.292 & 0.861 & -2.078 & -4.434 & 1.100 & 2.944 & 0.418 & -0.474 & 5.029 & 0.364 & 0.832 & 318 \\
\midrule
270 & 2 & -0.017 & -0.407 & 0.038 &  &  &  &  & -0.483 & -0.332 & 0.000 & 0.873 & 315 \\
& 3 & 0.125 & -0.004 & -0.462 & -0.543 &  &  &  & -0.709 & -2.559 & 0.000 & 0.236 & 315 \\
& 4 & 0.369 & -0.659 & -1.884 & 0.355 & 1.255 &  &  & -0.843 & 1.657 & 0.006 & 0.008 & 315 \\
& 5 & 0.378 & -1.180 & -1.739 & 1.771 & 1.122 & -0.804 &  & -0.894 & 1.122 & 0.003 & 0.611 & 315 \\
& 6 & -0.108 & 0.268 & 5.180 & -3.239 & -14.396 & 2.213 & 8.316 & -1.056 & 4.289 & 0.616 & 0.000 & 315 \\
\midrule
360 & 2 & -0.190 & -0.399 & 0.234 &  &  &  &  & -0.867 & 0.069 & 0.000 & 0.388 & 312 \\
& 3 & 0.265 & 0.544 & -1.232 & -1.463 &  &  &  & -1.381 & -6.310 & 0.000 & 0.008 & 312 \\
& 4 & 0.549 & -0.371 & -2.823 & 0.069 & 1.751 &  &  & -1.523 & 1.192 & 0.601 & 0.001 & 312 \\
& 5 & 0.566 & -0.808 & -2.697 & 1.294 & 1.641 & -0.695 &  & -1.570 & 0.772 & 0.116 & 0.682 & 312 \\
& 6 & 0.437 & -0.506 & -0.585 & 0.151 & -3.034 & 0.006 & 2.485 & -1.623 & 1.580 & 0.501 & 0.277 & 312 \\
\bottomrule
\end{tabular}
\begin{quote}
\small
\textit{Notes}: This table presents the estimation results of TVS SDF using the full sample period from January 1996 to February 2023.
See Table~\ref{table:SDFEstimationFullCVS} for the deﬁnitions of the columns and other settings.
\end{quote}
\end{table}

The SDFs estimated using the full dataset from January 1996 to February 2023 are reported in Tables \ref{table:SDFEstimationFullCVS} and \ref{table:SDFEstimationFullTVS}.
``Days'' refers to the maturity, ``Order'' to the polynomial degree of the estimated SDF, and ``$a_{i}$'' to the polynomial coefficients $a_{i}$.
``$f_{p,n}^{(1)}(-1)$'' and ``$f_{p,n}^{(1)}(1)$'' represent the derivative values at the lower and upper endpoints of the interpolation interval, respectively, and correspond to the slopes of the function in the extrapolation region.
``J-test'' reports the $p$-value of Hansen's $J$-test, while ``HC test'' gives the $p$-value for the null hypothesis that the highest-order coefficient equals zero.
Because GMM estimators are asymptotically normally distributed, the HC test is implemented that
the estimated highest-order coefficient is divided by its standard error, and the resulting statistic is evaluated under the asymptotic normal distribution to obtain the corresponding p-value.
All statistics of $J$-test and ``HC test'' are calculated using the method of \textcite{newey1987simple} with the lag length set to 1.5 times the maturity in months and rounded up.
``Data num'' denotes the number of periods used in the estimation.

At the 95\% significance level, the $J$-test indicates that if the value is smaller than 0.05,
the null hypothesis that the GMM moment conditions are satisfied is rejected, implying that the model is invalid.
For the CVS case, the $J$-test is rejected in all cases for maturities of 60-270 days, except for the order 4 at maturity of 270 days.
At maturity of 30 days, all specifications are accepted, and at maturity of 360 days, order 4 and higher are accepted.
For the TVS case,
for maturities of 30-360 days, the $J$-test is accepted for orders 3, 5, 5, 6, 5, 6, 4 and higher, respectively.
Exceptionally, even the order 3 is accepted at the maturity of 60 days.
A clearer tendency for higher order specifications to pass the $J$-test is observed for TVS than for CVS.
When we also note that only TVS passes the $J$-test for maturities from 60 to 180 days, TVS appears more consistent with market data as an SDF model.

Regarding the significance of the highest-order coefficient,
at the 95\% level the null hypothesis that the highest-order coefficient is zero is rejected if the $p$-value is smaller than 0.05.
For the HC test, no clear pattern can be observed -- for example, no tendency for significance at a given order to carry over to all lower orders; rather, certain orders are accepted only sporadically.
The CVS case shows such instances only at order 4 for maturities of 270 and 360 days, where both the $J$-test is accepted and the HC test indicates significance at that order.
For the TVS case, significance is found at order 5 for 60 days, order 5 for 90 days, order 6 for 120 days, orders 4 and 5 for 180 days, order 6 for 270 days, and order 4 for 360 days.
In conclusion, the TVS SDF is found to yield more valid estimates than the CVS SDF, as indicated by the results of the $J$-test and HC test.

With respect to the slopes of extrapolation, $f_{p,n}^{(1)}(-1)$ and $f_{p,n}^{(1)}(1)$,
the estimated SDF shows a downward slope at the left endpoint and an upward slope at the right endpoint for all polynomial orders of four and above.
The only exceptions appear at the 90 day maturity for the 5th and 6th orders.
This pattern suggests that both deep OTM puts and calls are priced relatively high.
It also points to the presence of fat tail aversion.
The result is consistent with Tables \ref{table:ActReturnCVS} and \ref{table:ActReturnTVS}.
For orders of three and below, the left endpoint always slopes downward.
The right endpoint, however, may slope upward or downward.
Thus, the degree of risk aversion on the call side is mixed.
Lower order polynomials have limited degrees of freedom to capture the overall shape of the SDF.
These degrees of freedom are likely used to fit the put side and the region around ATM.
As a result, the fit on the call side becomes weaker.
Therefore, a sufficiently high polynomial order is needed to obtain SDF shapes that align with Tables \ref{table:ActReturnCVS} and \ref{table:ActReturnTVS}.

\subsection{Equity Premium by Estimated Stochastic Discount Factor}

In this section, we evaluate the out-of-sample predictive performance of the daily EP calculated from estimated SDFs.
The SDFs are estimated under the expanding window period, where the estimation employs data up to the day immediately preceding each reference date.
From each estimated SDF, we compute the predicted EP using \eqref{formula:PhysicalExpectation}, where the SDF is adjusted to be consistent with the risk-free rate using \eqref{formula:AdjustedSDF} and \eqref{formula:SDFBeta}.
The EPs at the actual maturities surrounding the target maturity are calculated, and the EP at target maturity is interpolated by these values.
The EP values calculated from the CVS and TVS SDF are referred to as CVS EP and TVS EP, respectively.

For comparison of forecasting performance, we employ the statistic $R_{OS}^{2}$ proposed by \textcite{campbell2008predicting}.
Let the realized ER (excess return) be
\[
EPR_{t}=D_{t,T}^{-1}\left(\frac{S_{T}}{F_{t,T}}-1\right),
\]
the forecast EP be $EP_{t}$, and the benchmark EP be $\bar{EP}_{t}$.  
Then $R_{OS}^{2}$ is defined as
\begin{equation}
R_{OS}^{2}=1-\frac{\sum_{t}\left(EPR_{t}-EP_{t}\right)^{2}}{\sum_{t}\left(EPR_{t}-\bar{EP}_{t}\right)^{2}}.
\end{equation}
This statistic compares the mean squared predictive error (MSPE) of the forecast EP against the benchmark.  
Its values lie in $(-\infty,1]$, and a positive value indicates that the forecast EP outperforms the benchmark in terms of MSPE.
We use historical daily average (HA) of realized ERs from Jan 1996 until predicting point as benchmarks and compare them with the EPs computed from the estimated SDF.

\begin{table}[htbp]
\centering
\scriptsize
\caption{$R_{OS}^{2}$ of Equity Premium by SDF Type and Polynomoial Order}
\label{tab:R2OSOrderFullPeriod}
\begin{tabular}{lc|rrrrr|rrrrr}
\hline
&& \multicolumn{5}{c|}{CVS} & \multicolumn{5}{c}{TVS}\\
Term & Days & 2 & 3 & 4 & 5 & 6 & 2 & 3 & 4 & 5 & 6\\
\hline
Full& All & 4.01\% & \# & \underline{4.57\%} & 3.87\% & 3.21\% & 2.03\% & \# & \underline{7.04\%} & 5.71\% & 6.03\% \\
\cline{2-12}
& 30 & \underline{1.69\%} & 1.08\% & -0.01\% & -0.23\% & -0.31\% & 1.77\% & \underline{1.94\%} & 1.80\% & 1.86\% & 1.91\% \\
& 60 & 1.57\% & 1.27\% & 1.26\% & 3.10\% & \underline{3.21\%} & 1.21\% & 4.42\% & \underline{4.74\%} & 4.34\% & 4.08\% \\
& 90 & \underline{8.97\%} & 4.51\% & 3.79\% & 3.07\% & 2.61\% & 1.08\% & 7.19\% & \underline{7.49\%} & 5.67\% & 6.34\% \\
& 120 & 5.54\% & 5.46\% & \underline{9.72\%} & 9.67\% & 5.94\%  & 2.36\% & 8.46\% & \underline{12.05\%} & 10.40\% & 10.64\% \\
& 180 & 7.64\% & 7.30\% & \underline{10.78\%} & 9.99\% & 6.32\% & 3.64\% & 6.35\% & 11.07\% & \underline{12.74\%} & 11.78\% \\
& 270 & 7.78\% & 6.60\% & \underline{16.10\%} & 13.98\% & 13.80\% & 4.82\% & -1.38\% & 16.18\% & 14.51\% & \underline{18.77\%} \\
& 360 & 5.53\% & \# & \underline{17.90\%} & 7.91\% & 8.16\% & 3.43\% & \# & \underline{22.57\%} & 10.35\% & 11.38\% \\
\hline
1st/2&30 & -1.52\% & 0.39\% & 1.09\% & 1.51\% & \underline{1.69\%} & 1.21\% & 1.55\% & 1.55\% & 1.54\% & \underline{1.57\%} \\
& 60 & -2.14\% & 1.48\% & 1.69\% & \underline{3.24\%} & 2.18\% & -2.19\% & 3.69\% & \underline{3.82\%} & 3.97\% & 3.17\% \\
& 90 & 3.90\% & \underline{5.83\%} & 4.67\% & 4.73\% & 4.96\% & -5.17\% & 6.16\% & \underline{6.42\%} & 4.11\% & 4.98\% \\
& 120 & 3.29\% & 6.12\% & \underline{11.65\%} & 11.06\% & 11.56\% & -2.51\% & 7.72\% & \underline{12.13\%} & 10.13\% & 10.54\% \\
& 180 & 6.61\% & 7.48\% & \underline{10.63\%} & 9.56\% & 10.09\% & -1.38\% & 3.84\% & 10.69\% & \underline{12.30\%} & 11.16\% \\
& 270 & 7.19\% & 12.14\% & \underline{23.05\%} & 19.41\% & 21.92\% & -0.60\% & -4.59\% & \underline{21.03\%} & 16.86\% & 18.78\% \\
& 360 & 10.07\% & \# & \underline{26.20\%} & 5.89\% & 2.58\% & 4.27\% & \# & \underline{36.16\%} & 10.26\% & 10.58\% \\
\hline
2nd/2 & 30 & \underline{3.35\%} & 1.43\% & -0.57\% & -1.13\% & -1.36\% & 2.07\% & \underline{2.14\%} & 1.93\% & 2.03\% & 2.09\% \\
& 60 & 3.61\% & 1.16\% & 1.03\% & 3.02\% & \underline{3.78\%} & 3.07\% & 4.83\% & \underline{5.23\%} & 4.55\% & 4.58\% \\
& 90 & \underline{11.94\%} & 3.74\% & 3.28\% & 2.09\% & 1.23\% & 4.73\% & 7.80\% & \underline{8.12\%} & 6.57\% & 7.14\% \\
& 120 & 6.89\% & 5.07\% & 8.57\% & \underline{8.83\%} & 2.58\% & 5.26\% & 8.90\% & \underline{12.01\%} & 10.57\% & 10.70\% \\
& 180 & 8.29\% & 7.18\% & \underline{10.87\%} & 10.26\% & 3.97\% & 6.76\% & 7.91\% & 11.30\% & \underline{13.01\%} & 12.17\% \\
& 270 & 8.11\% & 3.45\% & \underline{12.14\%} & 10.89\% & 9.18\% & 7.90\% & 0.45\% & 13.42\% & 13.16\% & \underline{18.76\%} \\
& 360 & 3.01\% & \# & \underline{13.32\%} & 9.03\% & 11.24\% & 2.98\% & \# & \underline{15.06\%} & 10.40\% & 11.83\% \\
\hline
\end{tabular}
\begin{quote}
\textit{Notes}: This table reports the out-of-sample predictive $R^2$ ($R_{OS}^2$) based on the statistic proposed by \textcite{campbell2008predicting},
which evaluates the predictive performance of the SDF based EP against Historical Average (HA) EP:
A positive value of $R_{OS}^2$ indicates that the SDF based forecast outperforms the benchmark in terms of MSPE.
Each row corresponds to aggregation periods and maturity (days),
while each column represents the SDF type and polynomial order.
The underlined cell in each row indicates the best-performing specification through the polynomial order.
The left side of table compares the CVS SDF, and the right side of table compares TVS SDF.
``Term'' represent aggregation period. ``Full'' is from Dec 18, 2009 to data end. ``1st/2'' is from Dec 18, 2009 to Dec 17, 2015.
``2nd/2'' is from Dec 18, 2015 to Dec 15, 2017.
Data end means the corresponding date which maturity is Feb 28, 2023.
``All'' reports the $R_{OS}^{2}$ computed using data from all available days.
Since the variance of EP differs across maturities, the EP observations for each day are normalized by the standard deviation of $EPR_{t}$.
For the 360 days at order 3 (\#), the values are less than -500\% and treated as missing because the call side SDF is estimated to converge rapidly toward zero,
which causes the EP to become numerically unstable and fail to compute reliably.
\end{quote}
\end{table}

Table~\ref{tab:R2OSOrderFullPeriod} reports the $R_{OS}^2$ values aggregated across all out-of-sample periods, computed using the SDFs estimated under each functional form and polynomial order.
A positive value indicates that the SDF based forecast outperforms the benchmark (HA) in terms of MSPE.
Each row corresponds to aggregation periods and maturity (days),
while each column represents the SDF type and polynomial order.
The underlined cell in each row indicates the best-performing specification through the polynomial order.
The left side of table compares the CVS SDF, and the right side of table compares TVS SDF.
``Term'' represent aggregation period. ``Full'' is from Dec 18, 2009 to data end.
``1st/2'' is from Dec 18, 2009 to Dec 17, 2015.
``2nd/2'' is from Dec 18, 2015 to Dec 15, 2017.
Data end means the corresponding date which maturity is Feb 28, 2023.
``All'' reports the $R_{OS}^2$ computed using data from all available days.
Since the variance of EP differs across maturities, the EP observations for each day are normalized by the standard deviation of $EPR_{t}$.
The standard deviation is calculated by \textcite{newey1987simple} with the lag length set to 1.5 times the days to maturity.
For the 360 days at order 3 (\#), the value is treated as missing because the call side SDF is estimated to converge rapidly toward zero,
which can cause the EP to become numerically unstable and fail to compute reliably.

For both the TVS EP and the CVS EP, the fourth order specification consistently delivers the strongest and most stable out-of-sample performance across maturities and aggregation periods.
In the case of TVS, order 4 attains the highest $R_{OS}^2$ in the majority of maturities and remains positive throughout all periods, demonstrating a robust improvement over the benchmark.
The $R_{OS}^2$ computed using data from all maturities (``All'') is likewise maximized at order 4, further reinforcing its superior predictive accuracy.
Moreover, the TVS results exhibit a systematic maturity dependent pattern: lower order specifications tend to perform better at shorter maturities, whereas higher order specifications become more effective at longer maturities with several execption.
In contrast, the CVS EP displays greater variability in the optimal polynomial order across aggregation periods and maturities.
Although order 4 still most frequently achieves the highest $R_{OS}^2$ and maintains positive predictive power across nearly all maturities, the pattern is less systematic than in the TVS case.
The ``All'' results also show that order 4 yields the highest $R_{OS}^2$ for CVS, but the surrounding orders exhibit more fluctuation, reflecting the comparatively lower stability of the CVS specification.
Overall, these findings demonstrate that order 4 represents the most stable and consistently high-performing specification for both TVS and CVS, and it is therefore adopted in the subsequent analysis.
At the same time, the results indicate that TVS tends to model the SDF more stably than CVS, both in terms of robustness across maturities and in the systematic maturity dependent behavior of its optimal polynomial order.

\subsection{Performance of Equity Premium Implied by Option Prices}

\begin{figure}[htbp]
  \centering
  \setlength{\tabcolsep}{0pt}
  \caption{Time-Series Equity Premium}
  \label{fig:EPGraph}
  \begin{minipage}[t]{1.0\hsize}
    \begin{tabular}{cc}
        \begin{minipage}[t]{0.48\textwidth}\centering
        \includegraphics[width=\linewidth]{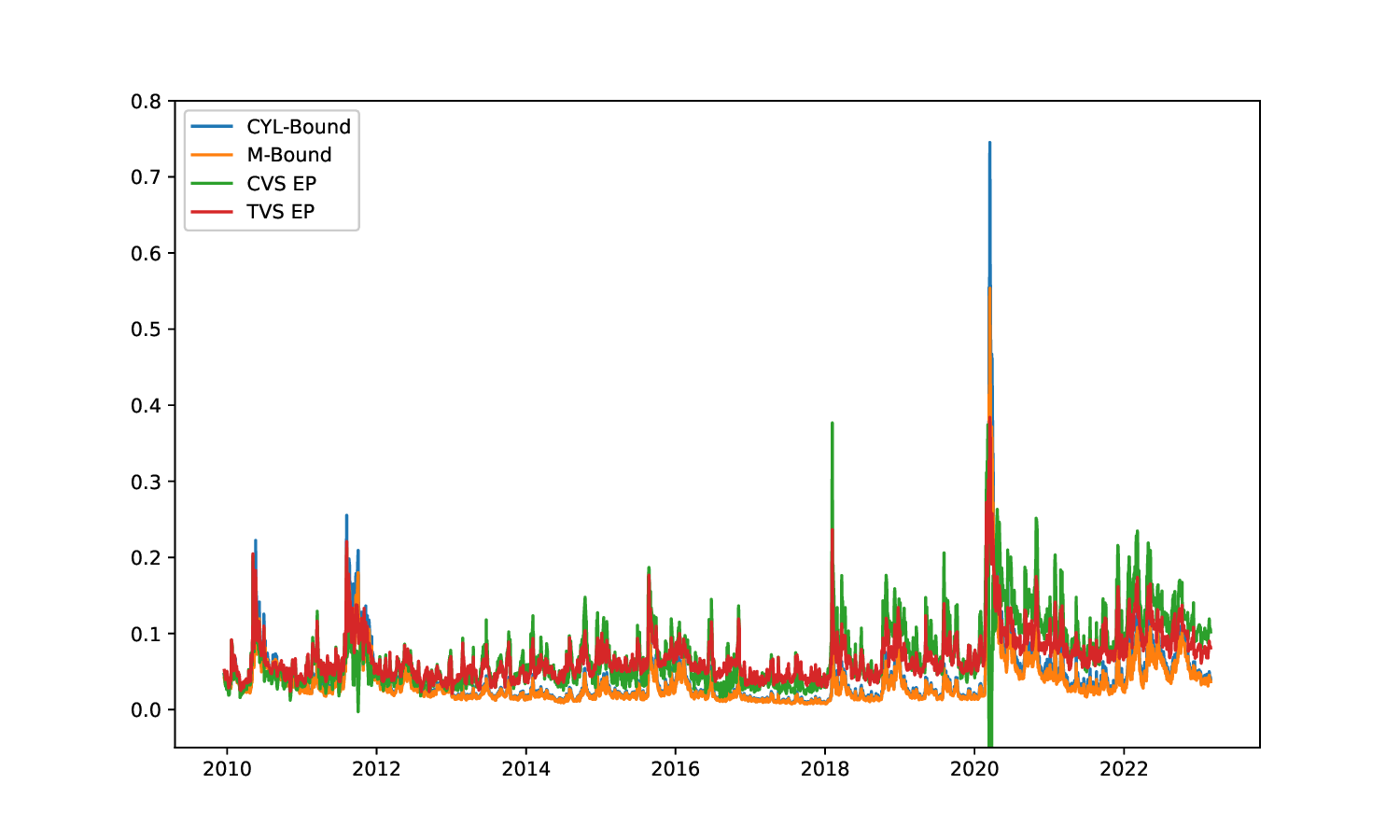}
        \vspace{0.25em}
        \footnotesize 30 days EP
        \end{minipage} &
        \begin{minipage}[t]{0.48\textwidth}\centering
        \includegraphics[width=\linewidth]{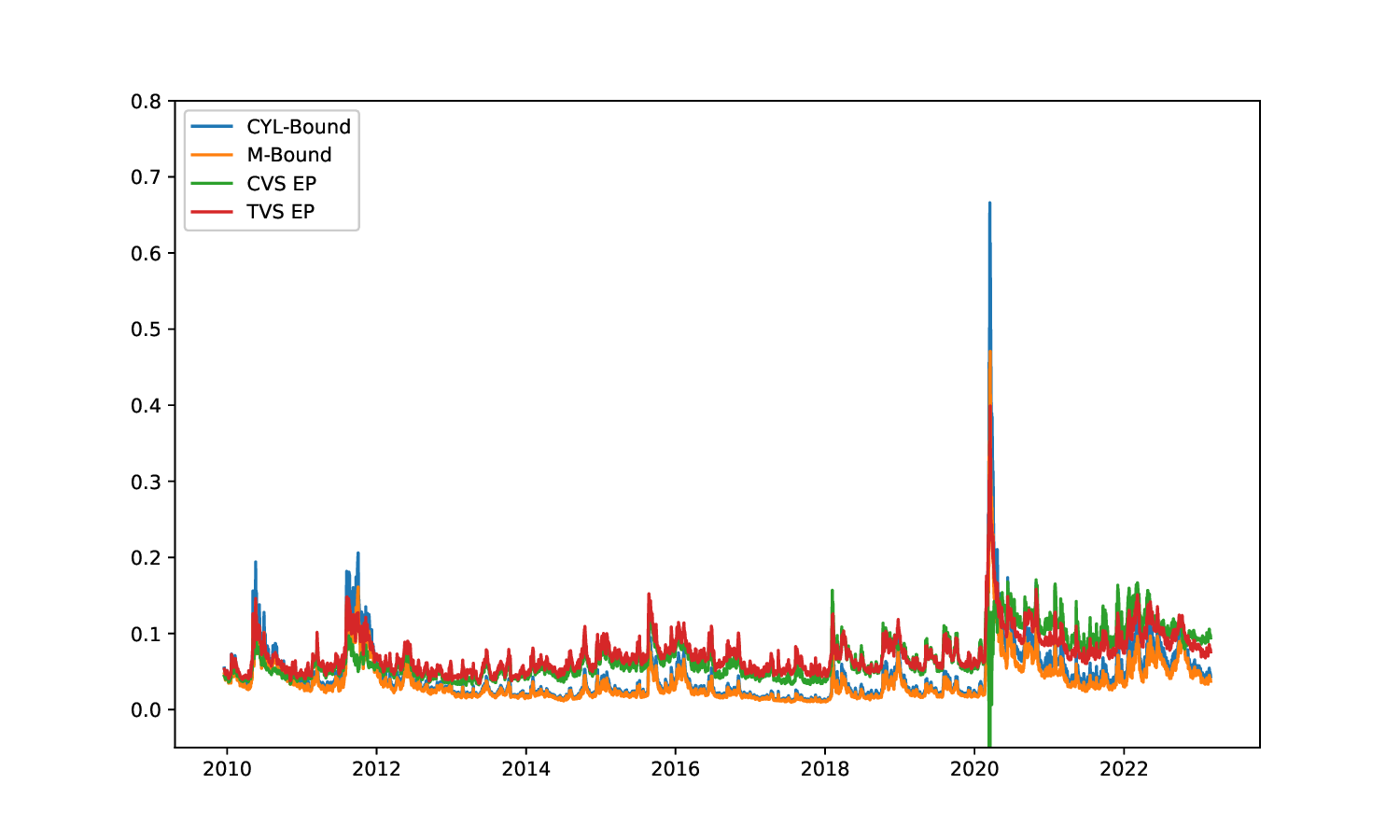}
        \vspace{0.25em}
        \footnotesize 60 days EP
        \end{minipage} \\
        \begin{minipage}[t]{0.48\textwidth}\centering
        \includegraphics[width=\linewidth]{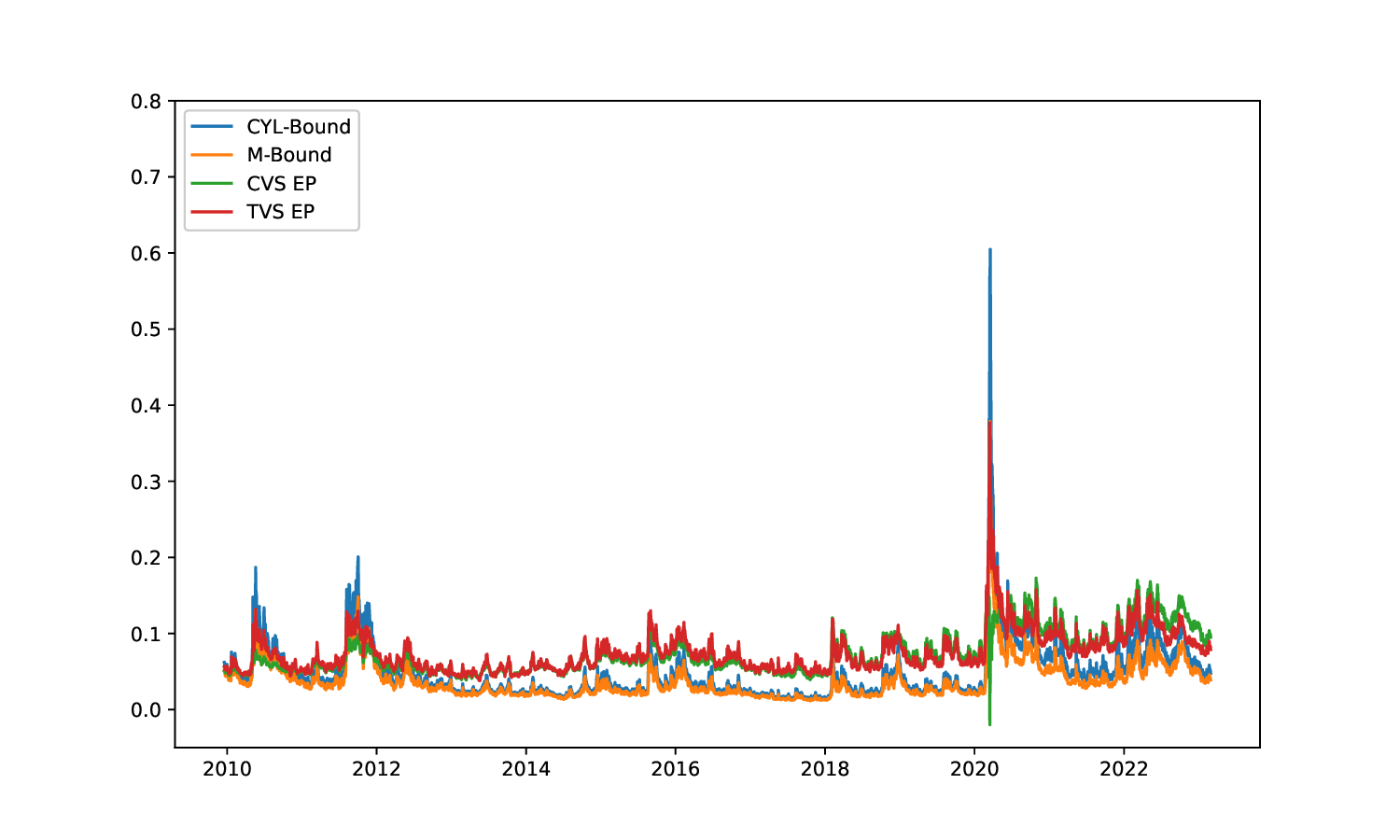}
        \vspace{0.25em}
        \footnotesize 90 days EP
        \end{minipage} &
        \begin{minipage}[t]{0.48\textwidth}\centering
        \includegraphics[width=\linewidth]{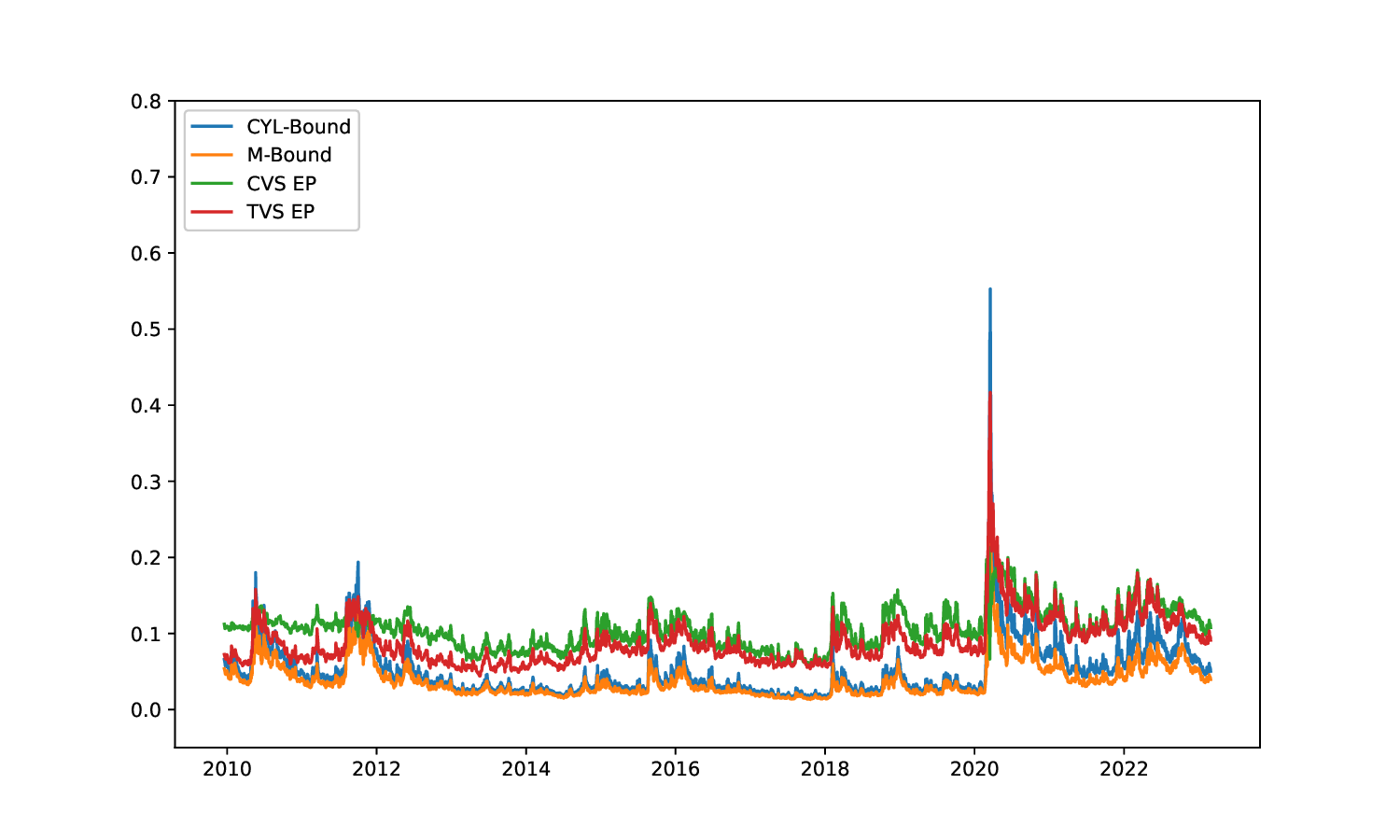}
        \vspace{0.25em}
        \footnotesize 120 days EP
        \end{minipage} \\
        \begin{minipage}[t]{0.48\textwidth}\centering
        \includegraphics[width=\linewidth]{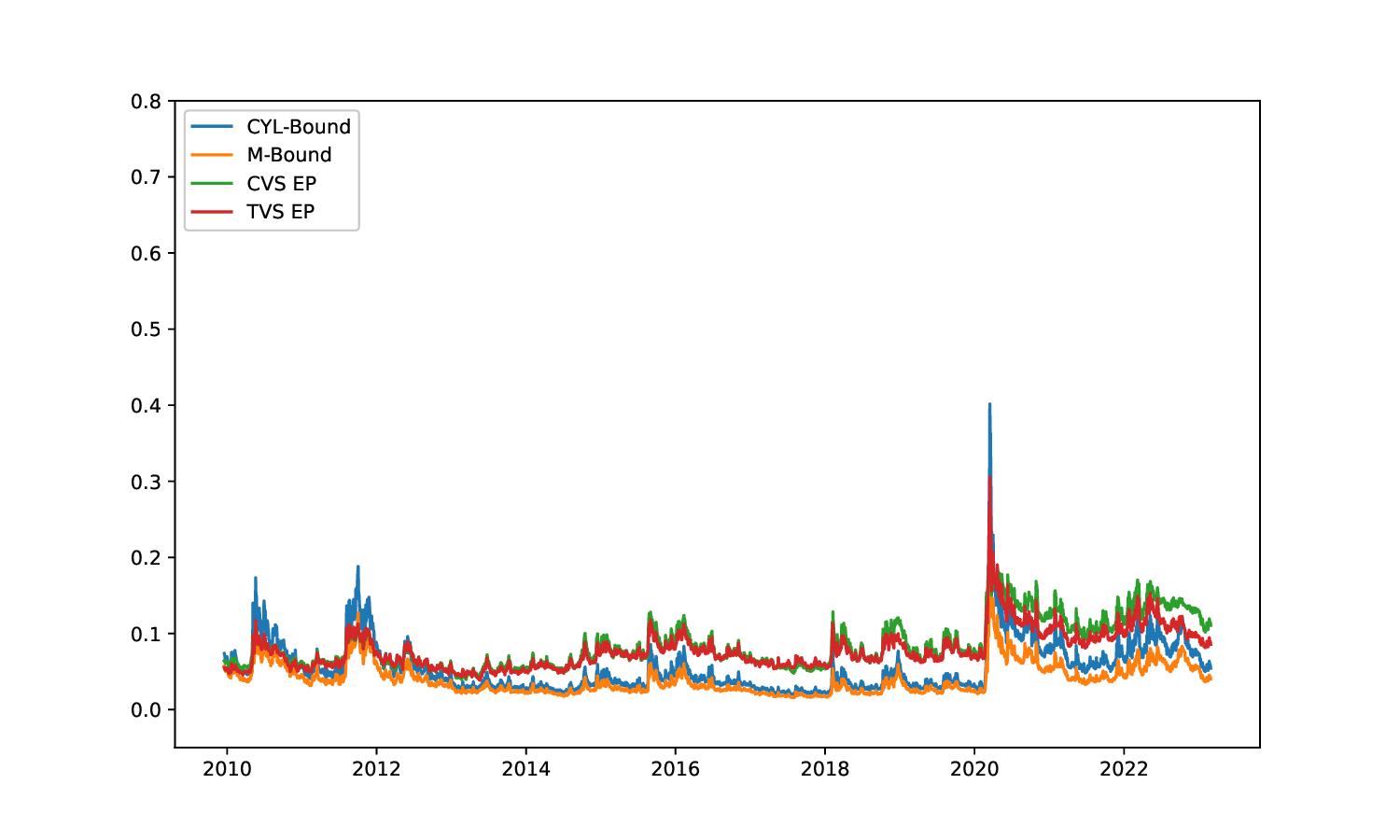}
        \vspace{0.25em}
        \footnotesize 180 days EP
        \end{minipage} &
        \begin{minipage}[t]{0.48\textwidth}\centering
        \includegraphics[width=\linewidth]{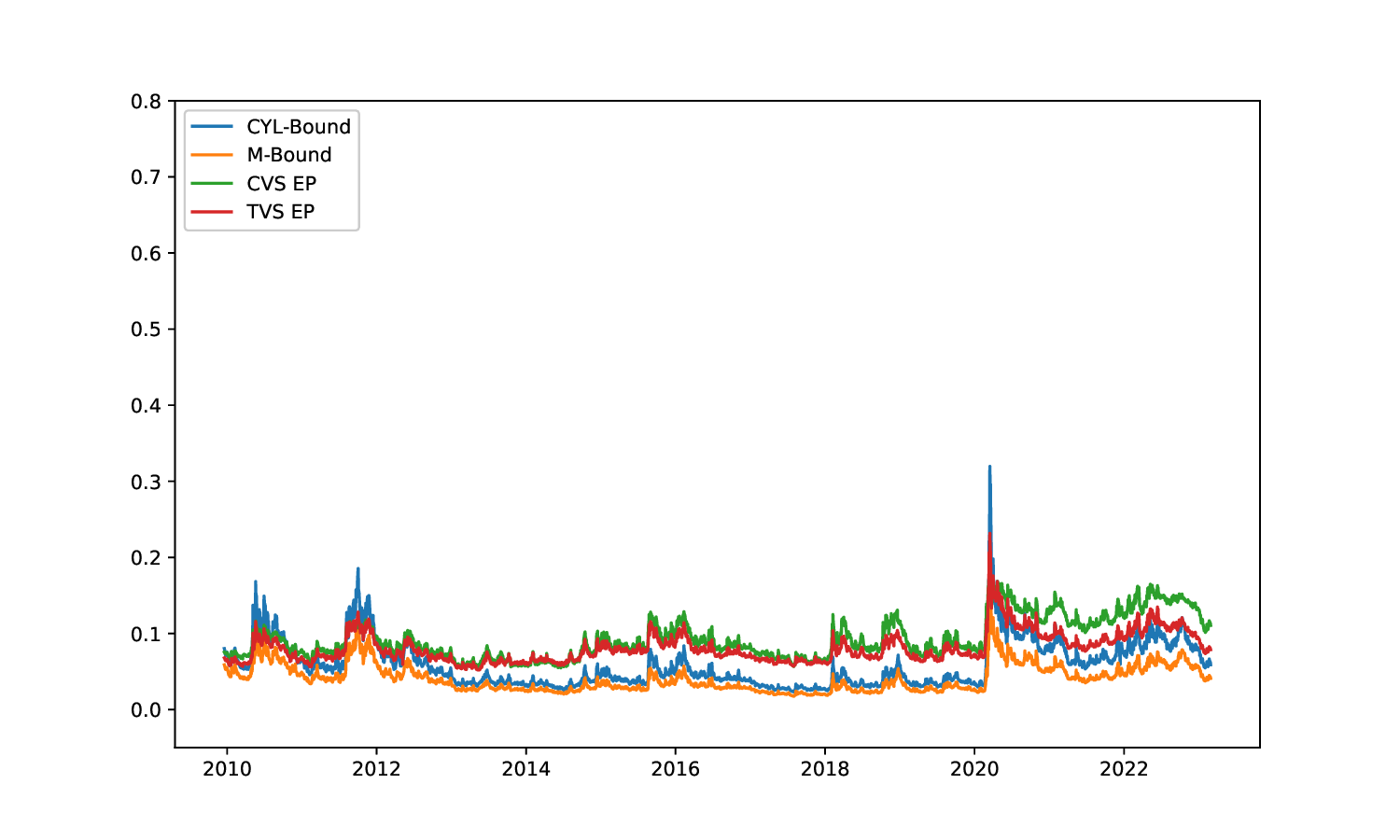}
        \vspace{0.25em}
        \footnotesize 270 days EP
        \end{minipage} \\
        \begin{minipage}[t]{0.48\textwidth}\centering
        \includegraphics[width=\linewidth]{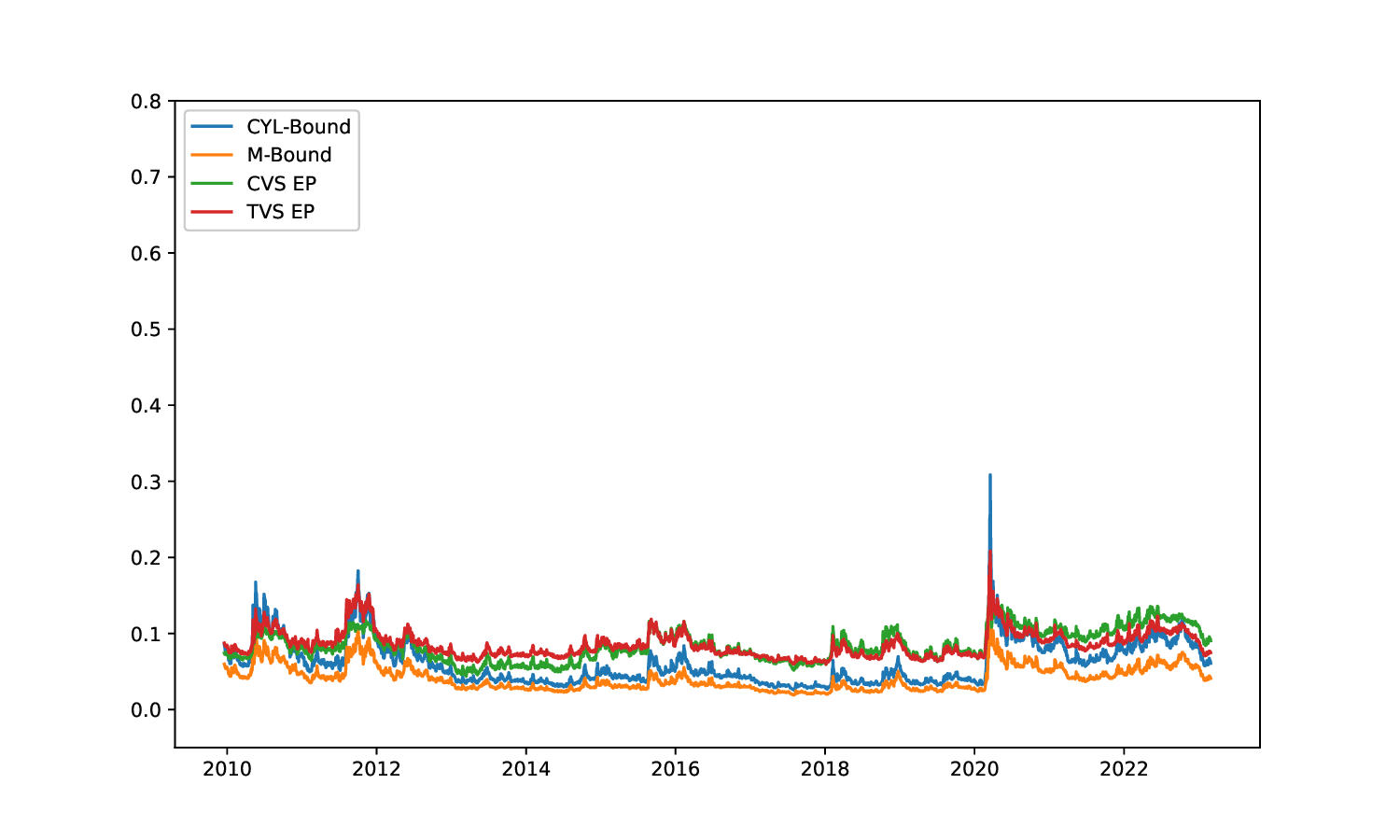}
        \vspace{0.25em}
        \footnotesize 360 days EP
        \end{minipage} &
        \begin{minipage}[t]{0.48\textwidth}\centering
        \vspace{0.25em}
        \end{minipage} \\
    \end{tabular}
    \begin{quote}
    \small
    \textit{Notes}: This table displays the time-series evolution of the daily EP for target maturities of 30, 60, 90, 120, 180, 270 and 360 days.
    The blue and orange lines denote the EP derived from the CYL-Bound and the M-Bound, respectively. The green and red line represents the out-of-sample CVS and TVS EPs.
    \end{quote}
  \end{minipage}
\end{figure}

Next, we examine the performance of out-of-sample EPs obtained from the estimated SDFs.
Based on the results in the previous section, we use EPs derived from the SDFs estimated with polynomial order 4, regardless of the maturity or estimation period.
For performance comparison, we employ the M-Bound and CYL-Bound.
The M-Bound and CYL-Bound are computed at the actual maturities surrounding the target maturity and are linearly interpolated.
We then plot the time-series evolution of the CVS and TVS EPs, along with the M-Bound and CYL-Bound, for each target maturity.

Figure \ref{fig:EPGraph} shows the time-series evolution of the daily EP for target maturities of 30, 60, 90, 120, 180, 270 and 360 days.
The blue and orange lines denote the EP derived from the CYL-Bound and the M-Bound, respectively. The green and red line represents the out-of-sample CVS and TVS EPs.

The EP calculated from the estimated SDF tends to be larger than both the M-Bound and CYL-Bound across most periods.
However, during periods of market instability such as 2020 (COVID-19 shock, the peak around 2020), the SDF based EPs fall below the M-Bound and CYL-Bound.
CVS EP and TVS EP exhibit similar levels for most periods across all maturities.
However, discrepancies between CVS and TVS EP are observed in certain instances, such as for the 2010-2014 of 120 days and 2020 onwards of the 270 days.
Furthermore, during periods of market instability such as 2020, low EPs -- including negative values -- are observed for short term maturities of CVS EP, representing a significant deviation from other EPs.
This is also inconsistent with previous studies, which generally find that higher EPs are observed during times of market instability.

Next, we compare the out-of-sample performance of the EPs calculated from the estimated SDFs against the Historical Average, as well as the EP derived from M-Bounds and CYL-Bounds.
Table~\ref{tab:oos_results_4} presents the $R_{OS}^{2}$ calculated under each set of conditions.
The left side of the table shows the results for CVS EP, while the right side displays the results for TVS EP.
The labels ``Full'', ``1st/2'', ``2nd/2'' represent the aggregation periods, consistent with the conditions defined in Table~\ref{tab:R2OSOrderFullPeriod}.
``Days'' refers to the time to maturity.
Each columns evaluates the predictive performance of the SDF based EP against three benchmarks:
Historical Average (HA) of realized ERs, M-Bound, and CYL-Bound.
$^{\ast}$, $^{\ast\ast}$ and $^{\ast\ast\ast}$ indicate significance of positiveness at the 10\%, 5\% and 1\% levels, respectively.
Following \textcite{clark2007approximately}, the statistics are computed using the method of \textcite{newey1987simple} with the lag length set to 1.5 times the days to maturity.

\begin{table}[htbp]
\centering
\scriptsize
\caption{Out-of-sample performance against Historical Average, M-Bound and CYL-Bound}
\label{tab:oos_results_4}
\renewcommand{\arraystretch}{1.2}
\begin{tabular}{cc}
\begin{tabular}{@{}cccc@{}}
\toprule
\multicolumn{4}{c}{\textbf{CVS EP}} \\ \midrule
\multicolumn{4}{c}{\textbf{Full}} \\ \midrule
Days & HA & M-Bound & CYL-Bound \\ \midrule
    30 & $-0.01\%$ & $-1.88\%$ & $-2.78\%$ \\
    60 & $1.26\%$ & $-1.39\%$ & $-3.64\%$ \\
    90 & $3.79\%^{**}$ & $0.50\%$ & $-3.71\%$ \\
    120 & $9.72\%^{**}$ & $6.58\%^{**}$ & $0.82\%$ \\
    180 & $10.78\%^{*}$ & $9.10\%^{*}$ & $0.52\%$ \\
    270 & $16.10\%^{**}$ & $17.09\%^{**}$ & $5.19\%^{*}$ \\
    360 & $17.90\%^{***}$ & $20.97\%^{***}$ & $6.14\%^{**}$ \\ \midrule
\multicolumn{4}{c}{\textbf{1st/2}} \\ \midrule
Days & HA & M-Bound & CYL-Bound \\ \midrule
    30 & $1.09\%^{**}$ & $0.49\%$ & $0.02\%$ \\
    60 & $1.69\%^{**}$ & $0.09\%$ & $-1.38\%$ \\
    90 & $4.67\%^{**}$ & $1.56\%$ & $-1.72\%$ \\
    120 & $11.65\%^{**}$ & $7.80\%^{*}$ & $2.51\%$ \\
    180 & $10.63\%^{**}$ & $7.06\%^{*}$ & $-2.81\%$ \\
    270 & $23.05\%^{**}$ & $21.54\%^{***}$ & $6.89\%$ \\
    360 & $26.20\%^{**}$ & $23.84\%^{***}$ & $4.01\%$ \\ \midrule
\multicolumn{4}{c}{\textbf{2nd/2}} \\ \midrule
Days & HA & M-Bound & CYL-Bound \\ \midrule
    30 & $-0.57\%$ & $-3.14\%$ & $-4.28\%$ \\
    60 & $1.03\%$ & $-2.21\%$ & $-4.91\%$ \\
    90 & $3.28\%$ & $-0.12\%$ & $-4.89\%$ \\
    120 & $8.57\%$ & $5.86\%^{*}$ & $-0.18\%$ \\
    180 & $10.87\%$ & $10.32\%$ & $2.49\%$ \\
    270 & $12.14\%$ & $14.68\%^{*}$ & $4.31\%$ \\
    360 & $13.32\%^{*}$ & $19.55\%^{***}$ & $7.11\%^{*}$ \\ \bottomrule
\end{tabular}
&
\begin{tabular}{@{}cccc@{}}
\toprule
\multicolumn{4}{c}{\textbf{TVS EP}} \\ \midrule
\multicolumn{4}{c}{\textbf{Full}} \\ \midrule
Days & HA & M-Bound & CYL-Bound \\ \midrule
    30 & $1.80\%^{**}$ & $-0.04\%$ & $-0.92\%$ \\
    60 & $4.74\%^{**}$ & $2.18\%^{**}$ & $0.00\%$ \\
    90 & $7.49\%^{**}$ & $4.33\%^{**}$ & $0.28\%$ \\
    120 & $12.05\%^{**}$ & $9.00\%^{**}$ & $3.39\%^{*}$ \\
    180 & $11.07\%^{*}$ & $9.39\%^{**}$ & $0.84\%$ \\
    270 & $16.18\%^{**}$ & $17.17\%^{***}$ & $5.28\%^{*}$ \\
    360 & $22.57\%^{***}$ & $25.46\%^{***}$ & $11.47\%^{***}$ \\ \midrule
\multicolumn{4}{c}{\textbf{1st/2}} \\ \midrule
Days & HA & M-Bound & CYL-Bound \\ \midrule
    30 & $1.55\%^{**}$ & $0.96\%$ & $0.49\%$ \\
    60 & $3.82\%^{**}$ & $2.26\%^{*}$ & $0.82\%$ \\
    90 & $6.42\%^{**}$ & $3.37\%^{*}$ & $0.15\%$ \\
    120 & $12.13\%^{**}$ & $8.30\%^{**}$ & $3.04\%$ \\
    180 & $10.69\%^{**}$ & $7.12\%^{*}$ & $-2.75\%$ \\
    270 & $21.03\%^{**}$ & $19.48\%^{***}$ & $4.44\%$ \\
    360 & $36.16\%^{***}$ & $34.12\%^{***}$ & $16.96\%^{*}$ \\ \midrule
\multicolumn{4}{c}{\textbf{2nd/2}} \\ \midrule
Days & HA & M-Bound & CYL-Bound \\ \midrule
    30 & $1.93\%^{*}$ & $-0.57\%$ & $-1.68\%$ \\
    60 & $5.23\%^{*}$ & $2.13\%$ & $-0.46\%$ \\
    90 & $8.12\%^{*}$ & $4.89\%^{*}$ & $0.36\%$ \\
    120 & $12.01\%$ & $9.41\%^{*}$ & $3.59\%$ \\
    180 & $11.30\%$ & $10.75\%^{*}$ & $2.96\%$ \\
    270 & $13.42\%$ & $15.92\%^{**}$ & $5.71\%^{*}$ \\
    360 & $15.06\%^{**}$ & $21.16\%^{***}$ & $8.97\%^{**}$ \\ \bottomrule
\end{tabular}
\end{tabular}
\begin{quote}
\small
\textit{Notes}: This table reports the out-of-sample predictive $R^2$ ($R_{OS}^2$) based on the statistic proposed by \textcite{campbell2008predicting},
which evaluates the predictive performance of the SDF based EP against three benchmarks: (1) Historical Average (HA) of realized ERs, (2) M-Bound, and (3) CYL-Bound.
A positive value of $R_{OS}^2$ indicates that the SDF based forecast outperforms the benchmark in terms of Mean Squared Predictive Error (MSPE).
The left side of the table shows the results for CVS EP, while the right side displays the results for TVS EP.
The labels ``Full'', ``1st/2'', ``2nd/2'' represent the aggregation periods, consistent with the conditions defined in Table~\ref{tab:R2OSOrderFullPeriod}.
``Days'' refers to the time to maturity.
Each columns evaluates the predictive performance of the SDF based EP against three benchmarks:
Historical Average (HA) of realized ERs, M-Bound, and CYL-Bound.
$^{\ast}$, $^{\ast\ast}$ and $^{\ast\ast\ast}$ indicate significance of positiveness at the 10\%, 5\% and 1\% levels, respectively.\end{quote}
\end{table}

Over the full sample period, the TVS EP exhibits consistently strong out-of-sample performance,
generally outperforming the HA, M-Bound, and CYL-Bound across most maturities.
This advantage becomes particularly pronounced at longer horizons, most notably at the 360 day maturity, where the predictive $R_{OS}^2$ attains high statistical significance at the 1\% level.
Although the short horizon results show weaker significance, the TVS EP still delivers positive $R_{OS}^2$ values at nearly all maturities, indicating that the model remains competitive even in noisier short term environments.
Across the full sample as well as in both sub-periods, the CYL-Bound consistently serves as the toughest benchmark to outperform,
as models generate the smallest incremental $R_{OS}^2$ relative to it.
Despite this, the TVS EP frequently outperforms the CYL-Bound demonstrating its robustness even against the most stringent benchmark.

A comparison between the CVS and TVS EP reveals broadly similar qualitative patterns.
Both models improve markedly at longer maturities, but the magnitude and statistical strength of the predictive gains are generally larger for the TVS EP.
The TVS specification not only produces higher $R_{OS}^2$ values across most maturities but also achieves statistical significance more frequently,
suggesting that incorporating time-varying volatility yields a more accurate and stable representation of the EP's risk-return trade-off.
The significance levels reported in the table further reinforce that the superior performance of the TVS EP is not attributable to sampling variation but is statistically well supported,
particularly at longer horizons where predictive accuracy peaks.
One of the most striking insights from the results is the clear horizon dependence of predictive power.
For both the CVS and TVS EPs, the $R_{OS}^2$ values tend to be near zero or even negative at the 30 day maturity but increase substantially as the horizon extends toward 360 days.
This pattern indicates that while SDF based EPs may struggle to filter out short term market noise,
they are highly effective at capturing the fundamental, long term drivers of the equity premium.

\begin{figure}[htbp]
  \centering
  \setlength{\tabcolsep}{0pt}
  \begin{minipage}[t]{1.0\hsize}
    \centering
    \caption{Cumulative Squared Prediction Error of TVS EP against Benchmarks}
    \label{fig:CSPE-TVS}
    \begin{tabular}{cc}
        \begin{minipage}[t]{0.45\textwidth}\centering
        \includegraphics[width=\linewidth]{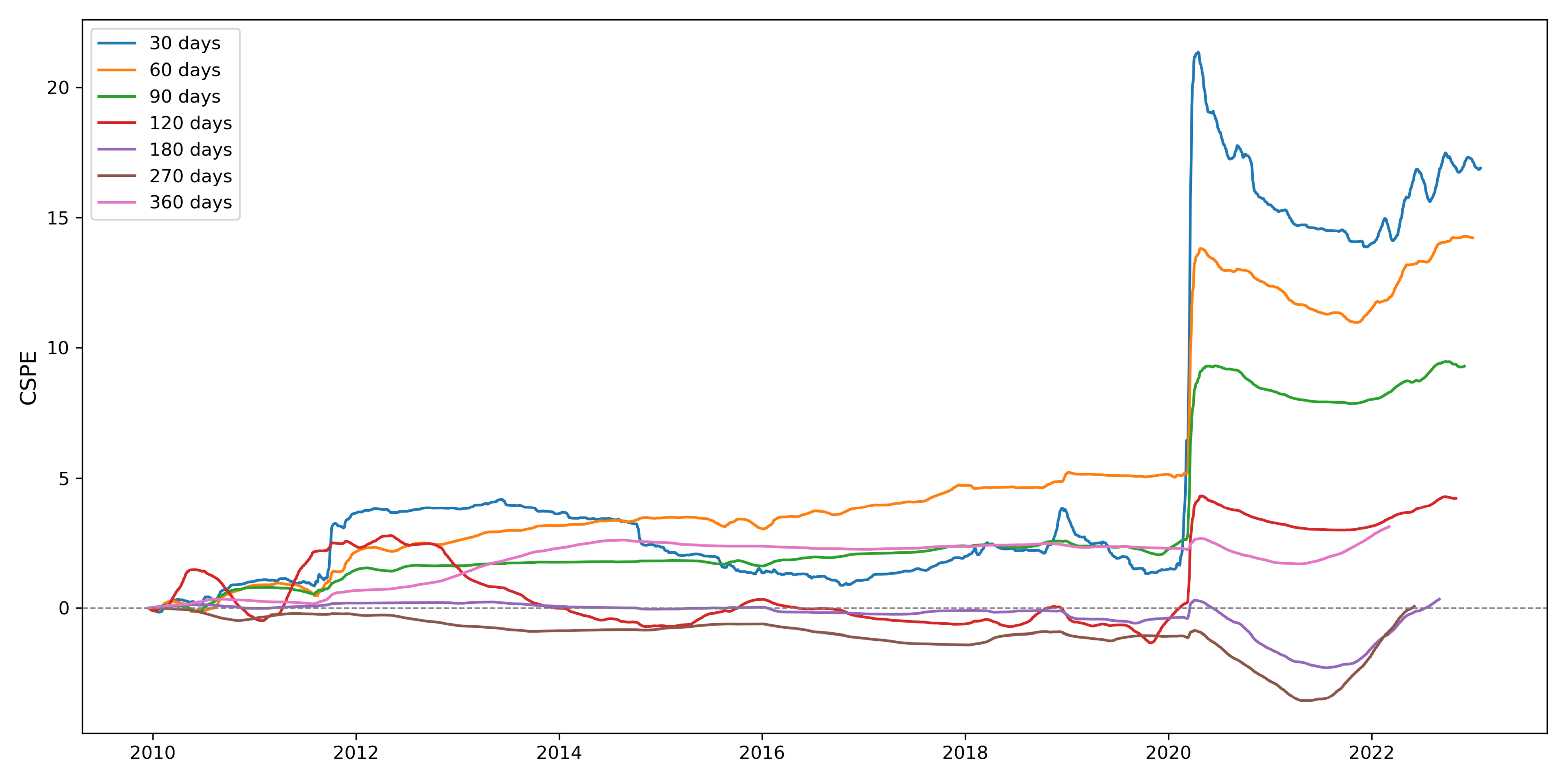}
        \footnotesize Against CVS EP
        \end{minipage} &
        \begin{minipage}[t]{0.45\textwidth}\centering
        \includegraphics[width=\linewidth]{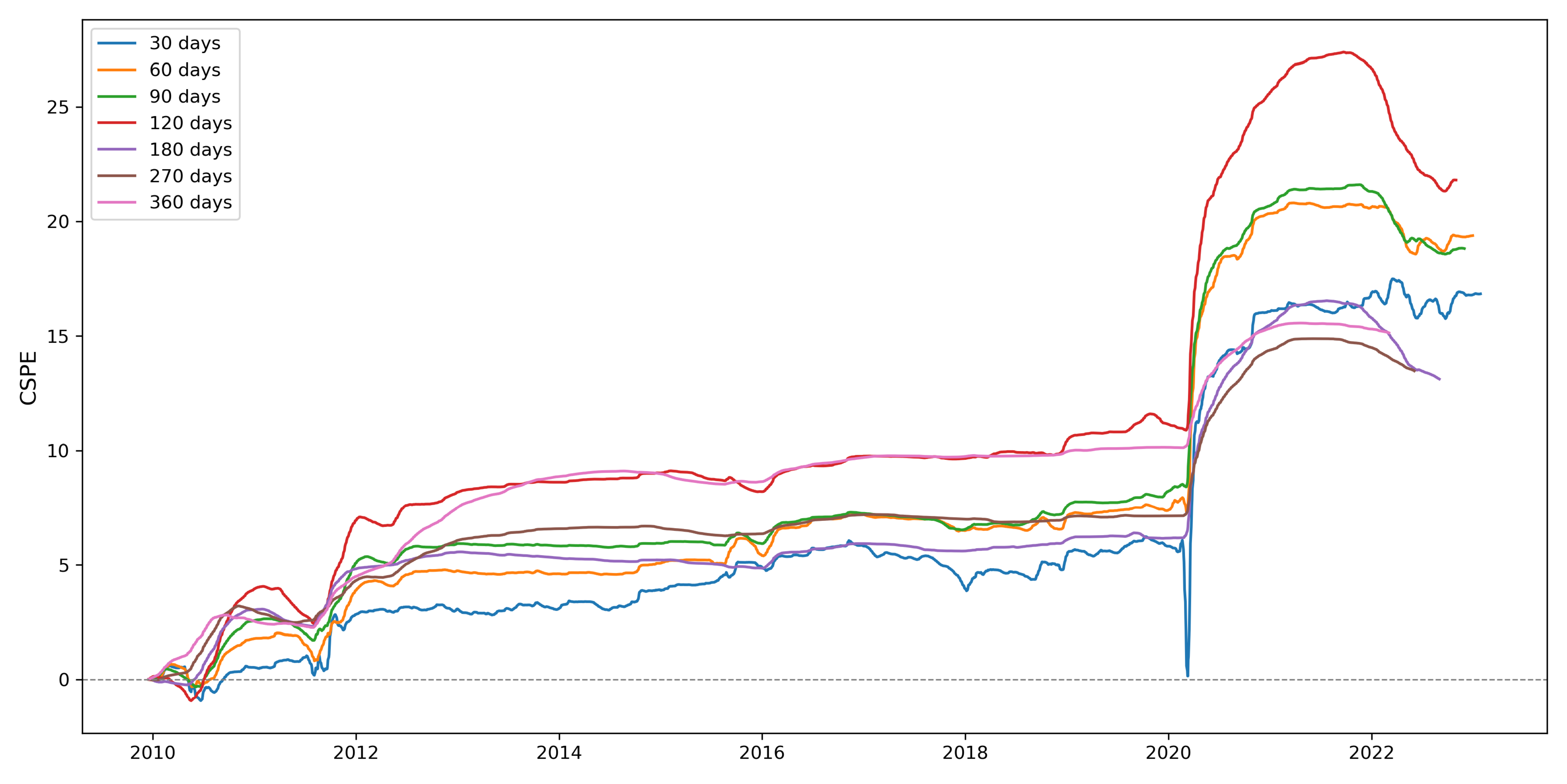}
        \footnotesize Against HA
        \end{minipage} \\\hspace{0.5em}
        \begin{minipage}[t]{0.45\textwidth}\centering
        \includegraphics[width=\linewidth]{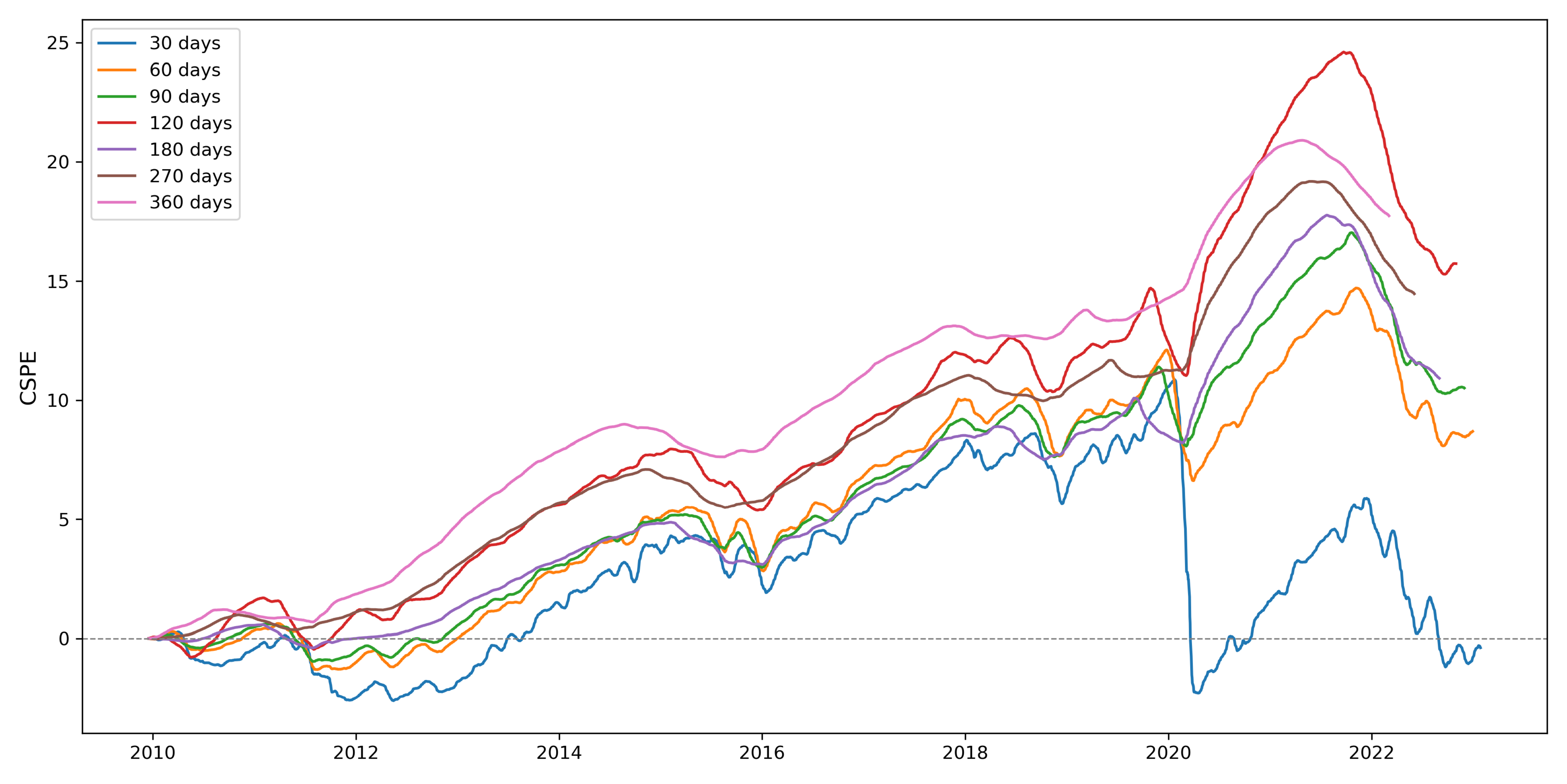}
        \footnotesize Against M-Bound
        \end{minipage} &
        \begin{minipage}[t]{0.45\textwidth}\centering
        \includegraphics[width=\linewidth]{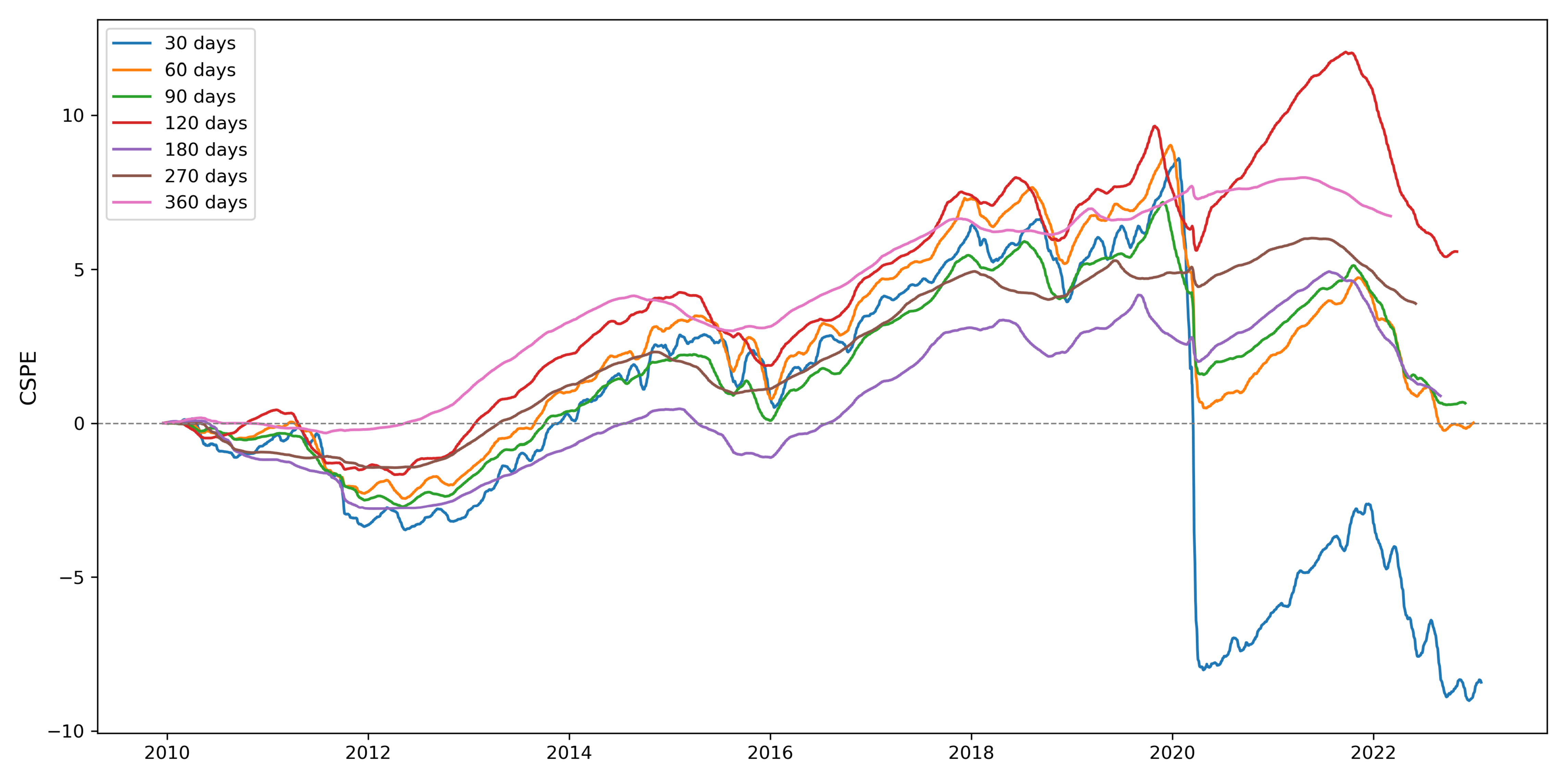}
        \footnotesize Against CYL-Bound
        \end{minipage} \\
    \end{tabular}
    \begin{quote}
    \small
    \textit{Notes}: The four panels plot the cumulative squared prediction error (CSPE) of the TVS EP relative to CVS EP, the historical average (HA) of realized excess returns, M-Bound, and CYL-Bound.
    Each line corresponds to a target maturity.
    At each date, CSPE is calculated as the cumulative difference between the squared prediction errors of the benchmark and the TVS EP.
    Thus, an increasing CSPE indicates that the TVS EP has a smaller cumulative squared prediction error than the corresponding benchmark.
    \end{quote}
  \end{minipage}
\end{figure}

To examine the evolution of the relative predictive performance of the TVS EP, we plot, following \textcite{goyal2003predicting}, the cumulative squared prediction error (CSPE)
\footnote{For benchmark $b$, let $EP_{t}^{(b)}$ denote the benchmark forecast. The CSPE through date $u$ is defined as $\mathrm{CSPE}^{(b)}_{u}=\sum_{t\leq u}\left\{\left(EPR_{t}-EP_{t}^{(b)}\right)^{2}-\left(EPR_{t}-EP_{t}^{\mathrm{TVS}}\right)^{2}\right\}$.}
relative to CVS EP, HA, M-Bound, and CYL-Bound in Figure \ref{fig:CSPE-TVS}.
An increasing CSPE indicates that the TVS EP outperforms the corresponding benchmark instantaneously.
The CSPE plots show the timing of relative gains and losses in predictive performance.

From 2010 to late 2011, the CSPE relative to HA is generally positive, whereas that relative to CYL-Bound is generally negative.
Thus, the TVS EP tends to outperform HA but underperform CYL-Bound during this period.
As shown in Figure \ref{fig:EPGraph}, this period contains two pronounced episodes--in mid-2010 and late 2011--during which the TVS EP, CVS EP, and the two bounds rise together.
Around these episodes, the CSPE relative to HA rises, whereas that relative to CYL-Bound declines.
From 2012 to late 2019, the CSPE relative to the M-Bound and CYL-Bound exhibits an overall upward trend across most maturities, despite occasional periods of decline.
By contrast, the CSPE relative to the CVS EP and HA fluctuates across maturities and does not show the clear pattern.
Figure \ref{fig:EPGraph} shows that the TVS EP and CVS EP are higher than the two bounds during most of this period.
The resulting CSPE paths indicate that the TVS EP achieved superior performance to the bounds for many maturities over this period.
Around the COVID-19 shock in early 2020, the CSPE paths change considerably.
Relative to CVS EP and HA, the CSPE rises for several maturities, while the 180 and 270 day series relative to CVS EP show a weaker response.
On the other hand, the CSPE relative to M-Bound and CYL-Bound declines, especially at short maturities.
In Figure \ref{fig:EPGraph}, the CYL-Bound increases sharply around the shock across maturities, while the CSPE relative to CYL-Bound declines during the same episode.
After the initial COVID-19 episode, the CSPE paths relative to the bounds are non-monotonic: several series recover before declining again later in the period.
Thus, the post-2020 results do not indicate that the relative performance of the TVS EP has been steadily weakening.

To summarize the time-series variation in the CSPE plots, Table \ref{tab:oos_results_4_3} reports the $R_{OS}^{2}$ values of the TVS EP and CVS EP relative to HA, M-Bound, and CYL-Bound for each subsample\footnote{For significance testing, the sample period is divided into approximately equal-length subsamples to make statistical power more comparable across subsamples.
For the 360 day maturity, realized excess returns are observed 360 days after the forecast date;
consequently, the third subsample contains 4 years and 2 months of observations.}.
In the ``1st/3'', the TVS EP has positive $R_{OS}^{2}$ point estimates relative to all three benchmarks at most maturities.
The results are more favorable at longer maturities, while the evidence relative to CYL-Bound is limited.
In the ``2nd/3'', the TVS EP has positive $R_{OS}^{2}$ values relative to M-Bound and CYL-Bound at every maturity, and most estimates are statistically significant at the 1\% level.
These results are consistent with the increase in CSPE relative to the bounds over the corresponding period.
In the ``3rd/3'', which summarizes performance over the later sample rather than the COVID-19 shock alone, the TVS EP continues to have positive $R_{OS}^{2}$ values relative to HA at every maturity.
Its performance relative to the bounds is less favorable, particularly at shorter maturities and relative to CYL-Bound.
Overall, the comparison with M-Bound and CYL-Bound is most favorable in the ``2nd/3'', whereas the results in the ``1st/3'' and ``3rd/3'' are weaker and vary across maturities.
Comparing the $R_{OS}^{2}$ values of the CVS EP and TVS EP relative to the same benchmark,
the TVS EP has higher values at most maturities in the ``1st/3'' and ``3rd/3'' subsamples,
whereas the results are mixed in the ``2nd/3'' subsample.
This finding is also consistent with the results relative to CVS EP in Figure \ref{fig:CSPE-TVS}.

\begin{table}[H]
\centering
\scriptsize
\caption{Out-of-sample performance against the Historical Average, M-Bound and CYL-Bound (1st/3, 2nd/3, 3rd/3)}
\label{tab:oos_results_4_3}
\renewcommand{\arraystretch}{1.2}
\begin{tabular}{cc}
\begin{tabular}{@{}cccc@{}}
\toprule
\multicolumn{4}{c}{\textbf{CVS EP}} \\ \midrule
\multicolumn{4}{c}{\textbf{1st/3}} \\ \midrule
Days & HA & M-Bound & CYL-Bound \\ \midrule
    30 & $-0.23\%$ & $-0.88\%$ & $-1.38\%$ \\
    60 & $1.28\%$ & $-0.35\%$ & $-1.98\%$ \\
    90 & $5.36\%^{**}$ & $1.75\%$ & $-1.98\%$ \\
    120 & $14.84\%^{***}$ & $10.14\%^{*}$ & $4.28\%$ \\
    180 & $13.21\%^{***}$ & $8.39\%^{**}$ & $-2.66\%$ \\
    270 & $28.25\%^{***}$ & $25.58\%^{***}$ & $9.83\%$ \\
    360 & $30.63\%^{***}$ & $27.05\%^{***}$ & $6.05\%$ \\ \midrule
\multicolumn{4}{c}{\textbf{2nd/3}} \\ \midrule
Days & HA & M-Bound & CYL-Bound \\ \midrule
    30 & $2.39\%^{*}$ & $6.35\%^{***}$ & $5.89\%^{***}$ \\
    60 & $0.62\%$ & $9.12\%^{***}$ & $7.70\%^{***}$ \\
    90 & $0.22\%$ & $15.11\%^{***}$ & $12.55\%^{***}$ \\
    120 & $8.06\%^{*}$ & $27.08\%^{***}$ & $23.77\%^{***}$ \\
    180 & $3.31\%$ & $28.90\%^{***}$ & $23.35\%^{**}$ \\
    270 & $10.00\%$ & $40.62\%^{***}$ & $32.95\%^{**}$ \\
    360 & $9.46\%^{*}$ & $40.47\%^{***}$ & $30.08\%^{***}$ \\ \midrule
\multicolumn{4}{c}{\textbf{3rd/3}} \\ \midrule
Days & HA & M-Bound & CYL-Bound \\ \midrule
    30 & $-0.43\%$ & $-4.28\%$ & $-5.48\%$ \\
    60 & $1.41\%$ & $-4.78\%$ & $-7.62\%$ \\
    90 & $3.74\%$ & $-4.19\%$ & $-9.21\%$ \\
    120 & $7.15\%$ & $-1.12\%$ & $-7.60\%$ \\
    180 & $10.89\%$ & $3.29\%$ & $-4.79\%$ \\
    270 & $10.56\%$ & $4.59\%$ & $-6.13\%$ \\
    360 & $12.47\%^{*}$ & $10.67\%^{*}$ & $-2.05\%$ \\ \bottomrule
\end{tabular}
&
\begin{tabular}{@{}cccc@{}}
\toprule
\multicolumn{4}{c}{\textbf{TVS EP}} \\ \midrule
\multicolumn{4}{c}{\textbf{1st/3}} \\ \midrule
Days & HA & M-Bound & CYL-Bound \\ \midrule
    30 & $1.24\%^{*}$ & $0.60\%$ & $0.11\%$ \\
    60 & $4.08\%^{**}$ & $2.50\%$ & $0.91\%$ \\
    90 & $7.68\%^{**}$ & $4.15\%^{*}$ & $0.51\%$ \\
    120 & $14.83\%^{***}$ & $10.14\%^{***}$ & $4.28\%$ \\
    180 & $13.39\%^{***}$ & $8.58\%^{**}$ & $-2.45\%$ \\
    270 & $24.92\%^{***}$ & $22.13\%^{***}$ & $5.65\%$ \\
    360 & $41.24\%^{***}$ & $38.21\%^{***}$ & $20.42\%^{*}$ \\ \midrule
\multicolumn{4}{c}{\textbf{2nd/3}} \\ \midrule
Days & HA & M-Bound & CYL-Bound \\ \midrule
    30 & $1.01\%$ & $5.02\%^{***}$ & $4.55\%^{***}$ \\
    60 & $3.36\%^{**}$ & $11.62\%^{***}$ & $10.24\%^{***}$ \\
    90 & $2.24\%$ & $16.83\%^{***}$ & $14.33\%^{***}$ \\
    120 & $5.02\%^{*}$ & $24.67\%^{***}$ & $21.26\%^{***}$ \\
    180 & $2.07\%$ & $28.00\%^{***}$ & $22.37\%^{**}$ \\
    270 & $4.49\%$ & $36.98\%^{***}$ & $28.84\%^{**}$ \\
    360 & $10.40\%^{***}$ & $41.09\%^{***}$ & $30.81\%^{***}$ \\ \midrule
\multicolumn{4}{c}{\textbf{3rd/3}} \\ \midrule
Days & HA & M-Bound & CYL-Bound \\ \midrule
    30 & $2.22\%^{*}$ & $-1.52\%$ & $-2.69\%$ \\
    60 & $5.37\%^{*}$ & $-0.57\%$ & $-3.29\%$ \\
    90 & $8.53\%^{*}$ & $0.99\%$ & $-3.78\%$ \\
    120 & $11.88\%$ & $4.04\%$ & $-2.12\%$ \\
    180 & $11.57\%$ & $4.03\%$ & $-3.99\%$ \\
    270 & $13.66\%$ & $7.90\%$ & $-2.45\%$ \\
    360 & $14.55\%^{**}$ & $12.79\%^{*}$ & $0.37\%$ \\ \bottomrule
\end{tabular}
\end{tabular}
\begin{quote}
\small
\textit{Notes}: This table reports the out-of-sample predictive $R^2$ ($R_{OS}^2$) based on the statistic proposed by \textcite{campbell2008predicting},
which evaluates the predictive performance of the SDF based EP against three benchmarks: (1) Historical Average (HA) of realized ERs, (2) M-Bound, and (3) CYL-Bound.
A positive value of $R_{OS}^2$ indicates that the SDF based forecast outperforms the benchmark in terms of Mean Squared Predictive Error (MSPE).
The left side of the table shows the results for CVS EP, while the right side displays the results for TVS EP.
``1st/3'' is from Dec 18, 2009 to Dec 17, 2013.
``2nd/3'' is from Dec 18, 2013 to Dec 15, 2017.
``3rd/3'' is from Dec 18, 2017 to Feb 28, 2023.
``Days'' refers to the time to maturity.
Each column reports the predictive performance of the SDF based EP against the three benchmarks.
$^{\ast}$, $^{\ast\ast}$ and $^{\ast\ast\ast}$ indicate that $R_{OS}^2$ is significantly positive at the 10\%, 5\% and 1\% levels, respectively.
\end{quote}
\end{table}

\section{Analysis of Estimated Stochastic Discount Factor and Equity Premium}
\label{sec:AnalysisOfEstimatedSDFAndEP}

\subsection{Shape of Estimated Stochastid Discount Factor}

\begin{figure}[htbp]
  \centering
  \setlength{\tabcolsep}{0pt}
  \begin{minipage}[t]{1.0\hsize}
    \centering
    \begin{tabular}{ccc}
        \begin{minipage}[t]{0.33\textwidth}\centering
        \includegraphics[width=\linewidth]{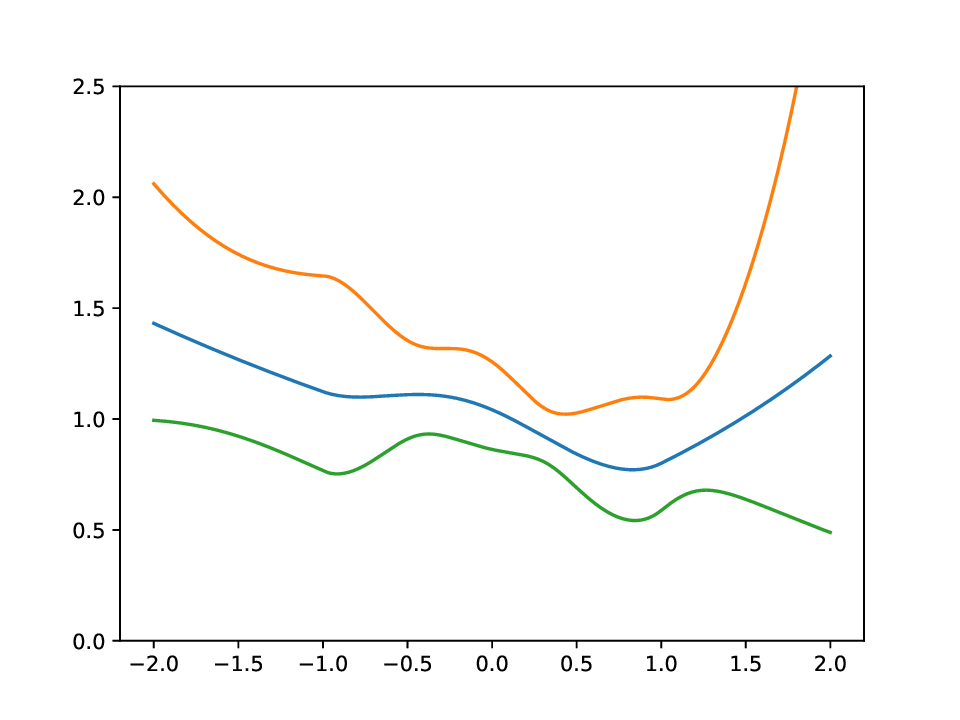}
        \footnotesize 30 days CVS SDF
        \end{minipage} &
        \begin{minipage}[t]{0.33\textwidth}\centering
        \includegraphics[width=\linewidth]{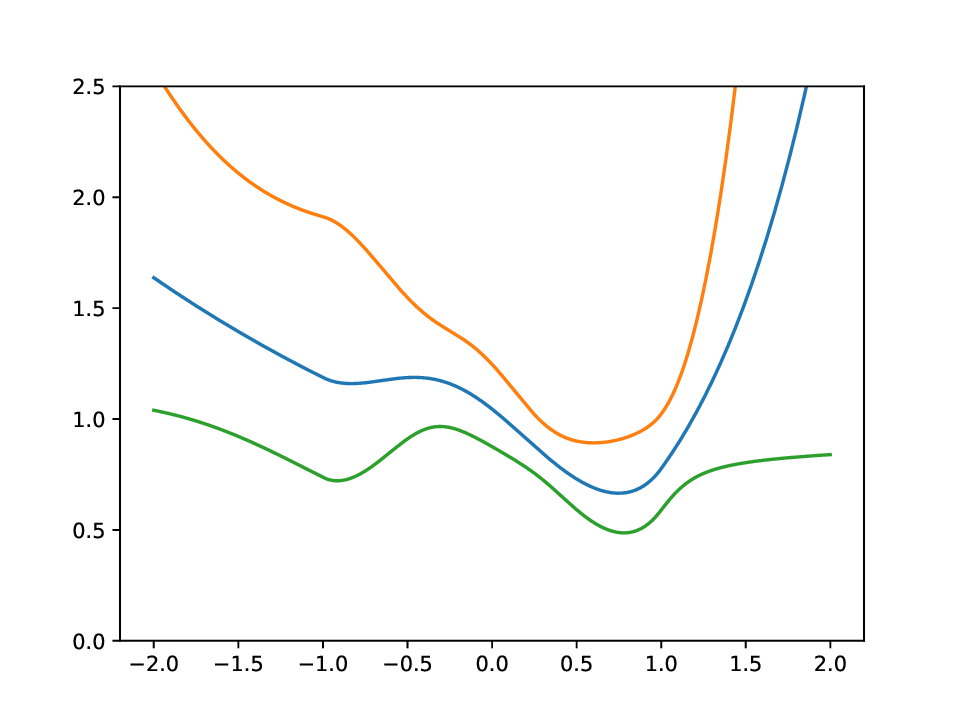}
        \footnotesize 60 days CVS SDF
        \end{minipage} &
        \begin{minipage}[t]{0.33\textwidth}\centering
        \includegraphics[width=\linewidth]{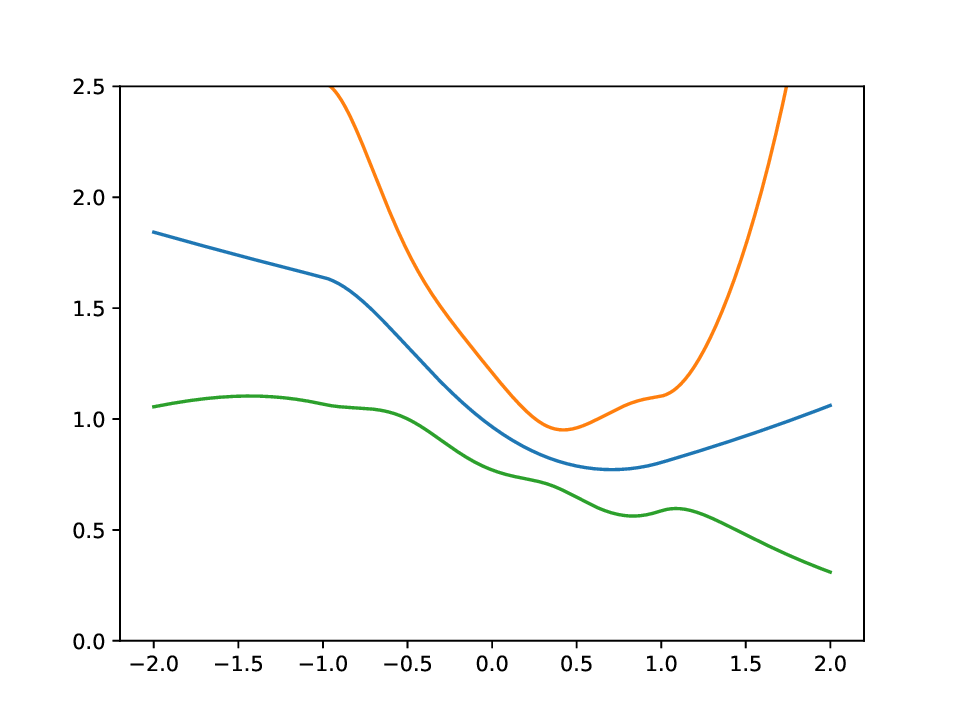}
        \footnotesize 90 days CVS SDF
        \end{minipage} \\
        \begin{minipage}[t]{0.33\textwidth}\centering
        \includegraphics[width=\linewidth]{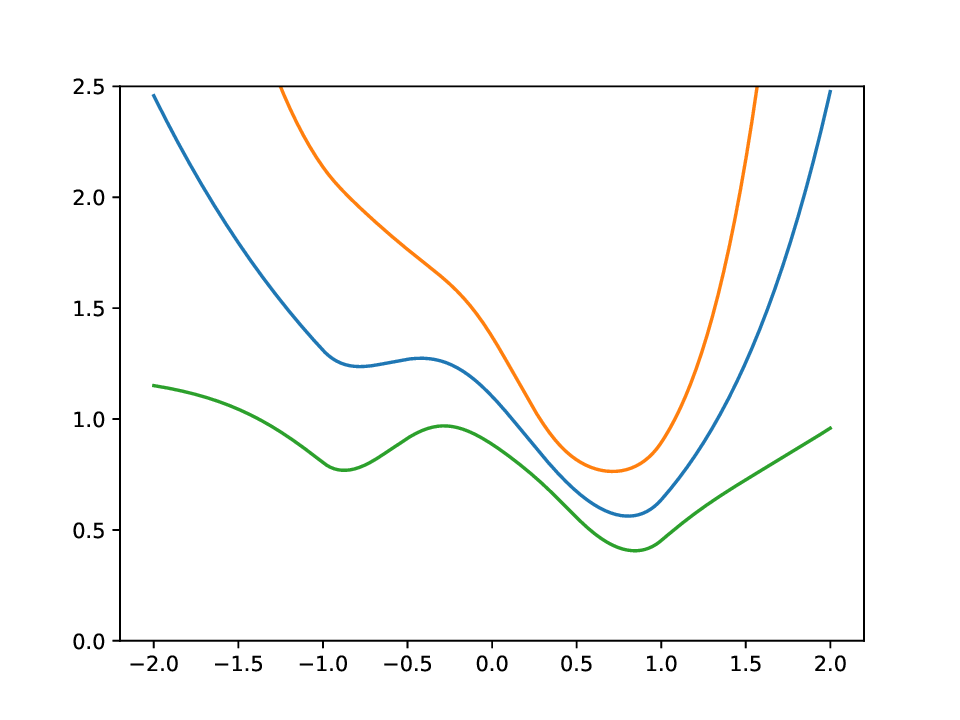}
        \footnotesize 120 days CVS SDF
        \end{minipage} &
        \begin{minipage}[t]{0.33\textwidth}\centering
        \includegraphics[width=\linewidth]{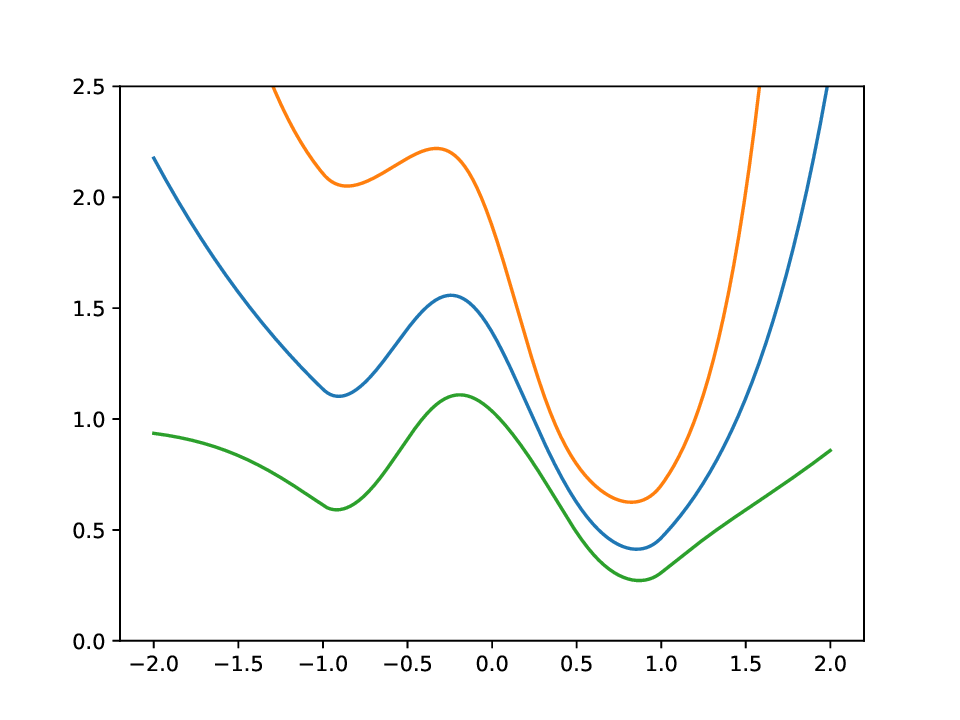}
        \footnotesize 180 days CVS SDF
        \end{minipage} &
        \begin{minipage}[t]{0.33\textwidth}\centering
        \includegraphics[width=\linewidth]{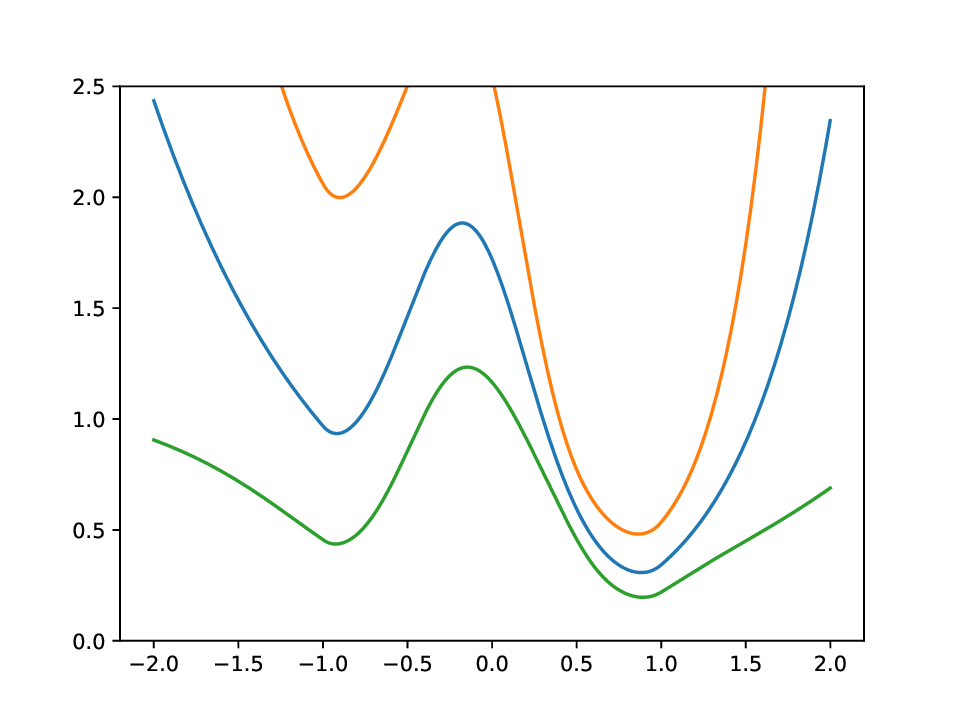}
        \footnotesize 270 days CVS SDF
        \end{minipage} \\
        \begin{minipage}[t]{0.33\textwidth}\centering
        \vspace{0.25em}
        \end{minipage} &
        \begin{minipage}[t]{0.33\textwidth}\centering
        \includegraphics[width=\linewidth]{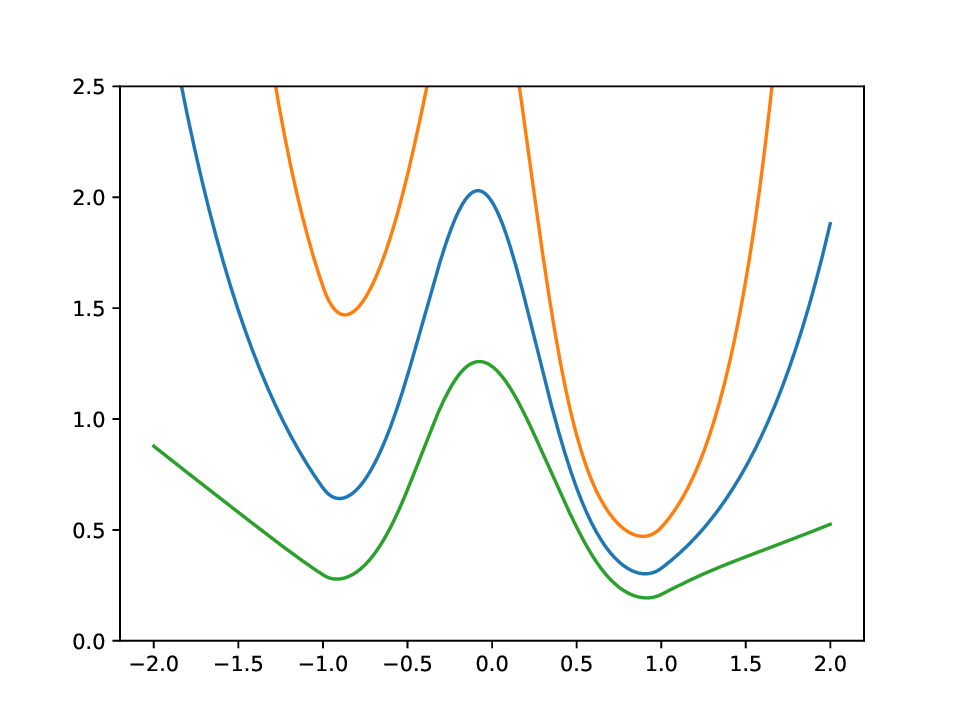}
        \footnotesize 360 days CVS SDF
        \end{minipage} &
        \begin{minipage}[t]{0.33\textwidth}\centering
        \vspace{0.25em}
        \end{minipage} \\
    \end{tabular}
    \caption{Estimated CVS SDF with Full-Sample Data}
    \label{fig:AllTermCVSSDF}
    \begin{quote}
    \small
    \textit{Notes}: This table displays the estimated functional forms of the CVS SDF, $\exp\{f_{p,n}(z)\}$, across different maturities.
    In each plot, the horizontal axis represents the normalized log return $z$.
    The blue line depicts the estimated SDF, while the orange and green lines represent the upper and lower bounds of the $2\sigma$ confidence interval, respectively.
    All results are based on the full sample data from Jan 1996 to Feb 2023.
    \end{quote}
  \end{minipage}
\end{figure}

\begin{figure}[htbp]
  \centering
  \setlength{\tabcolsep}{0pt}
  \begin{minipage}[t]{1.0\hsize}
    \centering
    \begin{tabular}{ccc}
        \begin{minipage}[t]{0.33\textwidth}\centering
        \includegraphics[width=\linewidth]{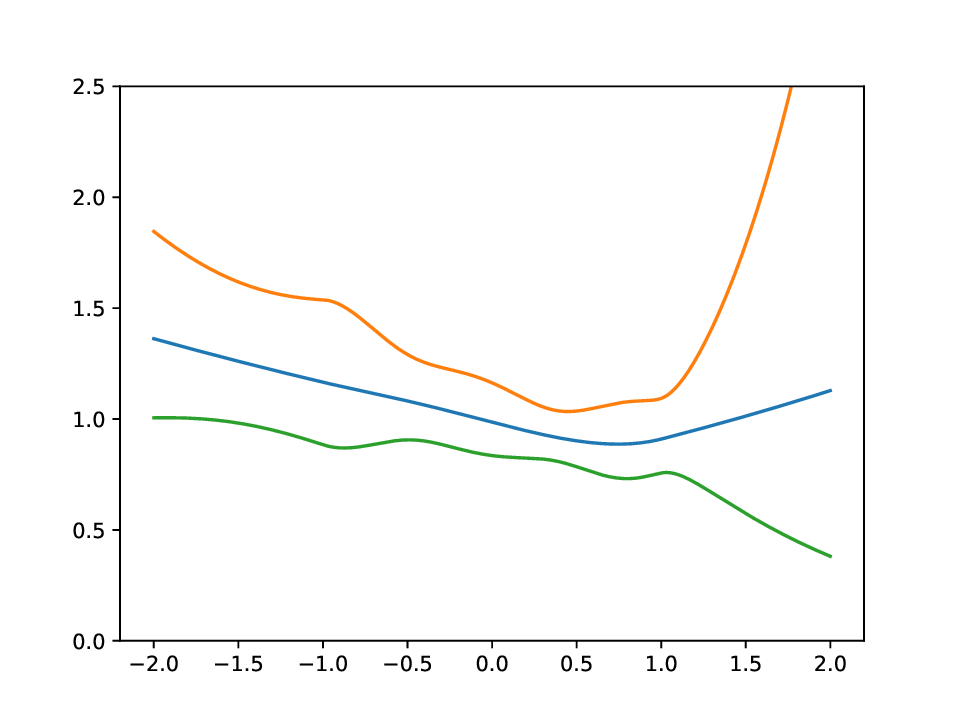}
        \footnotesize 30 days TVS SDF
        \end{minipage} &
        \begin{minipage}[t]{0.33\textwidth}\centering
        \includegraphics[width=\linewidth]{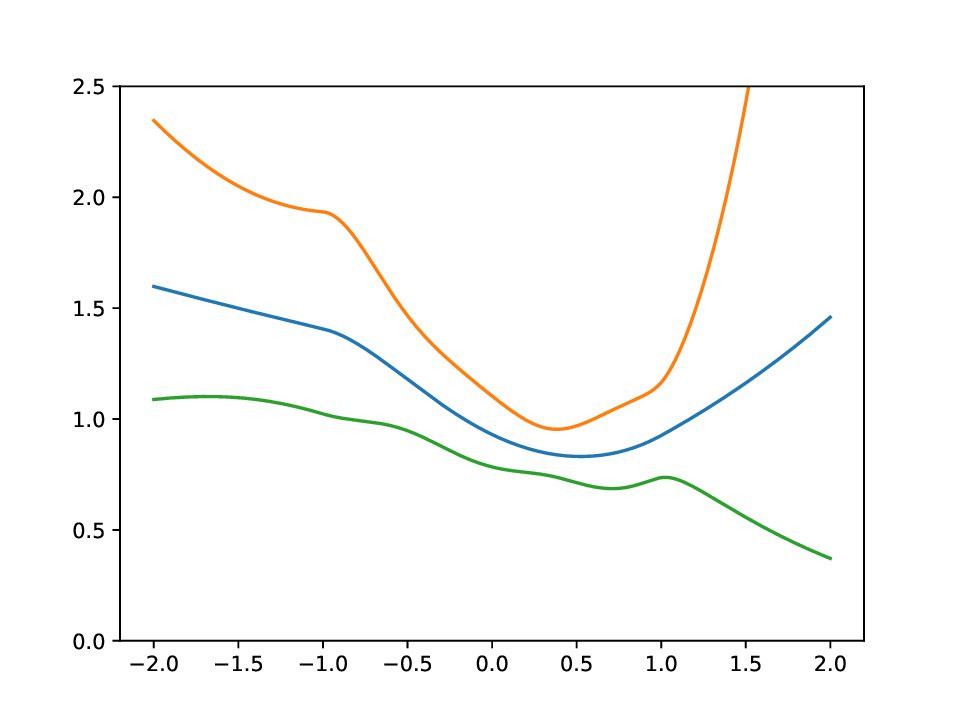}
        \footnotesize 60 days TVS SDF
        \end{minipage} &
        \begin{minipage}[t]{0.33\textwidth}\centering
        \includegraphics[width=\linewidth]{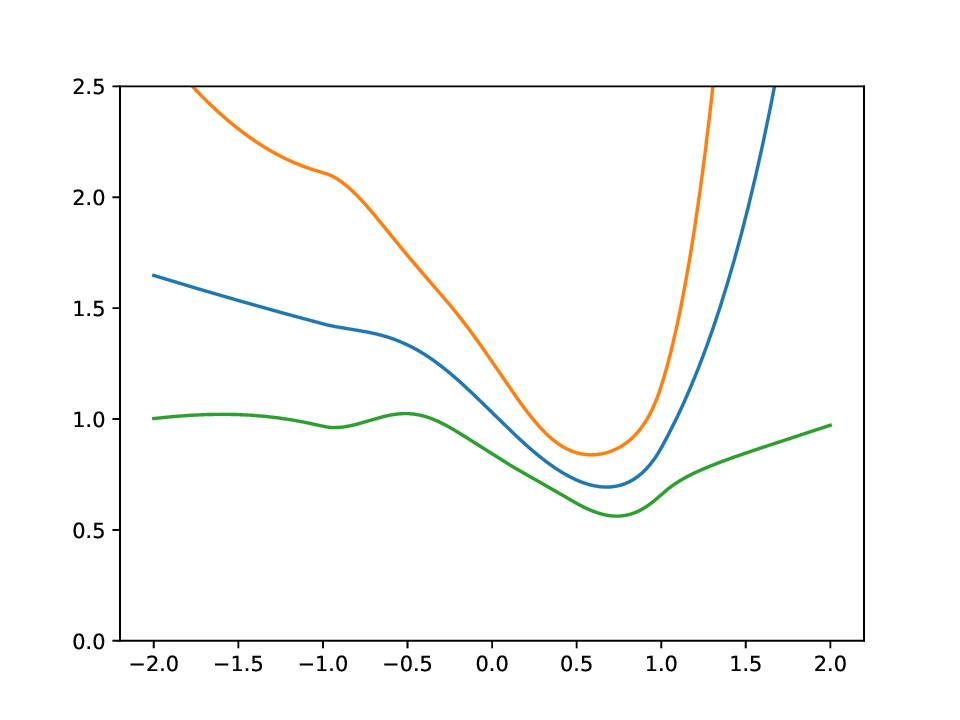}
        \footnotesize 90 days TVS SDF
        \end{minipage} \\
        \begin{minipage}[t]{0.33\textwidth}\centering
        \includegraphics[width=\linewidth]{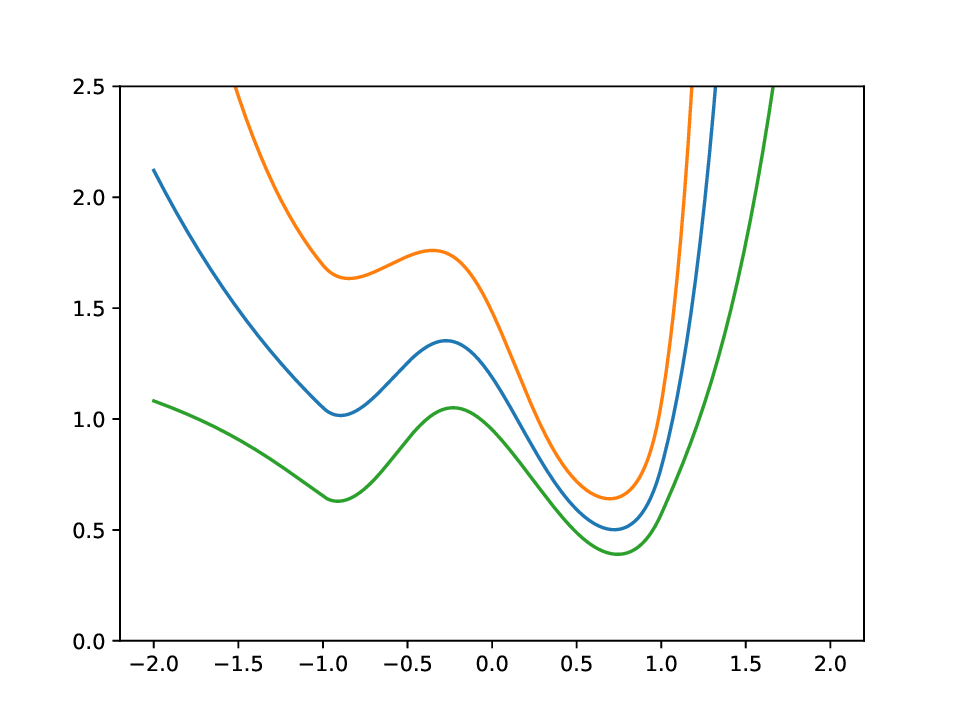}
        \footnotesize 120 days TVS SDF
        \end{minipage} &
        \begin{minipage}[t]{0.33\textwidth}\centering
        \includegraphics[width=\linewidth]{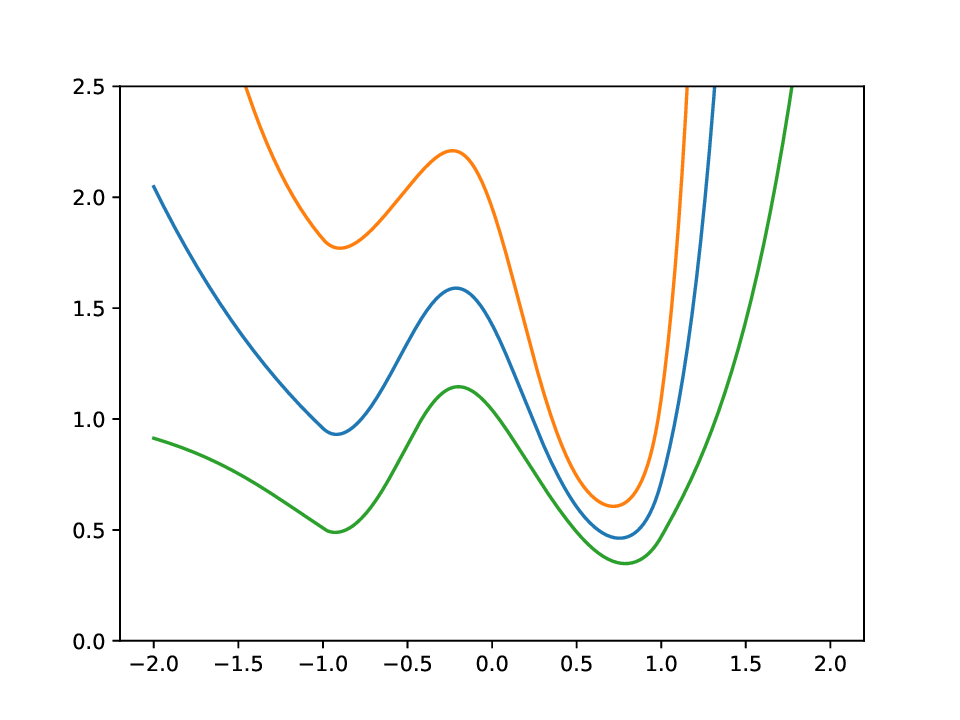}
        \footnotesize 180 days TVS SDF
        \end{minipage} &
        \begin{minipage}[t]{0.33\textwidth}\centering
        \includegraphics[width=\linewidth]{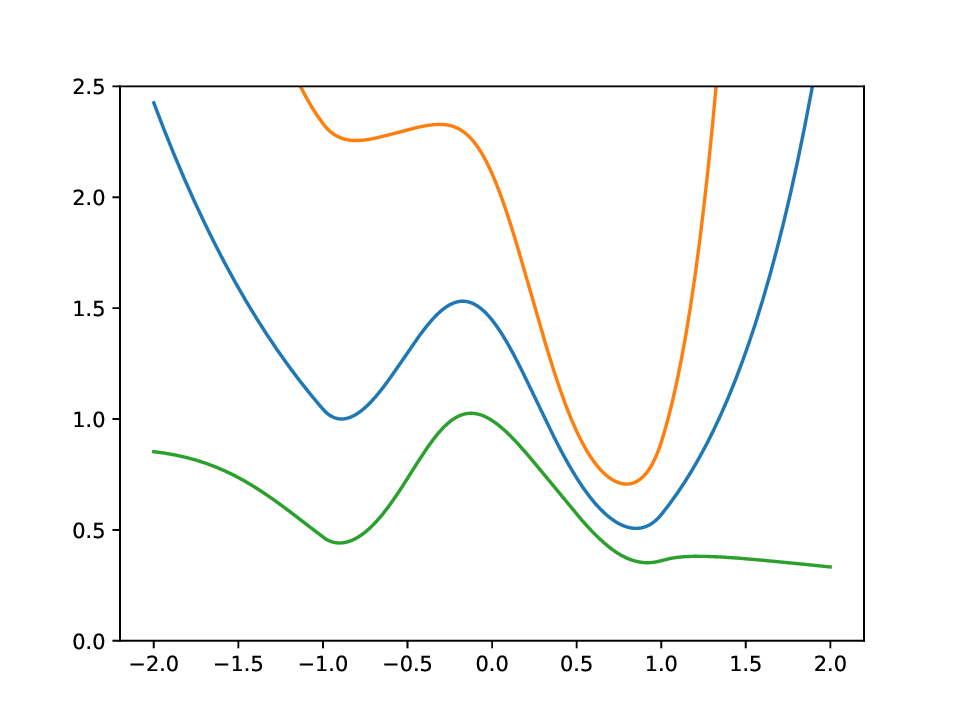}
        \footnotesize 270 days TVS SDF
        \end{minipage} \\
        \begin{minipage}[t]{0.33\textwidth}\centering
        \vspace{0.25em}
        \end{minipage} &
        \begin{minipage}[t]{0.33\textwidth}\centering
        \includegraphics[width=\linewidth]{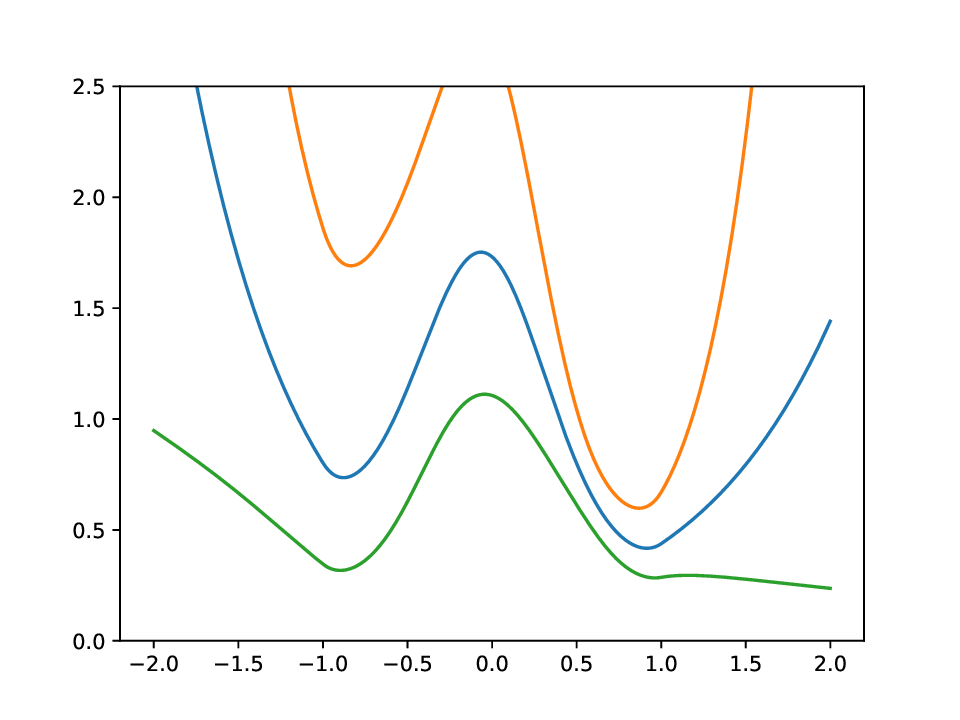}
        \footnotesize 360 days TVS SDF
        \end{minipage} &
        \begin{minipage}[t]{0.33\textwidth}\centering
        \vspace{0.25em}
        \end{minipage} \\
    \end{tabular}
    \caption{Estimated TVS SDF with Full-Sample Data}
    \label{fig:AllTermTVSSDF}
    \begin{quote}
    \small
    \textit{Notes}: This table displays the estimated functional forms of the TVS SDF, $\exp\{f_{p,n}(z)\}$, across different maturities.
    In each plot, the horizontal axis represents the normalized log return $z$.
    The blue line depicts the estimated SDF, while the orange and green lines represent the upper and lower bounds of the $2\sigma$ confidence interval, respectively.
    All results are based on the full sample data from Jan 1996 to Feb 2023.
    \end{quote}
  \end{minipage}
\end{figure}

Here, we examine and discuss the shape of the SDF.
The target SDFs are the estimated fourth-order polynomial specification identified in the previous chapter
as the most stable performer in forecasting realized ERs using the expected EP implied by the estimated SDF.
The estimated CVS and TVS SDFs of each maturity are illustrated in Figures \ref{fig:AllTermCVSSDF} and \ref{fig:AllTermTVSSDF}.  
In these figures, the horizontal axis shows the normalized log return $z$.
The blue line depicts the estimated functional form of SDF, $\exp\{f_{p,n}(z)\}$,
while the orange and green lines represent the upper and lower bounds of the $2\sigma$ confidence interval, respectively.
The $\Sigma$ for \eqref{formula:FuncVariance} is calculated using \textcite{newey1987simple} with the lag length set to 1.5 times the maturity in months and rounded up.

First, we confirm that both the CVS SDF and the TVS SDF exhibit similar estimated shapes.
Across all maturities, both SDFs display an upward-sloping pattern toward the outside on both the call and put sides.
From 30 to 360 days, the hump around the ATM and shallow-put region becomes more pronounced gradually,
and a clear W-shaped pattern emerges as maturity increases.
In addition, the slope of the put side tends to be flatten at short maturities.
These findings support the characterization of the SDF shape as W-shaped.
\textcite{cuesdeanu2018pricing} report that an SDF hump around ATM is observed during calm periods, whereas a U-shaped SDF appears in periods of high uncertainty.
Our results add to this literature by showing that W-shapes are more likely to be observed at longer maturities.

Looking at the estimated coefficients in Tables \ref{table:SDFEstimationFullCVS} and \ref{table:SDFEstimationFullTVS},
we can draw several economic and structural inferences regarding the shape of the SDF.
First, the first-order coefficient $a_1$ is negative across all maturities and for both CVS and TVS specifications.
Since $a_1$ primarily governs the global slope of the SDF, its consistently negative value indicates that the estimated SDF is generally decreasing in the return dimension,
which is consistent with the presence of risk aversion in the market.
Second, the second-order coefficient $a_2$ represents the local curvature of the log-SDF around the ATM region.
In Tables \ref{table:SDFEstimationFullCVS} and \ref{table:SDFEstimationFullTVS},
we observe that $a_2$ is negative for most maturities, except for the 90 day CVS and 60 day TVS.
This negative $a_2$ implies that the log-SDF is concave around the ATM,
indicating a lack of volatility aversion in this region and resulting in the hump shape observed in Figures \ref{fig:AllTermCVSSDF} and \ref{fig:AllTermTVSSDF}.
The formation of the W-shape, however, depends not only on the polynomial coefficients within the interval $[-1, 1]$ but also on the linear extrapolation beyond these endpoints.
While a positive $a_4$ contributes to the upward curvature within the interval, the tails of the W-shape are determined by the first derivatives at the boundaries,
$f_{p,n}^{(1)}(-1)$ and $f_{p,n}^{(1)}(1)$.
In our estimation, $f_{p,n}^{(1)}(-1)$ is consistently negative and $f_{p,n}^{(1)}(1)$ is consistently positive across almost all maturities.
This means that the log-SDF is decreasing for $z < -1$ and increasing for $z > 1$.
Combined with the ATM hump, these boundary slopes ensure that the SDF rises toward both the deep-put and deep-call sides,
structurally completing the W-shaped profile.
This configuration suggests a complex risk preference:
while the negative $a_2$ indicates a local departure from standard risk-averse behavior near the ATM,
the sign of the boundary derivatives confirms that the market demands a high risk premium for extreme tail events.

Several interrelated factors may explain why the SDF takes this shape.
Each of the explanations discussed below accounts for only a portion of the SDF's overall shape when considered individually,
but we believe that the combination of these mechanisms can collectively explain the full structure of the SDF.

Fundamentally, risk-averse investors have a downward-sloping SDF, which implies that the SDF increases on the put side.
In addition, we consider that the SDF exhibits a more complex shape as a result of the partial combination of the effects listed below.
For increasing of call side, \textcite{bakshi2010returns} adopt a model based on heterogeneity in investors' beliefs about future returns and argue that the increasing region of the SDF arises from the behavior of risk-averse investors who short the equity market.
For the hump of shallow put part, as shown for example in \textcite{bollen2004does},
investor demand is high for puts from ATM to shallow OTM and these options become relatively expensive.
In this situation the risk-neutral probability rises and the SDF increases as well.

\begin{figure}[htbp]
  \centering
  \begin{minipage}{0.48\textwidth}
    \centering
    \includegraphics[width=\textwidth]{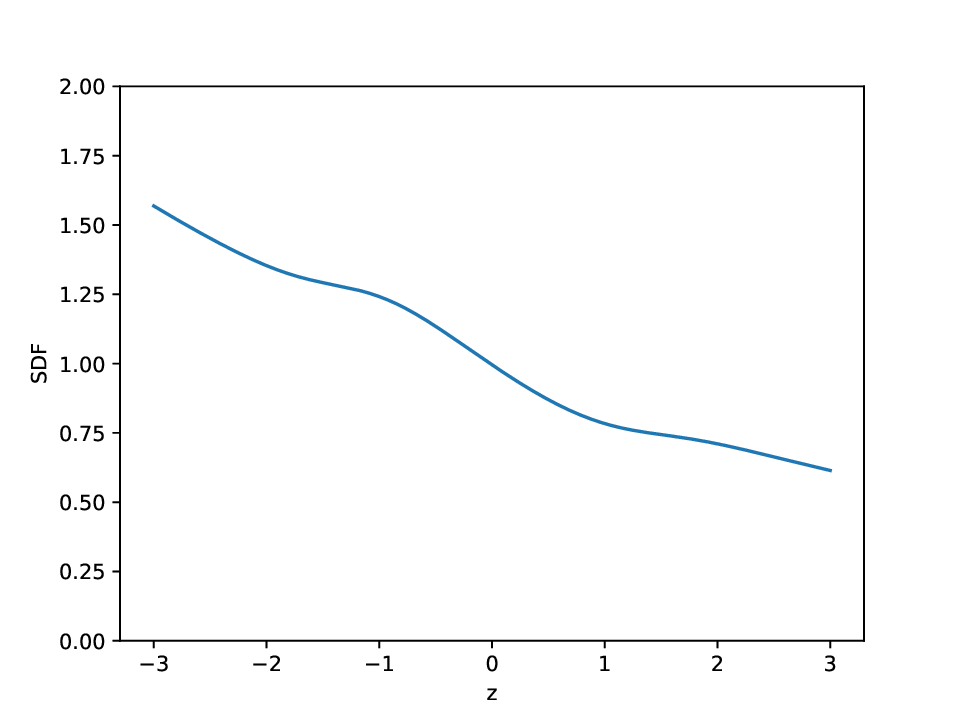}
    \footnotesize Narrow Volatility Dispersion
  \end{minipage}
  \hfill
  \begin{minipage}{0.48\textwidth}
    \centering
    \includegraphics[width=\textwidth]{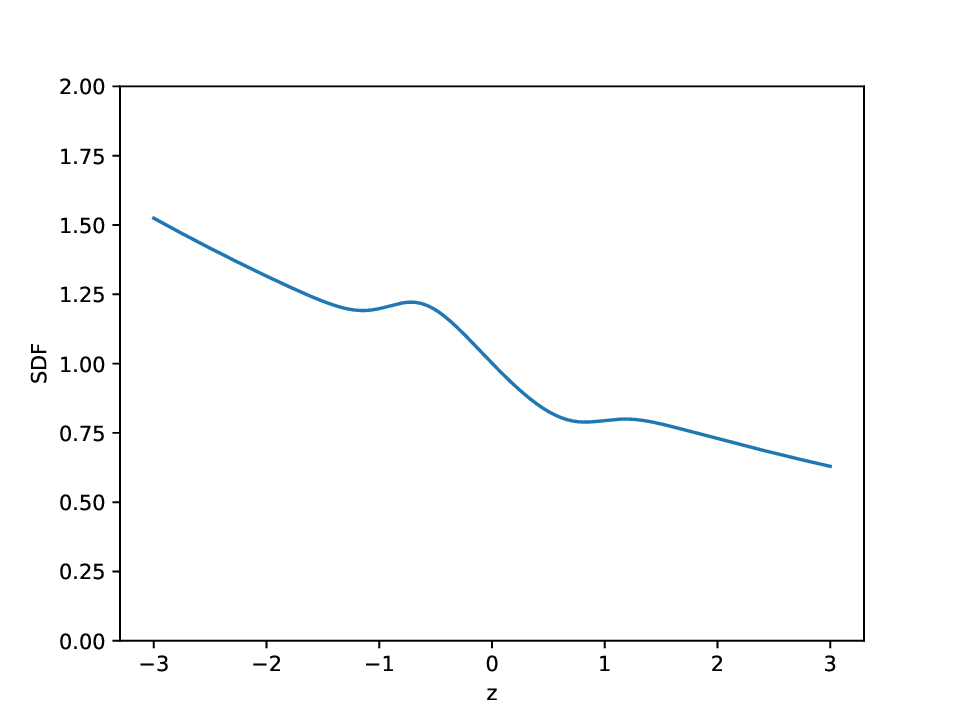}
    \footnotesize Wide Volatility Dispersion
  \end{minipage}
  \caption{Shape of the Model SDF by the dispersion of the volatility distribution}
  \label{fig:ModelSDF}
  \begin{quote}
  \small
  \textit{Notes}: 
  An example of the shape of the SDF implied by equation \eqref{eq:VarVolSDF} with $N=2$.
  The parameters of left figure are $\alpha_{1}=1.281$, $\alpha_{2}=0.6$,
  and those of right figure are $\alpha_{1}=1.356$, $\alpha_{2}=0.4$.
  The common parameters are $\Theta=0.2$, $\sigma_{t}=0.3$, $p_{1}=p_{2}=0.5$, $T-t=1$.
  We set $\alpha_{1}$ and $\alpha_{2}$ such that the variance risk premium becomes zero.
  In other words, they satisfy $\alpha_{1}^{2}\sigma_{t}^{2}p_{1} + \alpha_{2}^{2}\sigma_{t}^{2}p_{2} = \sigma_{t}^{2}.$
  When we consider small perturbations to $\alpha_{1}$ and $\alpha_{2}$, the similar hump still exists for both positive and negative variance risk premia.
  \end{quote}
\end{figure}

This hump can also be explained from another perspective: modeling the time variation in volatility using a one-period binomial model, which is inspired by the models which have additional state variables (e.g., \parencite{chabi2008state, lundtofte2010implied}).
We also assume a constant market price of risk, which is consistent with the argument by \textcite{SCHREINDORFER2025104106} that the impact of fluctuations in the price of risk is small.
We revisit the SDF within a stochastic process framework \eqref{eq:StochasticProcess} and introduce a simplified time-varying volatility settings.
Specifically, volatility transitions to $\alpha_{1}\sigma_{t}$ or $\alpha_{2}\sigma_{t}$ ($\alpha_{1}>\alpha_{2}>0$) with probabilities $p_{1}$ and $p_{2}=1-p_{1}$ at time $t+\Delta t$, and remains constant at that level until time $T$.
We define the normalized log spot rate as $z_{t,T}=\frac{1}{\sigma_{t}\sqrt{T-t}}\log\left(\frac{S_{T}}{F_{t,T}}\right)$.
Let $p_{t+\Delta t}(z_{t,T};\alpha_{i}), (i=1,2)$ denote the physical probability densities of $z_{t,T}$ immediately following the volatility transition.
Similarly, let $m_{t+\Delta t,T}(z_{t,T};\alpha_{i}), (i=1,2)$ represent the SDFs corresponding to each volatility state.
The projected SDF at time $t$ is then given by
\begin{eqnarray}
m_{t,T}(z_{t,T})&=&\frac{m_{t+\Delta t,T}(z_{t,T};\alpha_{1})p_{t+\Delta t}(z_{t,T};\alpha_{1})p_{1}+m_{t+\Delta t,T}(z_{t,T};\alpha_{2})p_{t+\Delta t}(z_{t,T};\alpha_{2})p_{2}}{p_{t+\Delta t}(z_{t,T};\alpha_{1})p_{1}+p_{t+\Delta t}(z_{t,T};\alpha_{2})p_{2}}.\label{eq:VarVolSDF}
\end{eqnarray}
Along with a constant market price of risk $(\lambda_{t}=\lambda>0)$, we obtain:
\begin{eqnarray}
p_{t+\Delta t}(z_{t,T};\alpha)&=&\frac{1}{\alpha}\phi\left(\frac{z_{t,T}}{\alpha}-\lambda\sqrt{T-t}+\frac{1}{2}\alpha\sigma_{t}\sqrt{T-t}\right)\\
m_{t+\Delta t,T}(z_{t,T};\alpha)&=&B_{t,T}^{-1}\exp\left\{\tfrac{1}{2}\lambda(\lambda-\alpha\sigma_{t})(T-t)\right\}
\exp\left\{-\frac{\lambda}{\alpha}z_{t,T}\right\}.
\end{eqnarray}
Since $p_{t+\Delta t}(z_{t,T};\alpha_{1}) \gg p_{t+\Delta t}(z_{t,T};\alpha_{2})$ in the region where $|z_{t,T}| \gg 1$, $m_{t,T}(z_{t,T})$ asymptotically converges to $m_{t+\Delta t,T}(z_{t,T};\alpha_{1})$.
Furthermore, both $m_{t+\Delta t,T}(z_{t,T};\alpha_{1})$ and $m_{t+\Delta t,T}(z_{t,T};\alpha_{2})$ are monotonically decreasing, with $m_{t+\Delta t,T}(z_{t,T};\alpha_{2})$ exhibiting a steeper slope than $m_{t+0,T}(z_{t,T};\alpha_{1})$.
Since the projected SDF $m_{t,T}(z_{t,T})$ is expressed as a weighted average of the post-transition SDFs, there exists a region near the ATM level where the slope is more negative (i.e., steeper) than that of $m_{t+0,T}(z_{t,T};\alpha_{1})$.
Consequently, when $\alpha_{1}\gg\alpha_{2}$, implying a significant dispersion in volatility between $t$ and $T$, there exists a region on the put side where the SDF becomes locally concave, potentially exhibiting a hump shape as illustrated in Figure \ref{fig:ModelSDF}.

A hump is observed in the shallow put region for the most figures in Figures \ref{fig:AllTermCVSSDF} and \ref{fig:AllTermTVSSDF}.
These patterns resemble put side of Figure \ref{fig:ModelSDF}.
Our simple time-varying volatility model implies that these shapes emerge due to variations in volatility over the period to maturity.
When the dispersion of the volatility distribution -- namely, the spread of possible volatility levels between $t$ and $T$ -- is small, the curve remains monotonic but exhibits a change in slope.
When this distribution becomes wide, the put side develops a tilde-shaped hump.
The estimated SDF indicates that the former case arises at short maturities, whereas the latter appears at long maturities.
This maturity dependence can be understood through volatility persistence.
Over short horizons, persistence keeps volatility close to its current level,
leaving the distribution of future volatility relatively narrow.
By contrast, a longer horizon allows uncertainty about future volatility to build up,
resulting in a wider distribution of possible volatility levels between $t$ and $T$.
Consequently, this maturity-driven widening of the volatility distribution provides a natural explanation for the transition from a monotonic shape to a hump-shaped pattern.
Although Figure \ref{fig:ModelSDF} corresponds to the case with a zero variance premium,
the continuity of the shape with respect to the parameters implies that this hump arises regardless of whether the variance premium is positive or negative.

This logic explains the shape of the SDF through volatility fluctuations between time $t$ and $T$, and is therefore not directly related to whether the SDF is scaled by the volatility observed at time $t$;
whether one uses the CVS SDF or the TVS SDF.
For this reason, the fact that both the CVS and TVS estimations yield similar shapes is fully consistent with this explanation.
However, because the TVS SDF applies volatility scaling in a more internally consistent manner,
it enables more accurate estimation, which in turn likely contributes to its superior performance relative to the CVS SDF.

\subsection{Consistency with Realized Returns and the Rational Expectations Hypothesis}

Next, we examine whether the EP obtained from estimated SDFs are consistent with realized ERs.
For each maturity, we regress the realized ERs on the explanatory variables given by the EP estimates from the M-Bound, CYL-Bound and out-of-sample CVS and TVS EPs of fourth-order polynomial.
The regression results are reported in Table \ref{tab:EPLinearRegression}.  
``Intercept'' and ``Slope'' denote the estimated regression coefficients.
To evaluate the statistical reliability of these estimates,
we report the upper and lower bounds of the 95\% confidence intervals (``CI Upper'' and ``CI Lower''), respectively.
The coefficient of determination $R^{2}$ is denoted by ``R2''.
All $t$-statistics of regression coefficients are calculated using the method of \textcite{newey1987simple} with the lag length set to 1.5 times the days to maturity.

\begin{table}
\centering
\scriptsize
\caption{Linear Regression of Realized Returns on Estimated EP (Out-of-Sample Data)}
\label{tab:EPLinearRegression}
\begin{tabular}{lllllllll}
\toprule
 &  & 30 & 60 & 90 & 120 & 180 & 270 & 360 \\
Type & Metric &  &  &  &  &  &  &  \\
\midrule
\multirow[t]{7}{*}{M-Bound} & Intercept & -0.015 & -0.014 & -0.024 & -0.027 & -0.021 & -0.011 & 0.006 \\
 & Intercept Upper & 0.055 & 0.061 & 0.064 & 0.070 & 0.083 & 0.081 & 0.073 \\
 & Intercept Lower & -0.085 & -0.089 & -0.112 & -0.124 & -0.125 & -0.103 & -0.062 \\
\cline{2-9}
 & Slope & 3.896 & 3.612 & 3.729 & 3.730 & 3.558 & 3.339 & 3.005 \\
 & Slope Upper & 5.179 & 4.939 & 5.573 & 5.951 & 5.978 & 5.412 & 4.292 \\
 & Slope Lower & 2.614 & 2.286 & 1.886 & 1.509 & 1.139 & 1.267 & 1.718 \\
\cline{2-9}
 & R2 & 0.066 & 0.098 & 0.136 & 0.162 & 0.175 & 0.208 & 0.256 \\
\cline{1-9}
\multirow[t]{7}{*}{CYL-Bound} & Intercept & -0.002 & 0.002 & -0.004 & -0.006 & -0.000 & 0.007 & 0.023 \\
 & Intercept Upper & 0.065 & 0.072 & 0.074 & 0.079 & 0.094 & 0.094 & 0.085 \\
 & Intercept Lower & -0.069 & -0.068 & -0.081 & -0.091 & -0.095 & -0.079 & -0.040 \\
\cline{2-9}
 & Slope & 3.095 & 2.621 & 2.520 & 2.435 & 2.220 & 2.022 & 1.759 \\
 & Slope Upper & 4.091 & 3.560 & 3.732 & 3.895 & 3.857 & 3.417 & 2.541 \\
 & Slope Lower & 2.100 & 1.681 & 1.308 & 0.975 & 0.583 & 0.627 & 0.978 \\
\cline{2-9}
 & R2 & 0.065 & 0.095 & 0.131 & 0.155 & 0.166 & 0.198 & 0.245 \\
\cline{1-9}
\multirow[t]{7}{*}{CVS EP} & Intercept & 0.127 & 0.097 & 0.056 & -0.033 & 0.047 & 0.070 & 0.057 \\
 & Intercept Upper & 0.288 & 0.244 & 0.250 & 0.256 & 0.280 & 0.360 & 0.317 \\
 & Intercept Lower & -0.033 & -0.051 & -0.137 & -0.323 & -0.186 & -0.219 & -0.203 \\
\cline{2-9}
 & Slope & -0.062 & 0.326 & 0.870 & 1.442 & 0.923 & 0.660 & 0.984 \\
 & Slope Upper & 2.388 & 2.705 & 3.814 & 4.349 & 4.080 & 4.184 & 4.335 \\
 & Slope Lower & -2.512 & -2.054 & -2.073 & -1.464 & -2.234 & -2.863 & -2.367 \\
\cline{2-9}
 & R2 & 0.000 & 0.001 & 0.007 & 0.026 & 0.023 & 0.012 & 0.019 \\
\cline{1-9}
\multirow[t]{7}{*}{TVS EP}  & Intercept & -0.028 & -0.089 & -0.077 & -0.063 & 0.010 & -0.066 & -0.142 \\
 & Intercept Upper & 0.112 & 0.086 & 0.125 & 0.150 & 0.251 & 0.210 & 0.046 \\
 & Intercept Lower & -0.167 & -0.263 & -0.280 & -0.275 & -0.231 & -0.343 & -0.331 \\
\cline{2-9}
 & Slope & 2.221 & 2.856 & 2.659 & 2.067 & 1.473 & 2.463 & 3.233 \\
 & Slope Upper & 4.495 & 5.203 & 5.490 & 4.612 & 4.929 & 6.177 & 5.439 \\
 & Slope Lower & -0.052 & 0.509 & -0.172 & -0.479 & -1.984 & -1.252 & 1.027 \\
\cline{2-9}
 & R2 & 0.018 & 0.052 & 0.073 & 0.084 & 0.043 & 0.093 & 0.203 \\
\cline{1-9}
\bottomrule
\end{tabular}
\begin{quote}
\small
\textit{Notes}: This table reports the results of out-of-sample linear regressions of realized ERs on the estimated EP: $R_{t,t+\tau} = \alpha + \beta \widehat{EP}_{t,t+\tau} + \epsilon_{t,t+\tau}$.
The evaluation period begins on December 18, 2009, following the first expanding-window SDF estimation.
``Intercept'' and ``Slope'' denote the estimated regression coefficients $\alpha$ and $\beta$.
``CI Upper'' and ``CI Lower'' represent the 95\% confidence interval bounds.
``R2'' is the coefficient of determination.
The EP estimates are derived from the M-Bound, CYL-Bound, CVS EP and TVS EP.
\end{quote}
\end{table}

For the all EPs, the intercept estimates are generally close to zero and their intervals typically include zero,
so there is no strong evidence of a systematic level bias.
For the CYL-Bound, the 95\% confidence intervals for the slope include one at horizons of 120 days and longer,
so the rational expectations hypothesis is not rejected for those maturities.
By contrast, the rational expectations hypothesis is rejected for the other specifications and horizons;
However, the lower bounds of the slope exceed one, indicating that predictability is present for those maturities.
The CVS EP shows virtually little predictive content over the evaluation period.
Its slopes are imprecisely estimated, confidence intervals routinely include zero, and $R^2$ values are near zero.
So it offers little evidence to reject rational expectations and predictability.
The TVS EP yields mixed results: except at the 360 day horizon,
the hypothesis $Slope=1$ is not rejected, so the rational expectations hypothesis is maintained.
However, although high slopes are estimated, the hypothesis $Slope=0$ is not rejected except at 60 and 360 days,
so evidence of predictability is inconclusive.
Also, $R^2$ is lower than for the M-Bound and CYL-Bound but higher than for the CVS EP.

Overall, CVS and TVS EPs produce relatively wide confidence intervals,
which substantially limits the precision of coefficient estimates and makes it difficult to draw sharp economic conclusions.

\section{Conclusion}

In this paper, we developed a novel framework for estimating the SDF from option prices and investigated its effectiveness in forecasting the equity premium.
Our primary contribution is the introduction of the time-varying volatility-scaled SDF (TVS SDF), which incorporates time-varying volatility as a scaling parameter to capture the dynamic nature of market risk preferences.
Furthermore, we applied a crucial adjustment to ensure that the estimated SDF is consistent with observed risk-free rates, thereby enhancing both the theoretical and empirical robustness of the model.
A significant strength of our methodology lies in the precise estimation of the SDF shape.
Our approach utilizes implied market data -- including the underlying asset price, risk-free rate, and dividend yield -- extracted exclusively from option prices.
This design is specifically intended to mitigate noise arising from observation lags and non-synchronicity between markets, thereby allowing us to extract the pure forward-looking expectations and risk attitudes of option market participants.

In the empirical analysis, we successfully recovered a stable SDF across various horizons by employing the TVS SDF.
The precisely estimated SDF exhibits a W-shape or a distinctive ``hump'' at shallow put strikes.
We confirmed that the SDF tends to transition from a hump-shaped to a W-shaped structure as maturity increases,
indicating that the time horizon is a key factor influencing the intensity of the central hump.
While the W-shape likely results from a combination of local factors, we demonstrate that the hump on the shallow put side can be theoretically explained by a model where the market price of risk remains constant while volatility is stochastic.
The empirical results confirm that the equity premium derived from our TVS SDF exhibits superior predictive power for future realized ERs compared to existing benchmarks,
such as the Martin's bound, in terms of out-of-sample $R^2$.
The regression evidence shows that while all EP measures yield intercepts consistent with zero,
the CYL-Bound satisfies the rational expectations restriction at longer horizons whereas the other specifications exhibit slopes whose lower confidence bounds exceed one,
CVS and TVS EPs suffer from wide confidence intervals and low explanatory power that limit inference about predictability,
and overall the imprecision of these estimates prevents drawing sharp economic conclusions.

In conclusion, our findings provide a robust bridge between the cross-sectional information in the options market and the time-series dynamics of the equity premium.
The improved estimation of the SDF shape offers important implications for asset pricing and risk management,
suggesting that properly accounting for market volatility is essential for decoding the risk-neutral signals embedded in option prices.

\printbibliography

\appendix

\section{Estimation of Implied Spot Rate, Interest Rate and Dividend}
\label{sec:RateEstimation}

Option prices move together with the underlying asset price and with variables such as the risk-free rate and the dividend yield.
When these quantities are recorded at different timestamps,
the mismatch can create artificial arbitrage signals such as violations of put-call parity and can distort empirical results.
To avoid this problem we build a consistent dataset by extracting implied rates directly from option prices so that all inputs are aligned in time.
One way to extract the implied risk-free rate from option prices is the box-spread approach (see, e.g., \parencite{van2022risk}).
The idea is to first create a synthetic forward position by combining a put and a call with the same strike,
which isolates the forward value of the underlying asset.
By taking long and short positions in such synthetic forwards at two different strikes,
the payoff becomes a fixed amount at maturity, effectively replicating a discount bond.
The implied risk-free rate is then obtained from the price of this synthetic bond.

For strikes $K_i < K_j$ $(1 \le i < j \le N)$ at which both call and put prices are observed,
\begin{eqnarray}
\tilde{D}_{i,j,t,T}
= \frac{\left(P_{t,T}(K_j) - C_{t,T}(K_j)\right) - \left(P_{t,T}(K_i) - C_{t,T}(K_i)\right)}{K_{j} - K_{i}},
\end{eqnarray}
has the same payoff as a discount bond.
We therefore treat this synthetic discount bond price as the implied risk-free rate.

In practice, the discount bond price implied by a box spread is sensitive to the choice of strike pairs $(K_{i}, K_{j})$.
To mitigate this instability, we consider the set $A$ of all admissible strike pairs $(i,j)$ and define
\begin{eqnarray}
\tilde{D}_{t,T}
= \operatorname{median}_{(i,j)\in A} \left( \tilde{D}_{i,j,t,T} \right),
\end{eqnarray}
using the median across strike pairs as our estimate of the discount bond price.
As \textcite{van2022risk} note, taking the median corresponds to the Theil--Sen estimator,
which is known to provide robust estimates even in the presence of large outliers in the data.
The implied risk-free rate is obtained as
\begin{eqnarray}
\tilde{r}_{t,T}
= -\frac{1}{T-t}\log(\tilde{D}_{t,T}).
\end{eqnarray}

To estimate the underlying asset price $S_t$, we use the relation
\begin{eqnarray}
\log(S_t) + q(T-t) = \log\!\left( C_{t,T}(K) - P_{t,T}(K) + K D_{t,T} \right),
\end{eqnarray}
which holds when the dividend yield $q$ is constant. Using option data with maturities $T$ between 10 days and 1.25 years and all available strikes $K_i$, we run a linear regression on the data pairs
$\left(T-t,\log\!\left( C_{t,T}(K_i) - P_{t,T}(K_i) + K_i \tilde{D}_{t,T} \right)\right)$ and take the exponential of the intercept as the estimate of the underlying asset price, denoted by $\tilde{S}_t$.

The dividend yield is obtained as follows. The quantity
\begin{eqnarray}
\tilde{D}^{q}_{i,j,t,T} = \frac{\frac{P_{t,T}(K_i) - C_{t,T}(K_i)}{K_i} - \frac{P_{t,T}(K_j) - C_{t,T}(K_j)}{K_j}}{\frac{1}{K_i} - \frac{1}{K_j}}
\end{eqnarray}
replicates a payoff of one unit of the underlying asset at maturity. Using the previously estimated underlying price $\tilde{S}_t$, we define
\begin{eqnarray}
\tilde{D}^{q}_{t,T}= \operatorname{median}_{(i,j)\in A}\left( \tilde{D}^{q}_{i,j,t,T} \right),
\qquad \tilde{q}_{t,T} = -\frac{1}{T-t}\log\!\left(\frac{\tilde{D}^{q}_{t,T}}{\tilde{S}_t}\right),
\end{eqnarray}
and treat $\tilde{q}_{t,T}$ as the implied dividend yield.

\section{Numerical Integration Scheme}
\label{sec:NumericalIntegrationScheme}

We consider a numerical integration scheme to approximate $\int_{x_{0}}^{x_{N}} g^{(2)}(x) f(x)\, dx.$
Let \(\Delta = \frac{x_{N}-x_{0}}{N}\), and define the equally spaced grid
$x_{i} = x_{0} + i\Delta \quad (i=-1,0,\ldots,N,N+1).$
Using central differences for the second derivative and the trapezoidal rule for integration, we obtain
\begin{equation}
\begin{aligned}
\int_{x_{0}}^{x_{N}} g^{(2)}(x) f(x)\, dx
&\simeq \sum_{i=1}^{N-1} \frac{g(x_{i+1}) - 2g(x_{i}) + g(x_{i-1})}{\Delta^{2}} f(x_{i}) \Delta \\
&\quad + \frac{1}{2}\left\{ \frac{g(x_{N+1}) - 2g(x_{N}) + g(x_{N-1})}{\Delta^{2}} f(x_{N})
+ \frac{g(x_{1}) - 2g(x_{0}) + g(x_{-1})}{\Delta^{2}} f(x_{0}) \right\}\Delta \\
&= \left\{\frac{g(x_{N+1}) - g(x_{N-1})}{2\Delta}\right\} f(x_{N})
- \left\{\frac{g(x_{1}) - g(x_{-1})}{2\Delta}\right\} f(x_{0}) \\
&\quad - \Delta \sum_{i=1}^{N} \frac{g(x_{i}) - g(x_{i-1})}{\Delta} \cdot \frac{f(x_{i}) - f(x_{i-1})}{\Delta}\\
&\simeq\bigl[g^{(1)}(x) f(x)\bigr]_{x_{0}}^{x_{N}} - \int_{x_{0}}^{x_{N}} g^{(1)}(x) f^{(1)}(x)\, dx.
\end{aligned}
\end{equation}
Therefore, even when $g$ is not twice differentiable, as long as both $f$ and $g$ are once differentiable,
this numerical integration scheme provides a valid approximation in the sense of integration by parts.

\end{document}